\documentclass[onecolumn,amsmath,amssymb,11pt,superscriptaddress,nofootinbib]{revtex4}
\pdfoutput=1
\usepackage[latin1]{inputenc}
\usepackage[english]{babel}
\usepackage{amssymb}
\usepackage{amsmath}
\usepackage{amsthm}
\usepackage[]{graphicx}
\usepackage[]{subfigure}
\usepackage{tensor}
\usepackage{color}
\usepackage{cancel}
\usepackage{setspace}
\usepackage{fancyhdr}
\usepackage{framed}
\usepackage{tikz}
\usepackage[bookmarks,linktocpage, colorlinks=true, plainpages = false, citecolor = blue,  linkcolor=blue, urlcolor = blue, filecolor = blue]{hyperref} 
\renewcommand\Re{\operatorname{Re}}
\renewcommand\Im{\operatorname{Im}}

\begin{document}

\allowdisplaybreaks
\begin{titlepage}

\title{No Rescue for the No Boundary Proposal: \\
Pointers to the Future of Quantum Cosmology}

\author{Job Feldbrugge}
\email{jfeldbrugge@perimeterinstitute.ca}
\affiliation{Perimeter Institute, 31 Caroline St N, Ontario, Canada}
\author{Jean-Luc Lehners}
\email{jlehners@aei.mpg.de}
\affiliation{Max--Planck--Institute for Gravitational Physics (Albert--Einstein--Institute), 14476 Potsdam, Germany}
\author{Neil Turok}
\email{nturok@perimeterinstitute.ca}
\affiliation{Perimeter Institute, 31 Caroline St N, Ontario, Canada}
\begin{abstract}
\noindent 
In recent work~\cite{Feldbrugge:2017kzv,Feldbrugge:2017fcc}, we introduced Picard-Lefschetz theory as a tool for defining the Lorentzian path integral for quantum gravity in a systematic semiclassical expansion. This formulation avoids several pitfalls occurring in the Euclidean approach. Our method provides, in particular, a more precise formulation of the Hartle-Hawking no boundary proposal, as a sum over real Lorentzian four-geometries interpolating between an initial three-geometry of zero size, {\it i.e}, a point, and a final three-geometry. With this definition, we calculated the no boundary amplitude for a closed universe with a cosmological constant, assuming cosmological symmetry for the background and including linear perturbations. We found the opposite semiclassical exponent to that obtained by Hartle and Hawking for the creation of a de Sitter spacetime ``from nothing". Furthermore, we found the linearized perturbations to be governed by an {\it inverse} Gaussian distribution, meaning they are unsuppressed and out of control.  Recently, Diaz Dorronsoro {\it et al.}~\cite{Dorronsoro:2017} followed our methods but attempted to rescue the no boundary proposal by integrating the lapse over a different, intrinsically complex contour. Here, we show that, in addition to the desired Hartle-Hawking saddle point contribution, their contour yields extra, non-perturbative corrections which again render the perturbations unsuppressed. We prove there is {\it no} choice of complex contour for the lapse which avoids this problem. We extend our discussion to include backreaction in the leading semiclassical approximation, fully nonlinearly for the lowest tensor harmonic and to second order for all higher modes. Implications for quantum de Sitter spacetime and for cosmic inflation are briefly discussed. 
\end{abstract}

\maketitle

\end{titlepage}

\tableofcontents

\section{Introduction}\label{sec:intro}

The no boundary proposal of Hartle and Hawking represents an attempt to explain the quantum origin of spacetime and provide an initial condition for cosmic inflation \cite{Hawking:1981gb,Hartle:1983ai}. All it apparently requires is:

(i) domination of the energy density by a positive cosmological constant or gently sloping scalar field potential, just as is assumed for inflation,

(ii) a closed (positively curved, compact) universe, and

(iii) that the quantum mechanical amplitude for a given three-geometry $\Sigma$ be given by the Feynman path integral over all compact four-geometries bounded only by $\Sigma$.  

The latter geometrical picture, in particular, offers to realize a hope dating back to the very beginnings of modern cosmology~\cite{Lemaitre:1931zzb,Tryon:1973xi,Brout:1977ix,Vilenkin:1982de}, that the unification of quantum mechanics and general relativity might resolve the big bang singularity and explain the beginning of the universe. The no boundary proposal has furthermore been influential well beyond cosmology, particularly in areas of mathematical physics including holography as well as conformal and topological field theory, where it has been used to motivate and define interesting quantum states. 

However, since its beginnings, the proposal has suffered from the lack of a precise mathematical formulation. In two recent papers, we attempted to rectify this shortcoming~\cite{Feldbrugge:2017kzv,Feldbrugge:2017fcc}. Our starting point is the {\it Lorentzian} path integral for quantum gravity, treated as a low energy, effective field theory within a semiclassical expansion. We argued that in the presence of a positive cosmological constant, the Lorentzian path integral propagator to evolve from a geometry of zero initial size to a given final three-geometry provides a mathematically meaningful definition of the no boundary amplitude. This is precisely the point of view adopted by Vilenkin in his early papers, although he never performed any path integral calculations. Instead, he imposed ``outgoing" boundary conditions on solutions of the Wheeler-DeWitt equation, a prescription which is, however, incomplete when perturbations are considered. Our Lorentzian path integral formulation in contrast allows us to simultaneously handle both the background and perturbations with no ambiguities.

By employing Picard-Lefschetz theory (apparently for the first time, in this context) we are able to uniquely express the Lorentzian path integral as a sum of absolutely convergent steepest descent path integrals. In our first papers, we provided evidence that this approach, based on transparent physical and well-defined mathematical principles, resolves many of the problems which have plagued semiclassical quantum gravity and quantum cosmology for decades. Surprisingly, in this new Lorentzian path integral formulation, Hartle and Hawking's no boundary proposal~\cite{Hawking:1981gb,Hartle:1983ai} and Vilenkin's tunneling proposal~\cite{Vilenkin:1982de,Vilenkin:1983xq} are equivalent~\cite{Feldbrugge:2017kzv,Feldbrugge:2017fcc}. 

\subsection{No boundary de Sitter: Picard-Lefschetz theory} 

We chose to focus on the simplest example of a quantum cosmology, namely the no boundary or ``tunneling from nothing" version of quantum de Sitter spacetime in the closed slicing, performing a semiclassical quantization of both the background and the perturbations. On the positive side, we found unique, well-defined results, free of the diseases such as the ``conformal factor" problem, which plague the Euclidean approach. However, we also found unexpectedly negative results concerning the semiclassical Hartle-Hawking state for quantum fields and fluctuations. 

We claim that the Lorentzian path integral amplitude for a closed universe with a positive cosmological constant $\Lambda$ to emerge ``from nothing"  into a period of de Sitter expansion, acquiring a frozen, dimensionless tensor perturbation $\phi_1$ on the final three-geometry, is given by
\begin{align}
\int {\cal D} g e^{i S[g] /\hbar} \propto e^{-\frac{12\pi^2}{\hbar \Lambda}+ \frac{3}{2\hbar \Lambda}l(l+1)(l+2)\phi_1^2} ,
\label{eq:LPLres}
\end{align}
with $S[g]$ being the usual Einstein-Hilbert-$\Lambda$ action taken in units where $8\pi G=1$. The path integral is taken over all compact four geometries bounded only by the final three-geometry. Here, we omit the functional determinant and a phase representing the late time classical evolution since we wish to focus on the semiclassical weighting factor, given by the real part of the semiclassical exponent. To avoid notational clutter, we also include just one tensor mode, with principal quantum number $l$ and amplitude $\phi_1$ on the final three-geometry, assuming that the physical wavelength of that mode is larger than the de Sitter radius at the final time, so that the mode has dynamically ``frozen out". To quadratic order in the perturbations, any number of modes may be included by simply replacing $l(l+1)(l+2)\phi_1^2$ with $\sum_{lmn} l(l+1)(l+2)\phi_{1,lmn}^2$, where $\phi_{1,lmn}$ are the coefficients in the expansion of the final tensor perturbation in real, orthonormalized spherical harmonics on the three-sphere, with quantum numbers $l,m,n$ (see, {\it e.g.}, \cite{Gerlach:1978gy}). 

The sign of the exponent in (\ref{eq:LPLres}) is the {\it opposite} of that usually associated with the Hartle-Hawking no boundary proposal. Hartle and Hawking obtained their result by considering the Euclidean action for quantum gravity, obtained by performing the usual Wick rotation on quantum fields (including tensor modes) and finding a saddle point representing a Euclidean four-sphere. This procedure recovers the usual Euclidean vacuum for quantum fields at short distances. 

Unfortunately, as we showed in \cite{Feldbrugge:2017kzv}, the Euclidean path integral for the relevant cosmological background is, in the case at hand, a meaningless divergent integral. We avoided that problem by {\it not} Wick rotating: instead, we evaluate the Lorentzian path integral directly. Our main tool is Picard-Lefschetz theory, a powerful and rigorous means of converting an oscillatory, conditionally convergent multidimensional integral into a sum of absolutely convergent, steepest descent integrals. In this case, we find perfectly unambiguous results for both the background and the perturbations. However, because the background scale factor has a kinetic term of the opposite sign to that of the perturbations (and other quantum fields), when we integrate out the background, this in effect imposes a Wick rotation of the opposite sign to the usual one, yielding an inverse Gaussian distribution for quantum fields and implying that the perturbations are out of control. On this basis, we claimed that the no boundary proposal (or its ``tunneling" equivalent) cannot in any way describe the emergence of a realistic cosmology. 

Let us now discuss the two terms in the exponent of (\ref{eq:LPLres}), since there is an interesting story behind each of them. 
The first term comes from integrating out the cosmological background. It is convenient to rewrite the usual FLRW cosmological line element $-dt^2 \bar{N}^2(t)+a(t)^2 d\Omega_3^2$ as $-dt^2 N^2/q(t)+q(t) d\Omega_3^2$, with $q(t)=a(t)^2$ and $d\Omega_3^2$ the standard metric on the unit three-sphere. Fixing a gauge in which the background lapse $N$ is a constant, the path integral over $q(t)$, being Gaussian, may be performed without difficulty. One is left with an ordinary, one-dimensional integral over $N$, given in equation (\ref{eq:prop}) below, with the exponent given in terms of the appropriate classical action (\ref{eq:S0cl}).

\begin{figure}[h]
\begin{center}
\begin{minipage}{0.5\textwidth}
\includegraphics[width=.97\textwidth]{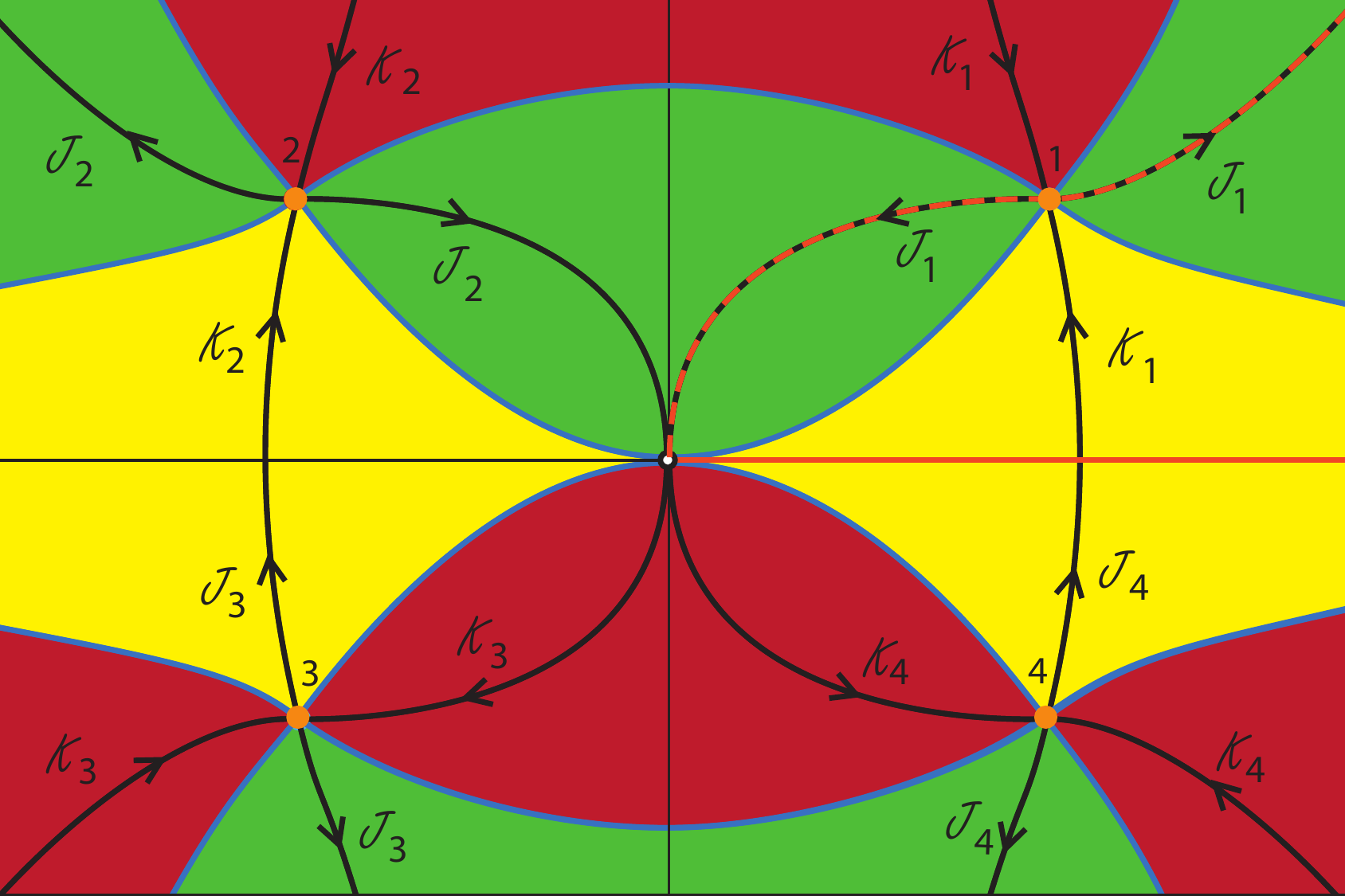}
\end{minipage}
\end{center}
\begin{minipage}{0.5\textwidth}
\includegraphics[width=.97\textwidth]{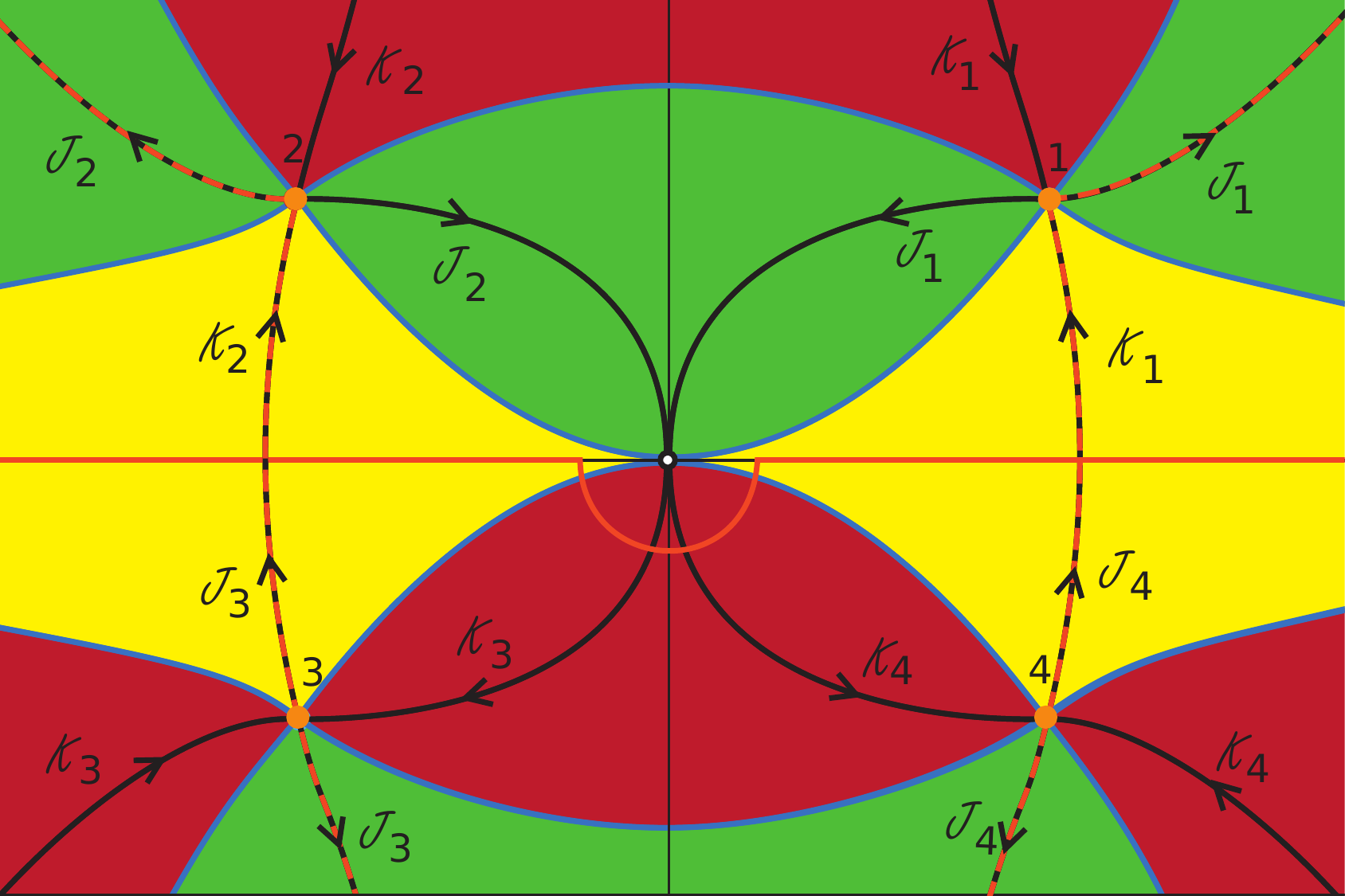}
\end{minipage}%
\begin{minipage}{0.5\textwidth}
\includegraphics[width=.97\textwidth]{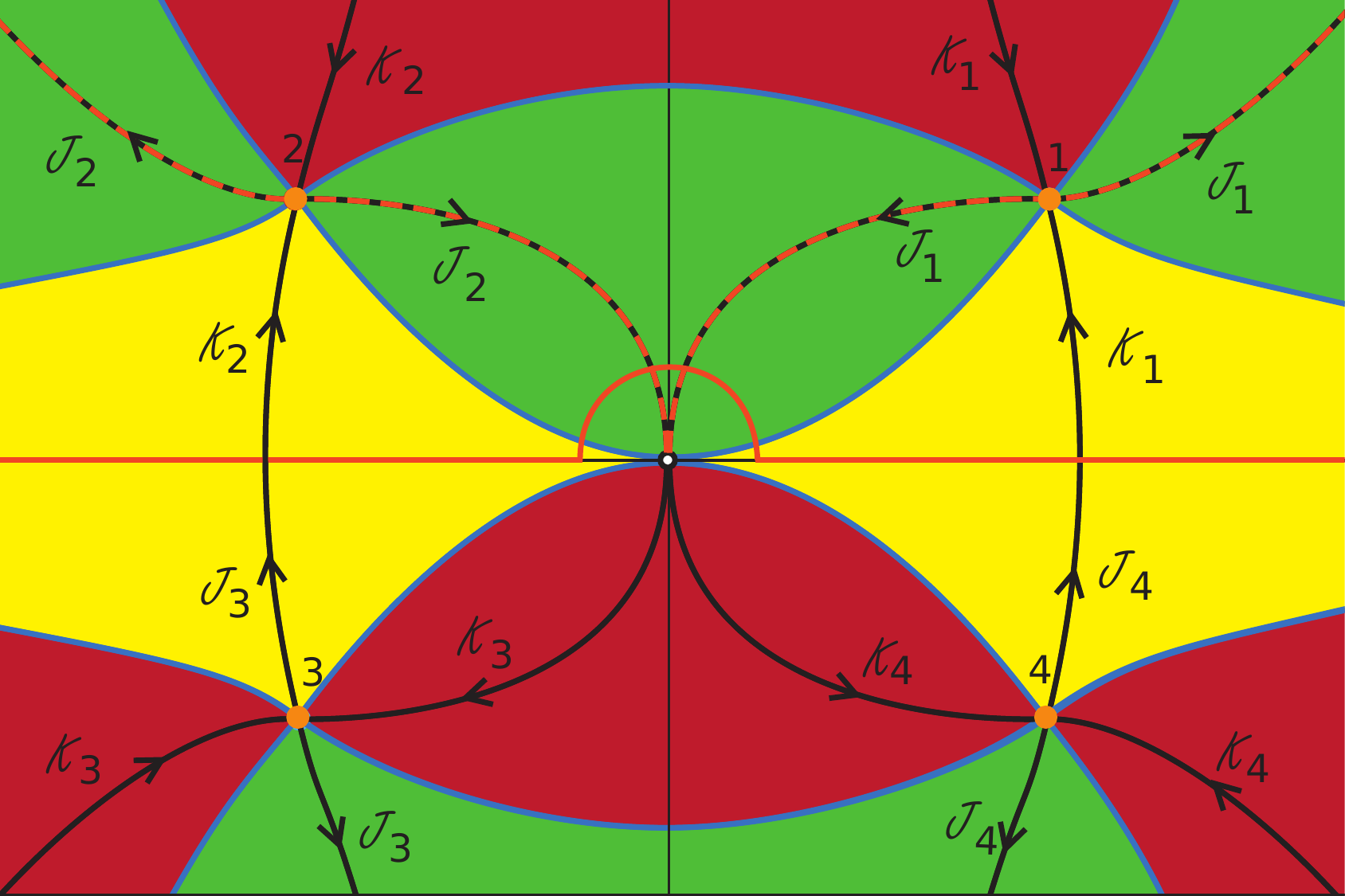}
\end{minipage}%
\caption{The Morse function for the background is plotted in the complex $N$-plane, for a closed, homogeneous and isotropic $\Lambda$-dominated cosmology. The solid orange line is the defining integration contour ${\cal C}$, and the dashed orange line is the corresponding deformed contour, passing along Lefschetz thimbles. As $N$ tends to infinity in the complex $N$-plane, the real part of the exponent in the integrand (the Morse function) tends to $+\infty$ in the red regions or $-\infty$ in the green regions.  It is constant along the blue contours. {\it Upper panel:} the real Lorentzian contour  $0^+<N<\infty$ used for the causal Lorentzian propagator. {\it Lower left panel:} the contour for $N$ running from $-\infty$ to $+\infty$ {\it below} the origin, as proposed by Diaz Dorronsoro \textit{et al.} \cite{Dorronsoro:2017}. {\it Lower right panel:} the real part of the causal propagator, equivalent to a continuous contour for $N$ running from $-\infty$ to $+\infty$ {\it above} the origin.}
\label{fig:background}
\end{figure}

Figure \ref{fig:background} exhibits the structure of the real part of the exponent (the Morse function) in the complex $N$-plane. The orange points indicate $N_\sigma$, the saddle points of the Morse function, which are also (by the Cauchy-Riemann equations) stationary points of the exponent, which is a holomorphic function of $N$. Since the effective action is real for real $N$, these saddles come in complex conjugate pairs. At these saddle points  the four-geometry is a completely regular solution of the complexified Einstein equations. Hartle and Hawking took the two lower saddles, labelled 3 and 4. However, when we define the contour ${\cal C}$ to be that appropriate to the causal Lorentzian propagator, or its complex conjugate, Picard-Lefschetz theory identifies the two conjugate saddles, 1 and 2 respectively, as the relevant ones. The classical action for the upper two saddles is the complex conjugate of that of the lower two saddles, $S_{cl}(N_\sigma^*)=S_{cl}^*(N_\sigma)$. 
Hence their semiclassical weighting factor $|e^{i S_{cl}/\hbar}|=e^{i  (S_{cl}-S^*_{cl})/(2\hbar)}$ is the inverse of the weighting for the two lower ones. 

Once the saddle points are identified, Picard-Lefschetz theory allows us to transform conditionally convergent integrals into absolutely convergent integrals as follows. Each saddle point $\sigma$ is generically the intersection of two contours -- one of steepest descent, labelled ${\cal J}_\sigma$, and one of steepest ascent, labelled ${\cal K}_\sigma$. The real part of the exponent (the Morse function) decreases monotonically on the former and increases monotonically on the latter. The steepest descent and steepest ascent contours are shown in black. The solid orange lines in the Figure show three possible choices we shall consider for the contour ${\cal C}$ over which the integral over $N$ may be taken. In all cases, we take ${\cal C}$ to run from one singularity where the Morse function diverges to $-\infty$  to another.  It is a general result of Picard-Lefschetz theory that in order for a saddle point to be relevant to an integral over ${\cal C}$, the steepest ascent contour from that saddle must intersect ${\cal C}$ (see, {\it e.g.} Ref.~\cite{Feldbrugge:2017kzv}). This being the case, provided the exponent is holomorphic in the relevant region of $N$ one can deform ${\cal C}$ into the complex $N$-plane so that it passes over  ${\cal J}_\sigma$, with Cauchy's theorem ensuring that the value of the integral is preserved. One must also be careful to check that any additional arcs introduced near the two limits of the integral (in our case, near $N=0$ and $N=\infty$) give a vanishing contribution. 

The upper panel in the Figure shows the defining contour for the causal Lorentzian propagator, $0^+<N<\infty$. One can deform this contour by ``sliding it down" the steepest ascent contour ${\cal K}_1$ onto the steepest descent contour ${\cal J}_1$, known as a ``Lefschetz thimble'' (the dashed orange line). In this way one obtains an equal, absolutely convergent integral over $N$. Since a saddle point is relevant if and only if its steepest ascent contour intersects the original integration contour, and since the classical action is real-valued on the real line (so the Morse function is zero there), it follows that the real part of the semiclassical exponent at any relevant saddle must always be negative. As we argued in \cite{Feldbrugge:2017fcc}, this argument is already sufficient to rule out the Hartle-Hawking result. 

How, then, did Hartle and Hawking reach the opposite conclusion? They took the saddle points in the lower-half $N$-plane to be the relevant ones, on the basis that one should reproduce the usual Wick rotation for quantum fields, but they never explicitly performed the path integral over the lapse. First, consider integrating $N_E=iN$ over the infinite real range $-\infty<N_E<\infty$. Any real Euclidean action obtained from a real Lorentzian action is necessarily odd in $N_E$.  Furthermore, if its equations of motion are time-reversal invariant, they are even in $N_E$. Hence, in the absence of any singular behavior,  integrating out the dynamical variables always leaves one with an effective Euclidean action for $N_E$ which is odd in $N_E$. If it diverges to $-\infty$ as $N_E\rightarrow +\infty$, then it diverges to $+\infty$ as $N_E\rightarrow -\infty$, and vice versa. So the semiclassical path integral over all $N_E$ always diverges. Therefore,  in any meaningful semiclassical Euclidean path integral, one simply cannot integrate $N_E$ over all real values. Note that this means that a semiclassical Euclidean path integral {\it cannot} be used to obtain a solution of the homogeneous Wheeler-DeWitt equation, or a "wave function of the universe," as Hartle and Hawking hoped. There are three available options: i) integrate $N_E$ over a half-range, should that integral converge; ii) leave the lapse real and Lorentzian, or iii) deform the lapse integral onto some other complex contour. We consider all three options in this paper, and show none is viable. 

Exploring the first option, if one integrates $N_E$ over positive values, the integral diverges at the origin (in the red region below it, shown in Fig.~\ref{fig:background}). This divergence is due to the essential singularity at $N=0$, which is nothing but the usual one for a quantum mechanical propagator in the limit of short times. For the Einstein action, with the condition that the initial three-geometry has zero size, at small $N$ the propagator behaves as $e^{-i 3\pi^2 q_1^2/(2 \hbar N)}$, where $q_1$ is the value of the scale factor squared on the final three-geometry. The minus sign is unusual and due to the negative kinetic term for the scale factor. Conversely, if one integrates $N_E$ over negative values, it diverges at $-\infty$ in the uppermost red region in the Figure. Hence, there is no Euclidean contour for the lapse which gives a meaningful result. Hence, in our work we reverted to option ii) and integrated over real $0^+<N<+\infty$. As we shall explain, integrating over all real $N$ yields the real part of our answer, so one obtains (\ref{eq:LPLres}) once again. 

\subsection{Hartle-Hawking rescued?} 

In their recent paper, Diaz Dorronsoro {\it et al.} attempted to recover the predictions of the original Euclidean formulation of the no boundary proposal path integral by following our Lorentzian-Picard-Lefschetz approach. They claim to identify a new contour for the lapse, shown as the solid orange contour in the lower left panel of Figure \ref{fig:background}, which recovers both the original Hartle-Hawking weighting for the background and a Gaussian distribution for the fluctuations. Their contour runs from $N=-\infty$ to $N=+\infty$, passing {\it below} the essential singularity at $N=0$. It is immediately apparent that their contour cannot be deformed onto the real $N$-axis, to make the spacetime four-metric real and Lorentzian, because the integrand diverges as one approaches the origin from below. Hence their contour cannot be legitimately termed Lorentzian. 

Diaz Dorronsoro {\it et al.} emphasize that the path integral along their contour is real, despite the contour being complex. This is indeed the case because the Lorentzian action is odd in $N$ and their contour is even under $N\rightarrow -N^*$. Second, they stress that it solves the homogeneous Wheeler-DeWitt equation, whereas our causal Lorentzian propagator does not. Combining these points in a rhetorical flourish, they emphasize that their construction provides a ``real" wavefunction. In section II of this paper, we discuss the basic physical principles underlying the causal Lorentzian propagator in quantum gravity, explaining why it is complex, like the Feynman propagator for a relativistic particle or a string, and why, when the Hamiltonian is applied, it yields $-i$ times a delta function, rather than zero. Should we wish to, we may trivially obtain a ``real" wavefunction (in both senses) from our causal Lorentzian propagator merely by taking its real part. This is equivalent to using the orange contour illustrated in the lower right panel of Figure \ref{fig:background}. Since the integrand becomes exponentially small as the origin is approached from the upper half $N$-plane, one may equivalently describe the contour in terms of two disconnected pieces, $-\infty<N<0^-$ and $0^+<N<\infty$, or over a single complex contour running from $-\infty$ to $\infty$ passing above the origin. The contour proposed by Diaz Dorronsoro {\it et al.}  has no ``real" advantage over the causal, Lorentzian propagator (or its real part) in these terms. As we discuss in Section II, solutions of the homogenous Wheeler-DeWitt equation are arbitrary without further information about the quantum state. In contrast, the causal ``no boundary" propagator as we have described it is in principle unique. 

Nevertheless, let us further investigate their proposed wavefunction. It is not hard to see that the Hartle-Hawking saddles 3 and 4 are indeed relevant to Diaz Dorronsoro {\it et al.}'s proposed contour, as they claim. All we need to do is follow the steepest ascent contours, ${\cal K}_3$ and ${\cal K}_4$ from the saddle points towards the essential singularity at $N=0$. Since they intersect the small orange semicircle below the origin (where the integrand is diverging), Picard-Lefschetz theory tells us they are relevant. The real part of the semiclassical exponent is indeed allowed to be positive, precisely because their defining contour is not Lorentzian. 
However, if we follow the steepest ascent contours ${\cal K}_1$ and ${\cal K}_2$ from the upper two Lorentzian-Picard-Lefschetz saddles, we see that these {\it also} intersect their contour and thus all four saddles are relevant to their contour. As a consequence, their wavefunction includes contributions of the form of (\ref{eq:LPLres}), bringing along with them unsuppressed perturbations. 

The idea of using more general contours for the lapse goes back many years. Halliwell and Louko, in particular, investigated steepest descent contours in de Sitter minisuperspace models, producing contour diagrams very similar to ours~\cite{Halliwell:1988ik}. Halliwell and Hartle further developed the idea, realizing also that certain saddle point solutions would lead to unacceptable quantum field theory distributions~\cite{Halliwell:1989dy}. In fact, they used this very argument against the tunneling proposal. (For a related discussion, from the Wheeler-DeWitt point of view, see also \cite{Rubakov:1999qk}). They seemingly did not, however, appreciate our point here, which is that any contour for $N$, running from one singularity of the Morse function to another and yielding a convergent integral, {\it inevitably} includes contributions from the unacceptable saddle points.  In Section VI of this paper, we shall prove by simple enumeration that {\it no} contour for $N$ avoids the problem of unsuppressed fluctuations. 

\subsection{Perturbation conundrum} 

The observant reader will have noticed a logical conundrum raised by the above arguments. If Picard-Lefschetz theory tells us that the real part of the semiclassical exponent is always negative, how do we explain the dependence of (\ref{eq:LPLres}) on the perturbation amplitude? Clearly, by increasing $\phi_1$, one can make the real part of the exponent arbitrarily positive. Of course, linear perturbation theory breaks down at large $\phi_1$, so one might hope that nonlinearities somehow prevent the second term from ever overcoming the first. This, we shall show in Section V, is not the correct explanation. Instead, something more subtle and interesting is going on. The point is that general relativity is not a regular theory. In particular, time evolution generically allows for the development of singularities. Around these singularities, the perturbations develop unusual, non-analytic behavior. We shall show that this introduces branch cuts into the effective action for the lapse $N$ and breaks the analyticity assumptions underlying the use of Picard-Lefschetz theory.  For the de Sitter model with perturbations, the singularities do not occur on the Picard-Lefschetz thimble for the higher-dimensional theory. Hence they do not introduce any ambiguity into our results. Ratherm they occur on the original, defining contour for the Lorentzian path integral, if one integrates out the background and the perturbations, at real, off-shell values for the lapse $N$, {\it i.e.}, for real, Lorentzian off-shell and singular four-geometries. For the perturbations (and only for the perturbations!), this breakdown of analyticity allows the effective exponent for $N$ to gain a positive real part as one approaches the real $N$-axis from above.  

Before delving into further detail, let us outline the basic steps in our approach. We fix a convenient gauge, in which the lapse $N$ is a constant in time and the perturbations are taken to be transverse-traceless to eliminate unphysical degrees of freedom. Then we perform the Lorentzian path integral for Einstein-$\Lambda$ gravity in three steps. First, we integrate out the radius squared $q(t)$ of the spherical background universe. As we have mentioned, this is a Gaussian path integral presenting no difficulties. Next,  we integrate out the perturbations, treated to quadratic order in general relativistic linear perturbation theory. Finally, we integrate over the lapse $N$. 

Picard-Lefschetz theory plays an important role in ensuring these calculations make sense. Let us start by assuming that we can represent all variables appearing in the path integral as finite sums over Fourier modes in the cosmological time $t$ and, furthermore, that the answer is independent of any UV cutoff appearing in this sum. In this way, the path integrals become ordinary integrals, albeit in high dimension. Picard-Lefschetz theory (and Cauchy's theorem) may now be rigorously used to deform the original, highly oscillatory integral into an equivalent, absolutely convergent integral over a many-dimensional Picard-Lefschetz thimble. Once this has been done, Fubini's theorem (see, {\it e.g.}, \cite{AG:1996}) assures us we may evaluate the high-dimensional integral iteratively as a series of one dimensional integrals, and that the final result will be independent of the order in which those integrals are performed. If, on the contrary, we do not distort all contours to the Picard-Lefschetz thimble before integrating out some variables, we can easily generate singularities on the original, real contour for the remaining variables. An example is provided in Appendix A, showing how such singularities are generated and, equally, how they are avoided in the higher-dimensional Picard-Lefschetz procedure. 

We followed this Picard-Lefschetz  procedure in calculating the causal, Lorentzian propagator in the approximation where we neglected backreaction of the perturbations on the background. It led unambiguously to the result (\ref{eq:LPLres}). How, then, did it generate a positive real, semiclassical exponent where Picard-Lefschetz flow arguments would appear to forbid one? The explanation lies in the fact that, for a range of real but off-shell values of $N$, the background metric develops singularities which lead to non analytic behavior of the perturbations.  The consequence is that integrating out the perturbations generates a pair of finite branch cuts on the real $N$-axis. These give a positive real part to the semiclassical exponent on the upper side of the real $N$-axis, including the points where the steepest ascent contours from saddles 1 and 3 meet the real axis and therefore allowing for those saddle points to contribute positively to the semiclassical exponent. 

\subsection{Resolution: real strong singularities}

It is important to emphasize that the singularities we are discussing only occur at real values of $N$ and {\it do not} occur on the Picard-Lefschetz thimble ${\cal J}_1$ relevant to the causal Lorentzian propagator. Therefore, they introduce no ambiguity in our result (\ref{eq:LPLres}). However, they {\it do} occur on the defining contour for the Lorentzian path integral, after the perturbations are integrated out. In that sense, they are similar to the singularities in the prefactor described in Appendix A, which again occur after a partial integration is performed. However, and this is a key point, in any finite-dimensional Gaussian integral of the form $\int d\vec{x} e^{i (\vec{x}^T O \vec{x} +\vec{u}^T\vec{x})/\hbar}$, where the matrix $O$ is real, symmetric and nonsingular and the vector $\vec{u}$ is real, any number of partial integrations will not alter the imaginary nature of the exponent. Each such integration may be performed as a saddle point integral, and the integrated out variable always assumes a real value at its unique saddle point value. Hence, after any number of partial integrals, the real part of the remaining semiclassical exponent is always equal to zero. Furthermore, if the matrix elements of $O$ are merophorphic in some variable -- in our case, the lapse $N$ -- the exponent will remain meromorphic in that variable. One can never generate branch cuts in the exponent by performing partial integrations. 

\begin{figure}[h]
\includegraphics[width=.8\textwidth]{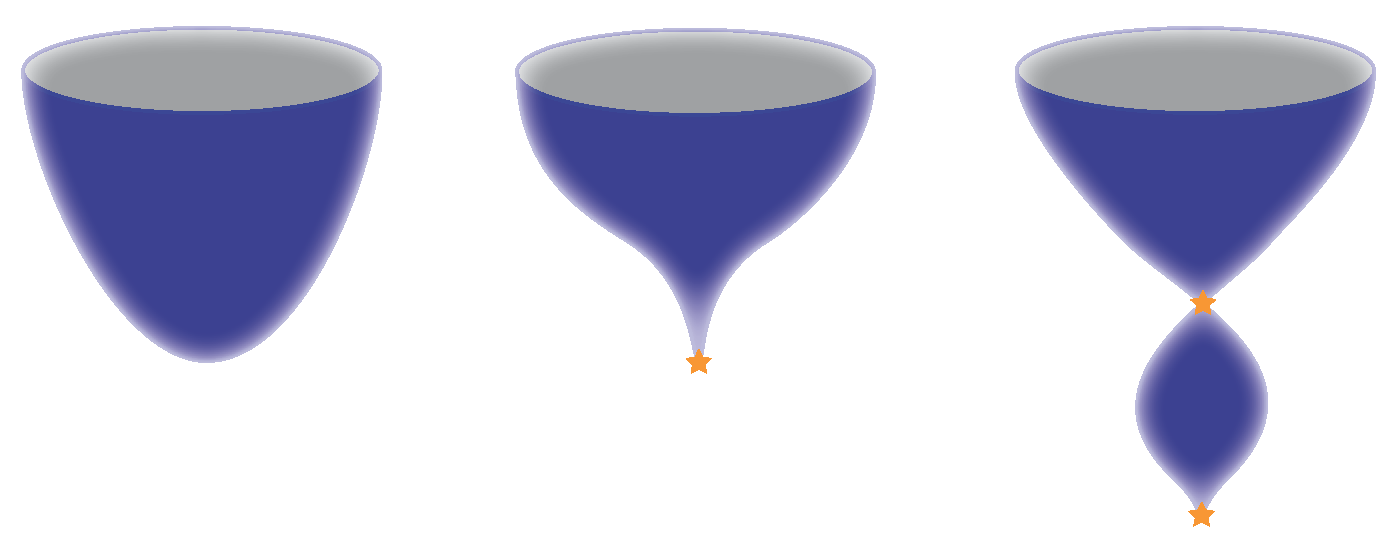}
\caption{The classical background geometries appearing in the no boundary path integral. {\it Left:} The regular, complex saddle point geometry. {\it Middle:} A real Lorentzian off-shell background geometry appearing at $N_-^2\leq N^2\leq N_\star^2$, possessing one strong singularity. {\it Right:} The real Lorentzian geometry appearing at $N^2> N_\star^2$, possessing two strong singularities.}
\label{fig:onoffshell}
\end{figure}

In our case, something different and inherently infinite dimensional takes place. At off-shell, real values of the lapse $N$, the background develops strong singularities. By this we mean that once we integrate out the background variable $q$, for a range of real $N$ the quadratic operator appearing in the action of the perturbations becomes singular. The left panel of Figure \ref{fig:onoffshell} illustrates the complex but regular geometry which appears as a full saddle point of the path integral. However, at real $N$ (and only at real $N$) the background geometries -- stationary in $q$  but off-shell in $N$ -- may exhibit either one strong singularity, if $|N|$ exceeds a critical value $N_-$  or two strong singularities, if it exceeds an even larger value $N_\star$. These cases are illustrated in the middle and right panels of Figure \ref{fig:onoffshell}, respectively. 

To discuss what happens near the singularity at $t=0$, consider momentarily setting the cosmological constant $\Lambda$ to zero. Then the solution to the second order equation of motion for $q$ is simply $q=q_1 t$, where $0\leq t\leq 1$ and $q_1$ denotes the final value. The background line element is $-dt^2 N^2/(q_1 t)+(q_1 t) d\Omega_3^2$, and the background action is given in (\ref{bact}) below. 
For a tensor perturbation mode $\phi$, with principal quantum number $l$ on the 3-sphere, it is convenient to rewrite the action (given in general form in (\ref{eq:S2}) below) and the associated equation of motion in terms of the canonically normalized variable $\chi(t) \equiv q(t) \phi(t)$ as follows:
\begin{align}
S^{(2)}=\int_0^1 dt {1\over 2 N} \left(\dot{\chi}(t)^2+{\gamma^2-1\over 4 \, t^2} \chi(t)^2\right)-{1\over 2 N}\left[{\dot{q}\over q} \chi^2\right]_0^1, \qquad  -\ddot{\chi}(t)  +{\gamma^2-1\over 4 \,t^2} \chi=0, 
\label{schop}
\end{align}
where (for $\Lambda=0$) $\gamma=\sqrt{1-4 l(l+2) N^2/q_1^2}$. Notice the perturbation action and equation of motion are both meromorphic in $N$. However, the two solutions to the equation of motion, $\chi_\pm=t^{{1\over 2}(1\pm \gamma)}$, are not, because $\gamma$ has a branch cut in $N$. First consider real $N$ satisfying $N^2<q_1^2/(4 l(l+2)$, so $\gamma$ is real and smaller than one. While both solutions for $\chi$ vanish at $t=0$, only $\chi_+$ has finite action. Therefore, we take this to be the relevant saddle point solution. Normalizing it to obtain $\phi=\phi_1$ at $t=1$, the classical action for the perturbation is $(\gamma-1)q_1^2\phi_1^2/(4 N)$. We now consider analytically continuing in $N$ to other values in the complex $N$-plane. Evidently, the action has branch points at $N_\pm=\pm q_1/(2\sqrt{l(l+2)})$. It is convenient to draw the branch cuts to run to $\pm\infty$ respectively. As we increase $N$ along the real axis, we must either pass above or below the branch cut. Passing above, the real part of $\gamma$ remains positive but the imaginary part becomes negative. Thus, the real part of the semiclassical exponent, $i S_{cl}(N)/\hbar $ is positive on the real $N$-axis, above the branch cut. Conversely, it is negative below the branch cut. Examining the perturbation solutions for imaginary values of $\gamma$ one sees they undergo an infinite number of oscillations as $t$ tends to zero. This means they cannot be approximated with any finite sum of Fourier modes in $t$, and there is no contradiction with the argument given in the opening paragraph of this subsection.  Notice also that the non-analyticity in the partially integrated exponent arises precisely at values of $N$ on the branch cuts, where the perturbation action fails to select a particular perturbation mode. It is plausible that this is precisely the edge of the wedge of convergence associated with the higher-dimensional Picard-Lefschetz contour, at the limit where the original path integral ceases to be absolutely convergent and hence cannot be performed iteratively. 
 
\begin{figure}[h]
\centering
\includegraphics[width=0.75\textwidth]{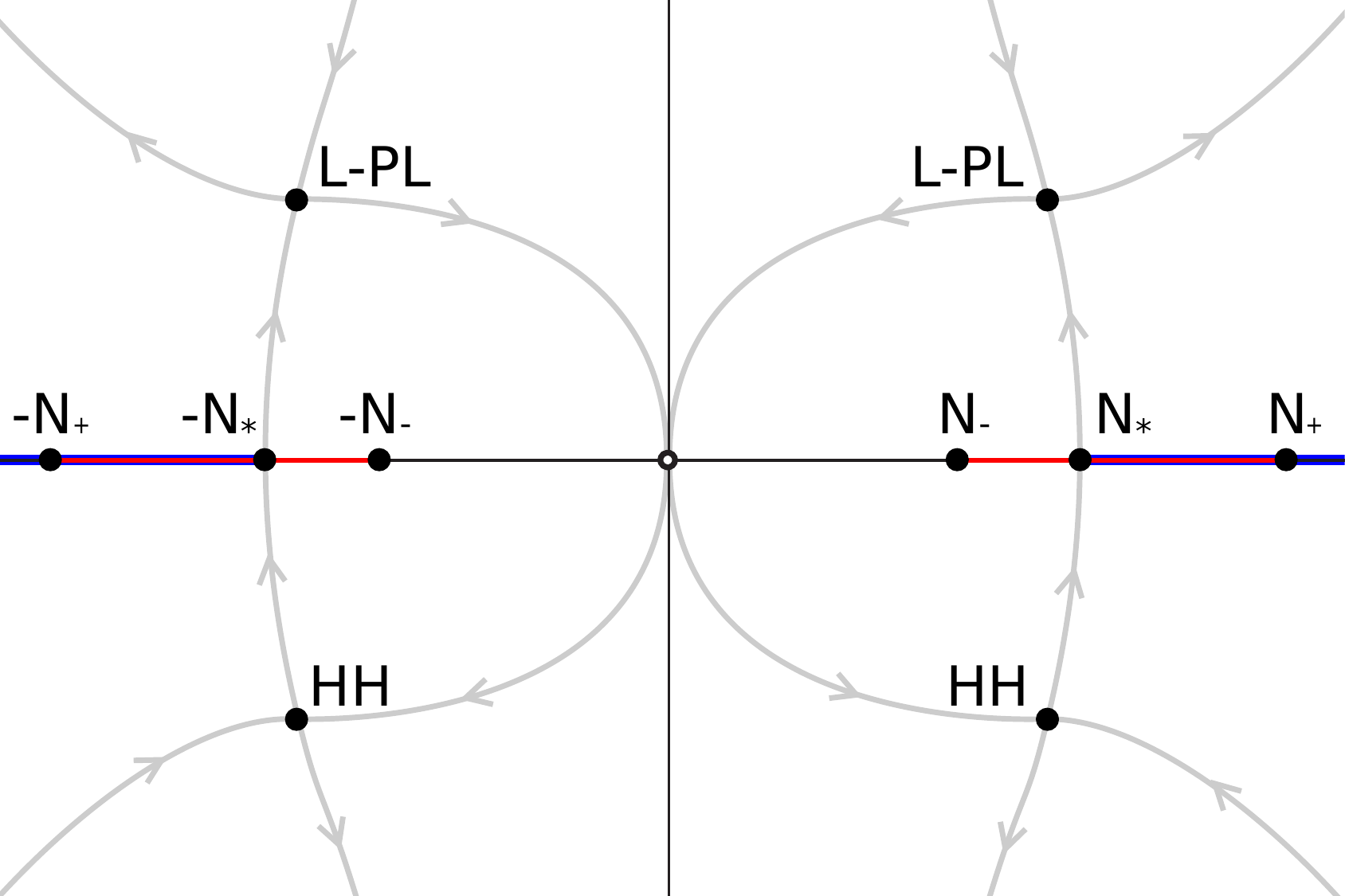}
\caption{The branch cuts (in red) on the real $N$-axis, for $-N_+< N < -N_-$  and $N_-< N < N_+$,  form impenetrable barriers for Picard-Lefschetz theory. The classical scale factor squared $q$ crosses zero for a second time (as in the right panel of Fig.~\ref{fig:onoffshell}) on the blue lines. The Hartle-Hawking and Lorentzian-Picard-Lefschetz saddles are indicated HH and L-PL respectively. The gray lines are the lines of steepest ascent and descent emanating from the four saddle points, with the arrows indicating directions of descent.}
\label{fig:Barrier}
\end{figure}

The above simplified case exemplifies the mechanism operating in our path integral for de Sitter. In the full situation, with $\Lambda>0$, in the vicinity of $t=0$ the perturbations are still described by (\ref{schop}) but the power $\gamma$ appearing in the asymptotic behavior of modes near $t=0$ has a more intricate structure in $N$, given in equation \eqref{eq:gamnpm} below. It possesses four square root branch points instead of two, at values $N=\pm N_\pm$ with $N_+>N_->0$, with two finite branch cuts connecting them on the real $N$-axis (red lines in Figure \ref{fig:Barrier}). Furthermore, $\gamma$ develops simple poles at $\pm N_\star$, where $N_\star\equiv \sqrt{N_+ N_-}$, requiring separate analysis (see Appendix D). In Appendix B we prove 
that in the full problem, just as in our simplified case, the real part of $\gamma$ is positive for all complex $N$ away from the branch cuts. This means that the mode $\chi_-$ has infinite action and must be excluded. Furthermore, as pictured in  Figure \ref{fig:onoffshell}, for real $N$ satisfying $|N|>N_\star$, a second Lorentzian singularity forms, at a value $t=t_s$, where $0<t_s<1$. The behavior of the perturbations near this second singularity is similar to that near $t=0$, but this time it turns out that that $\chi_+$ has divergent action, and hence it must be eliminated. Thus, {\it no} perturbation mode has finite action for real $N$ with $|N|>N_+$. This finding shall be important in our analysis of how to deform Diaz Dorronsoro {\it et al.}'s proposed contour into one over which the $N$-integral becomes absolutely convergent.

The subtle, and inherently infinite-dimensional phenomenon just described turns out to explain why it is possible to obtain a positive real term in the semiclassical exponent for the fluctuations, and still remain consistent with Picard-Lefschetz flow away from the branch cuts on the real $N$-axis, where the effective action for $N$ is still analytic. For example, in our treatment of the background, the original steepest  ascent contour from saddle 1 intersects the real $N$-axis at precisely $N_\star$, the value at which the geometry becomes doubly singular, as indicated in Figure \ref{fig:onoffshell}. As we have described, integrating out the perturbations generates a positive real part of the exponent proportional to $\phi_1^2$ on the upper side of the branch cut. Therefore, although our saddle point contribution (\ref{eq:LPLres}) grows exponentially with increasing $\phi_1^2$, so does the real part of $i S_{cl}(N)/\hbar$ on the upper side of the $N$-axis, where the steepest ascent contour from our saddle meets it. There is therefore no inconsistency with Picard-Lefschetz flow: even as we increase $\phi_1$, saddle point 1 remains relevant to the Lorentzian causal propagator.  

In Appendix C, we analyse the no boundary path integral for $\Lambda=0$ in detail, showing that, in that case, both the Euclidean and the Lorentzian contours make sense as defining contours for $N$. In the former case, one must take $N$ to run from just above the origin to $+i\infty$. In the latter, it runs over $0^+<N<\infty$. The result of this analysis, however, is that the introduction of gravity inverts the Euclidean vacuum distribution of the quantum fields, because the background ``chooses the wrong Wick rotation," as was explained in Ref.~\cite{Feldbrugge:2017fcc}. This result holds equally for the Euclidean or Lorentzian definitions of the no boundary path integral. 

The situation is more subtle with the contour proposed by Diaz Dorronsoro {\it et al}, because in this case the background Picard-Lefschetz thimble descending from a Hartle-Hawking saddle intersects the branch cut at $N_\star$. Strictly speaking, Picard-Lefschetz flow fails at this point when we integrate out the perturbations since the effective action for $N$ is no longer analytic in $N$. However, Cauchy's theorem still holds. By distorting the background steepest descent contour in $N$ to run around the branch cut, and continue along the original contour in the upper half complex $N$-plane, {\it before} we even integrate out the perturbations, we can maintain the absolute convergence of the integral, as well as the validity of Cauchy's theorem. At first sight, it appears that there might be two ways to circumnavigate the cut in Figure \ref{fig:Barrier}, namely on the side nearest to the origin or farthest from it. But here, the second singularity in the background geometry imposes an additional constraint. As mentioned above, since no perturbation mode has finite action for real $N$ satisfying $|N|>N_+$, we cannot go around the branch cut on that side of it.  So, as it turns out, when perturbations are included there is only one way to go round the branch cut -- on the side nearest the origin. This is the unique choice for distorting the Picard-Lefschetz thimbles associated with the Hartle-Hawking saddles, and it is illustrated in Figure \ref{fig:GW} below. 

The fact that we obtain a unique result even for (what we regard as) an unphysical choice of the defining contour for the lapse and in a situation which is inherently infinite-dimensional is a sign in favor of the mathematical validity of our approach. Nevertheless, one should emphasize that Lorentzian Einstein gravity becomes singular at these off-shell values $|N|\geq N_\star$, with the background geometry developing a second singularity as illustrated in Figure \ref{fig:onoffshell}. Furthermore, the infinite oscillations developed in the interval $|N|>N_-$ might also lead one to doubt the validity of the Einstein action as a correct low energy effective theory, this far off-shell. While the analytic continuation we perform to avoid the branch cuts is, we believe, an entirely natural definition of the off-shell, low energy theory, we cannot rule out the possibility that new degrees of freedom enter and significantly alter the result. Were this true, however, it would presumably invalidate the no boundary proposal.

If we distort the Picard-Lefschetz background thimbles as described above, the real part of the exponent $i S_{cl}(N)/\hbar$ becomes more and more positive on the upper side of the  branch cuts and, by symmetry, more and more negative on the lower side (see Fig.~\ref{fig:MorseCut}). Thus, even as we increase the perturbation amplitude $\phi_1$ and the height of the Hartle-Hawking saddle point falls, its steepest descent contour ${\cal J}_4$ still runs down to hit the branch cut. The integral around the branch cut yields an additional contribution to the path integral which, again, gives an inverse Gaussian in the perturbations, $e^ {+\frac{3}{2\hbar \Lambda}l(l+1)(l+2)\phi_1^2}$ but this time without the $e^{-\frac{12\pi^2}{\hbar \Lambda}}$ suppression factor associated with the Lorentzian-Picard-Lefschetz saddle (\ref{eq:LPLres}). 

The conclusion of this analysis is quite striking. Namely, if one wishes to include the Hartle-Hawking saddles in the Lorentzian path integral, then Cauchy's theorem and the choices we are forced to make to obtain an absolutely convergent integral imply there are additional non-perturbative contributions giving unsuppressed fluctuations. In Section VI we prove that {\it no} contour in $N$ can avoid such contributions. As we discuss in the conclusions, this has potentially profound implications for quantum de Sitter spacetime and for inflation. 

As a final remark, note that throughout this section we have only treated the perturbations to quadratic order in the action. This is at best a partial representation of the complete theory, and one might wonder whether higher order effects might significantly alter the analyticity properties of the effective action for the lapse near singularities such as those we encountered on the real $N$-axis. Fortunately, at the semiclassical level we are working at, it is not difficult to study nonlinear backreaction using numerical methods.  We do so in Section V, with the conclusion that the basic picture we have obtained using general relativistic linear perturbation theory remains unchanged. 

\subsection{Wider implications}

\begin{figure}[h]
\begin{minipage}{0.5\textwidth}
\includegraphics[width=.97\textwidth]{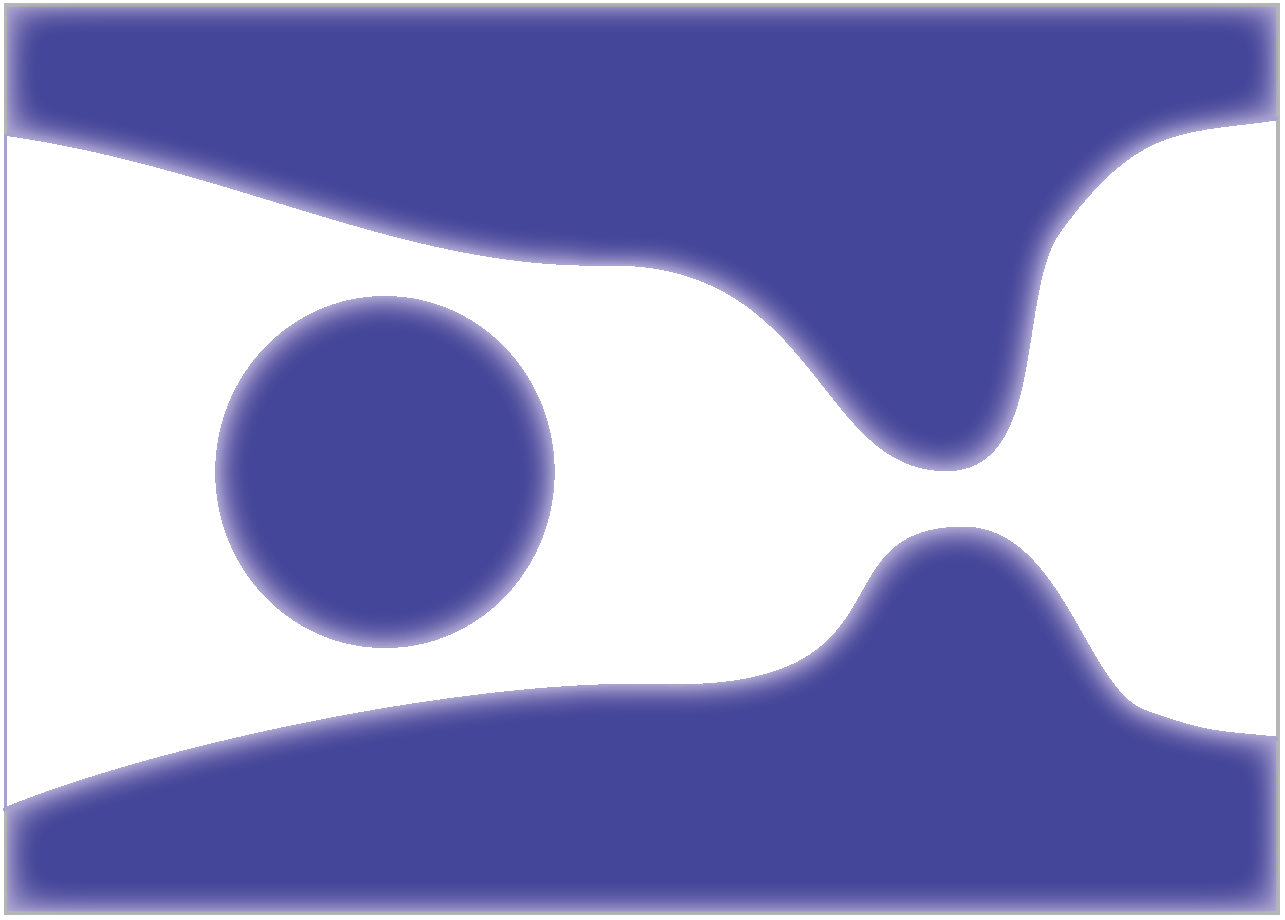}
\end{minipage}%
\begin{minipage}{0.5\textwidth}
\includegraphics[width=.97\textwidth]{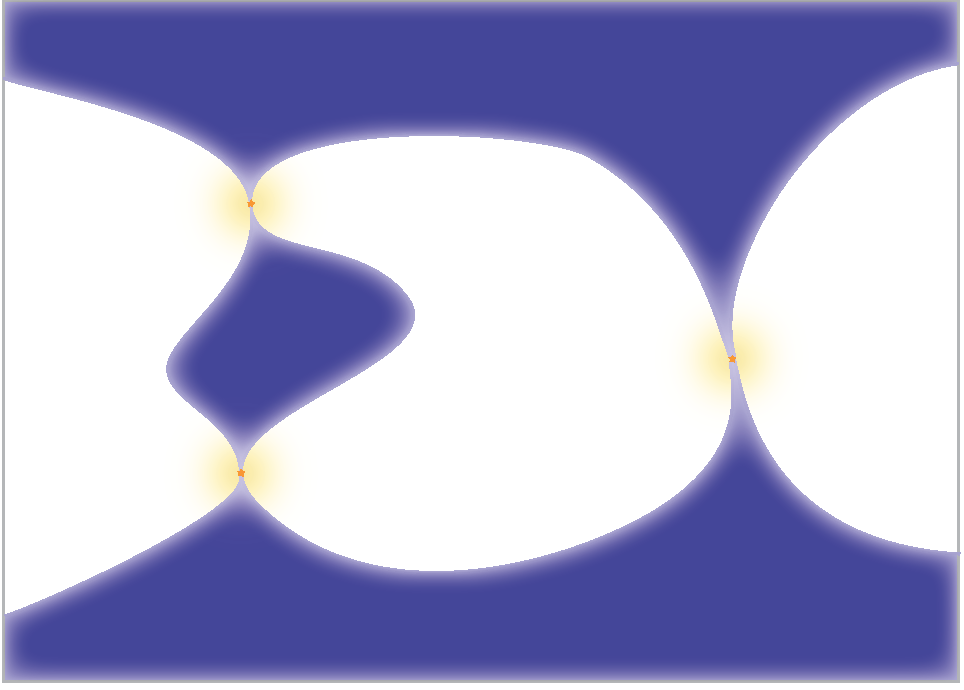}
\end{minipage}%
\caption{Our results imply that a smooth, $\Lambda$-mediated topology changing transition is ill-defined. Thus topology change should not be thought of as illustrated in the left panel (where the physical regions of spacetime are blue, space is horizontal and time vertical). Rather, topology change most likely requires passage through a singularity, where massless degrees of freedom play a crucial role in enabling the transition and extensions of the semi-classical methods employed in the present paper are needed.}
\label{fig:topology}
\end{figure}

We believe that our results have implications well beyond the no boundary proposal. For instance, the fact that the no boundary amplitude is out of control has a bearing on the question of topology change. Smooth, topology changing transitions can be thought of as combinations of no boundary amplitudes -- see the left panel in Fig.~\ref{fig:topology}. Our analysis suggests that such transitions must be disallowed. This does not mean that topology change cannot occur, but it indicates that any topology changing transition would have to proceed via a singular, more quantum transition. That such a transition might be feasible is supported by earlier studies indicating that a singular, radiation-dominated bounce appears to be possible, and appears not to suffer from unsuppressed fluctuations~\cite{Gielen:2015uaa}. This finding actually resonates, to some extent, with the description of singularity resolution in string theory, where it is typically found that new, massless degrees of freedom appear which are crucial in regularizing topology change. Likewise, it is in accordance with what we know from observations about the standard big bang cosmology, that the early universe was dominated by radiation. Even if one is interested in inflationary scenarios, our findings suggest that a ``beginning" with only inflationary potential energy is not allowed. An earlier phase such as a radiation-dominated phase may have been required prior to inflation. \\

This paper is structured as follows. In Section II, we briefly review the broad physical and mathematical principles of Lorentzian path integral quantum cosmology. We emphasize that the basic definition of the theory involves an integration over a real, nonzero lapse functions $N$, although it is mathematically convenient to deform that integration contour to an equivalent complex one using Cauchy's theorem, in order to improve the integral's convergence. In Section III, we review the path integral for the background de Sitter cosmology, comparing the contour for the Lorentzian propagator with that proposed by Diaz Dorronsoro {\it et al.}. We show that the latter contour cannot be deformed to the real $N$-axis. In Section IV, we include perturbations, treated in linear perturbation theory. In Section IV, we include nonlinear backreaction, to all orders for the lowest tensor mode and up to second order for higher modes, showing that it has little effect on our conclusions. In Section V we prove by simple enumeration a new theorem, that no possible choice of the integration contour for  the lapse -- whether physically motivated or not -- rescues the no boundary proposal from the problem of unsuppressed perturbations. Finally, we summarize our main findings and briefly comment on implications for quantum de Sitter spacetime and for inflation. In Appendix A, we illustrate some features of higher-dimensional Picard-Lefschetz theory in a simple two-dimensional oscillatory integral. In Appendix B, we prove that the quantity $\gamma$ governing the perturbations near background singularities of the types discussed above, obeys $\Re[\gamma]>0$ for all complex $N$ except on the special intervals noted. In Appendix C, we discuss the no boundary path integral for $\Lambda=0$, showing that both the Euclidean and the Lorentzian contours make sense as defining contours for $N$. However, unsuppressed perturbations are obtained in both cases. Finally, in Appendix D, we examine the point $N=N_\star$, showing the precise behavior of the modes and the classical action at that special value.

\section{Basic physical and mathematical principles}\label{sec:principles}
 
The novelty of our work is to combine, for the first time, two theoretical strands each over three decades old. The first is the work of C. Teitelboim (now C. Bunster) in formally developing a Feynman path integral for quantum gravity  \cite{Teitelboim:1982,Teitelboim:1983,Teitelboim:1983fk} based on ideas tracing back to Bryce DeWitt, John Wheeler and Richard Feynman. The second is an area of pure mathematics known as Picard-Lefschetz theory, aimed at the evaluation of oscillatory integrals, in any finite number of dimensions, via contour deformation exploiting Cauchy's theorem and saddle point/steepest descent approximations. Although there was an upsurge of interest in semiclassical quantum gravity effects in the early 1980's, with the computation of scalar and tensor quantum fluctuations in inflation as well as Hartle and Hawking and Vilenkin's proposals for the beginning of the universe, it is surprising to us that Teitelboim's foundational work seems to have attracted only casual reference. It is likewise remarkable that Picard-Lefschetz theory, as a rigorous and highly appropriate mathematical tool, seems to have been altogether overlooked. (For a brief review, with applications to Chern-Simons theory, see \cite{Witten:2010cx}. For recent and more general applications to quantum field theory, see \cite{Cherman:2014ofa,Behtash:2015zha})

Teitelboim's goal was to develop the theory of quantum geometrodynamics, a program initiated by Wheeler, DeWitt and others. The basic quantity of interest in this program is the quantum mechanical propagator: the amplitude for obtaining a final three-geometry $\Sigma_1$ from a given initial, three-geometry $\Sigma_0$, represented by a Hamiltonian path integral over all possible intervening four-geometries. As we explained in Refs.~\cite{Feldbrugge:2017kzv,Feldbrugge:2017fcc}, the no boundary proposal is most naturally formulated in this framework as the amplitude for obtaining $\Sigma_1$ when $\Sigma_0$ is taken to have zero size. When framed in these terms, we showed that the no boundary proposal becomes equivalent to Vilenkin's ``tunneling" proposal \cite{Vilenkin:1982de,Vilenkin:1983xq}, and that the relevant Lorentzian no boundary propagator, for general relativity with a positive cosmological constant, is a relatively well-defined mathematical object, whereas the Euclidean propagator is not. 

\subsection{Why Lorentzian?}

It is worth spelling out why we base our approach on the Lorentzian rather than the Euclidean path integral. Obviously, to do so is more conservative: we take the real-time classical theory as fundamental and try to ensure our quantum theory recovers it's successes in the relevant physical regimes. In fact, the motivation for performing a Wick rotation in gravity appears to have been a misguided belief that Lorentzian path integrals are too oscillatory to be well defined. Hartle and Hawking state in the second paragraph of their paper {\it ``The oscillatory integral in (the usual nonrelativistic path integral propagator) is not well defined but can be made so by rotating the time to imaginary values"} \cite{Hartle:1983ai}. This statement is incorrect: the integrals appearing in real time (Lorentzian) path integrals are typically conditionally, although not absolutely, convergent. If suitable saddle points exist, as they do very generally, then the path integral can be made absolutely convergent by deforming the integration contour to run along the appropriate steepest descent contour. This is precisely what Picard-Lefschetz theory accomplishes. 

In general relativity, rotating the time coordinate to imaginary values is problematic in several ways. The kinetic term for the conformal factor has the wrong sign - the well-known ``conformal factor problem", making the Euclidean action unbounded below. Gibbons, Hawking and Perry proposed to remove that divergence by rotating the conformal factor to imaginary values~\cite{Gibbons:1978ac}. Unfortunately, this rotation does not respect the boundary conditions in geometrodynamics, which involve the initial and final three-geometry. Instead, in the examples which follow, we will first perform the path integral over the scale factor and perturbations, leaving us with an ordinary integral over the lapse $N$ which we perform using steepest descent methods. When we consider the saddle point solution for the spacetime metric, it is complex and includes a ``Euclidean" region. But any complex deformations of the contours are, in our method, chosen by the theory and not the theorist. It is worth mentioning that some of our arguments regarding the improved behavior of the Lorentzian, as opposed to the Euclidean, path integral were anticipated by earlier discussions, for example by Giddings~\cite{Giddings:1989ny,Giddings:1990yj} and particularly by Sorkin~\cite{Sorkin:2009ka}, although with less general methods. 

\subsection{Wavefunction or propagator?}

In contrast, Hartle and Hawking took the Euclidean path integral to be fundamental. This seems to be the basis for their belief that the wavefunction has to be real. Second, they gave a formal argument that the Euclidean path integral satisfies the homogeneous Wheeler-DeWitt equation, and in follow-up papers, {\it e.g.},~\cite{Halliwell:1984eu} claimed that the Euclidean path integral provides boundary conditions for the wavefunction on the boundary of superspace.  Diaz Dorronsoro {\it et al.} \cite{Dorronsoro:2017} emphasize that their proposed wavefunction is both real and solves the homogeneous Wheeler-DeWitt equation, and they implicitly criticize our Lorentzian propagator because it is not real, and yields $-i$ times a delta functional on the right hand side of the Wheeler-DeWitt equation. However, in Ref.~\cite{Feldbrugge:2017kzv}, we explicitly demonstrated that for Einstein gravity with a positive cosmological constant, the Euclidean path integral is divergent. Our arguments above show it cannot provide boundary values for solutions of the Wheeler-DeWitt equation, as was hoped. Therefore there seems little motivation for insisting that the wavefunction should be real. In fact, as we shall discuss momentarily, a real wavefunction presents problems with recovering local quantum field theory unitarity. In contrast, the Lorentzian formulation provides a natural and mathematically meaningful way to formulate the no boundary amplitude, as the path integral propagator for obtaining a given final three-geometry starting from an initial three-geometry of zero size, a viewpoint emphasized by Vilenkin~\cite{Vilenkin:1983xq}. In Ref.~\cite{Feldbrugge:2017kzv}  we showed that the Lorentzian no boundary propagator is well defined, and furthermore that the dominant saddle point contribution for the background is a regular complex four-geometry with the final three-geometry as its only boundary, exactly the semiclassical picture Vilenkin, and Hartle and Hawking, had anticipated. 

If, on the contrary, the no boundary proposal is reduced to choosing some particular solution of the Wheeler-DeWitt equation, all geometrical justification or uniqueness disappears. As a simple example of this ambiguity, our Lorentzian propagator trivially provides a real (in both senses) solution of the homogeneous Wheeler-DeWitt equation, just by taking its real part. Diaz Dorronsoro {\it et al.}'s wavefunction, based on a complex contour for the lapse, with an appropriate symmetry, provides another. By taking linear combinations of the two, one obtains an infinite number of ``real" wavefunctions with no obvious means to choose between them.  

\subsection{Causality or gauge invariance?}

Before proceeding any further, it may be helpful to undertake a short excursion in order to explain why the Lorentzian path integral propagator necessarily {\it does not} satisfy the homogeneous Wheeler-DeWitt equation, and why this in no sense undermines its utility as a fundamental amplitude in the theory. This was actually understood a long time ago in a beautiful series of papers by C. Teitelboim, emphasizing the tension between gauge invariance and causality. We particularly recommend the brief summary article, Ref.~\cite{Teitelboim:1983}.

Schematically, the Lorentzian path integral over all four-geometries bounded by an initial three-geometry $\Sigma_0$ and a final three-geometry $\Sigma_1$, is given by 
\begin{equation}
\langle 1| 0\rangle =\int {\cal D} N \int {\cal D} N^i \int_{\Sigma_0}^{\Sigma_1}{\cal D} h^{(3)}_{ij}  {\cal D} \pi^{(3)ij}  e^{{i\over \hbar} S[h^{(3)}_{ij};\pi^{(3)ij};N]}
\label{prop}
\end{equation}
where the Lorentzian four-geometry is studied in a $3+1$ split with $N$ being the lapse function, $N^i$ the shift, $h^{(3)}_{ij}$ the $3$-metric, $\pi^{(3) ij} $ is its conjugate momentum, and $S= \int_0^1 \mathrm{d}t \int \mathrm{d}x^3 [\pi^{(3) ij} \dot{h}^{(3)}_{ij} - N^i H_i - N H]$ the action for general relativity expressed in first order Hamiltonian form. The path integral is taken over all four-geometry bounded by $\Sigma_0$ and $\Sigma_1$. Here for simplicity we have neglected the ghosts and BVF formalism needed to ensure general covariance, which were worked out by Teitelboim, Henneaux and others, and generalized to supergravity, in the 1980's \cite{Henneaux:1992ig}. 

Although the expression (\ref{prop}) for the propagator is still formal, the ranges of integration for all but one of the variables to be path-integrated over are fairly clear. At each $t$ and at every spatial point one integrates over all possible real three-metrics and momenta. Likewise one integrates over all real values of the shift in order to enforce the Einstein three-momentum constraint (the $G_{ti}$ Einstein equation) at every spacetime point. 

The integration over the lapse $N$ is more subtle. As Teitelboim argued, it is generally possible to choose a gauge in which $N$ depends only on the spatial coordinates. The value of $N$ at some point then controls the total proper time between the initial and final three-geometries, and the path integral measure over $N$ becomes an infinite number of ordinary integrals. The question arises whether one should integrate over all real values of $N$ or only over positive values. Classically, $N$ and $-N$ represent the same spacetime geometry, suggesting that it would be overcounting to include both. Teitelboim argued that integrating $N$ only over one of these choices -- positive values, for example -- is to be preferred, since it allows one to introduce a primitive notion of causality into the theory, independently of the existence of any classical spacetime. His remarks echo Feynman's earlier discussion, in his first papers on quantum electrodynamics, where he obtained his famous propagator as the quantized amplitude for a relativistic particle, rather than from any consideration of quantum fields (see Appendix A of Ref.~\cite{Feynman:1950ir}).  Because Feynman employed the same, primitive, ``world-line" notion of causality, his propagator is still referred to as the ``causal" propagator. 

In quantum geometrodynamics, it is the causality constraint of integrating only over positive $N$ which enables one to globally distinguish an ``in" from an ``out" state, and to meaningfully define quantum mechanical transition amplitudes. It also prevents one from considering histories (four-geometries) where the final three-geometry crosses the initial three-geometry creating a region where the two reverse roles. However, there is a tension between diffeomorphism invariance and the primitive causality constraint. Through the Lie derivative (and the corresponding Poisson bracket algebra), timelike diffeomorphisms may be used to push the initial three-surface backward or forward. If the final three-surface is held fixed (as it is, in the propagator), as the initial three-surface approaches it one must exclude diffeomorphisms which would push the initial three-surface ahead of the final one. That is, diffeomorphism invariance becomes retricted to half of the usual space of diffeomorphisms. As Teitelboim puts it \cite{Teitelboim:1983}, the causality constraint $N>0$ ``disrupts the group structure of the four-dimensional diffeomorphisms". Hence, one should not be surprised that the causal propagator is  {\it not} annihilated by the Hamiltonian and, in this sense, is no longer completely invariant under the generator of time-like diffeomorphisms. 

One can see this very well in lower-dimensional examples of quantum geometrodynamics, such as the quantized relativistic particle, or the quantized free relativistic string. In these cases, as is well known (see {\it e.g.} \cite{Green:1987}), integrating over positive $N$ is precisely what is required to construct the Feynman propagator, used in perturbative calculations of unitary scattering amplitudes (or, for the string, for defining vacuum states). In these examples, the propagator is formally given by
\begin{equation}
\langle 1| 0\rangle =\int_{0^+}^{\infty} dN \langle 1|e^{-i N H/\hbar}| 0\rangle =-i \hbar\, \langle 1|H^{-1}| 0\rangle 
\label{prop1}
\end{equation}
where $H$ is the Hamiltonian: $H=p^2+m^2$ for a free particle or $H=L_0-1$ for a free open string. In the Picard-Lefschetz approach, we do not actually need to include the usual $i\epsilon$ to ensure convergence of the integration over $N$ ~\cite{Gielen:2016fdb}. Note that we define the integral over $N$ to run only over positive real values. This is because in the examples of interest, the integrand possesses singular behavior at small $N$, so that the integral over $N$ is only defined as its lower, real limit is taken to zero. This singular behavior is no accident. It is generated in passing from the Hamiltonian to the Lagrangian formalism: at $N=0$ the momenta cannot be expressed in terms of the velocities. In our work, we shall take the Hamiltonian formulation, in which all fields including $N$ are real, to be the fundamental definition of the theory. The restriction to $N>0$ (or $N<0$) is then necessary for a well-defined passage to the Lagrangian formulation.

It follows from (\ref{prop1}) that the propagator is {\it not} annihilated by the Hamiltonian constraint, even though the latter is required to vanish on all physical states. Indeed, it follows from (\ref{prop1}) that $H \langle 1| 0\rangle$ equals $-i \hbar$ times a matrix element of the identity operator. For the free relativistic particle in $d$ spacetime dimensions, in the coordinate representation one obtains
\begin{equation}
 H_{x_1} \langle x_1|x_0 \rangle=(-\hbar^2 \Box_{x_1}+m^2) \langle x_1|x_0 \rangle=-i \hbar \delta^d(x_1-x_0),
\label{prop2}
\end{equation}
the usual equation satisfied by the Feynman propagator $\Delta_F(x_1-x_0)\equiv \langle x_1|x_0 \rangle$. 

Within four-dimensional quantum geometrodynamics, one expects something similar: the Hamiltonian applied to the causal propagator yields a delta functional which is zero unless the initial and final three-geometries (the analogs of the initial and final spacetime coordinates of the relativistic particle) are identical. Exactly solvable minisuperspace examples are worked out in detail  in \cite{Gielen:2015uaa,Gielen:2016fdb}. For the no boundary Lorentzian propagator, the delta functional occurring on the right hand side of the Wheeler-DeWitt equation is nonzero only when the final three-geometry degenerates to a point. 

\subsection{Recovering unitarity}


Teitelboim ends his short paper \cite{Teitelboim:1983} as follows:  {\it ``Therefore, it appears that in both gravity and supergravity one is faced with the alternative of preserving either gauge invariance or causality. It is the opinion of this author that one should preserve causality. In the case of positron theory, this turns out to be the correct choice ultimately because only by using the Feynman propagator does one obtain a unitary amplitude.(...) Whether or not a similar situation will arise for the quantized gravitational field remains to be seen."} ~\cite{Teitelboim:1983}. We believe the same issue indeed arises, as follows.  It is presumably a fundamental constraint on any theory of quantum cosmology that for scales and times much shorter than the Hubble length and time, and much longer than the Planck length and time, we should recover local quantum field theory, along with unitarity of scattering amplitudes in the quantum field theory sense.  Consider the Lorentzian path integral propagator between two large three-universes, the final one slightly larger than the initial one, with a local quantum field such as a gravitational wave in a stationary state such as the vacuum, or some fixed number of freely propagating quanta. The path integral will have a classical saddle point solution at positive real $N$ representing an expanding universe with the corresponding quantum field state. Because of the symmetry of the classical theory under $N\rightarrow -N$ there will inevitably also be a saddle point representing a contracting universe. If we integrate both positive and negative values of $N$, we cannot avoid picking up both saddle points. We thus obtain a superposition of amplitudes for the same quantum field state, within an expanding universe and its time reverse respectively. The inferred Schr\"odinger wavefunctional for the quantum field will combine field wavefunctionals in which the stationary state is evolved both forward and backward in the ``time" as represented by the size of the universe. Such evolution is not unitary. Therefore, integrating over both signs of $N$ seems to be inconsistent, at a basic level, with recovering perturbative quantum field theory unitarity in a description of local processes. 

One may say this even more strongly as follows. A real wavefunction, as advocated by Hartle and Hawking and Diaz Dorronsoro {\it et al.} has no chance of directly recovering unitarity which, at a fundamental level, rests upon quantum mechanical amplitudes being complex. This is particularly obvious for stationary states: the norm of $e^{-i E t/\hbar}$ is preserved but the norm of $\cos(E t/\hbar)$ is not. In the quantum cosmology literature, this problem is sometimes side-stepped by regarding the expanding and contracting parts of the Hartle-Hawking wavefunction as describing two ``decoherent histories," which should be studied separately. In effect, to describe an expanding universe, one throws half of the Hartle-Hawking wavefunction away. This seems, at best, uneconomical: if one integrates only over positive $N$ in the first place, and takes the causal propagator to the basic amplitude in the theory, there is no such redundancy and no projection is required. 

\section{The Background}\label{sec:MiniSuperSpace}

In order to be self-contained we briefly summarize the calculation of the path integral for the background. More details, and references to older literature, are provided in our earlier paper \cite{Feldbrugge:2017kzv}. 

\subsection{The propagator for a de Sitter cosmology}\label{sec:Baction}

For a homogeneous, isotropic background four-geometry, the gauge fixed Feynman propagator for the scale factor of the universe $a$ reduces to
\begin{equation}
G[a_1;a_0] =\int_{0^+}^\infty \mathrm{d}N \int_{a(0)=a_0}^{a(1)=a_1} \mathcal{D}a\, e^{i S[a;N]/\hbar}\,,
\label{inta}
\end{equation}
where $S$ is the Einstein-Hilbert-$\Lambda$ action.
Throughout this paper, our focus will be on carefully calculating semiclassical exponents, {\it i.e.}, contributions to the propagator proportional to $e^{i S_{cl}/\hbar}$ with $S_{cl}$ some classical action. We shall ignore terms in the exponent of higher order in $\hbar$ associated, for example, with operator ordering ambiguities in the quantum Hamiltonian on superspace (see the discussion above equation (18) in Ref.~\cite{Feldbrugge:2017kzv}). Nor shall we keep track of Jacobian factors associated with redefinitions of variables in the path integral measure. We shall proceed by transforming the action $S$ into a convenient form and then simply adopting the canonical phase space measure for these variables. A more careful treatment would include Jacobian and ordering corrections as well as Fadeev-Popov factors  associated with the constraints and gauge fixing conditions. 

As outlined in the introduction, we consider a positively curved Friedman-Lema\^{i}tre-Robertson-Walker (FLRW) universe containing only a positive cosmological constant. It is convenient to write the background metric as follows:  
\begin{equation}
\mathrm{d}s^2 = - \bar{N}^2 \mathrm{d} t^2 +a^2 \mathrm{d}\Omega_3^2\equiv   - \frac{N^2}{q}\mathrm{d}t^2 + q \mathrm{d}\Omega_3^2\,,
\end{equation}
where the first expression is the usual FLRW metric, with $\mathrm{d}\Omega_3^2$ the metric on the unit $3$-sphere. The second expression is a convenient rewriting, with $q=a^2$ representing the size modulus for the three-geometry and $N=a \bar{N}$ the redefined lapse. This form has the advantage that the Einstein-Hilbert-$\Lambda$ action (with $\Lambda$ the cosmological constant) is quadratic in $q$ \cite{Halliwell:1988wc},
\begin{equation}
S^{(0)}= 2 \pi^2 \int_0^1 \left[-\frac{3}{4N} \dot{q}^2 + N(3-\Lambda q)\right]\mathrm{d}t\,.
\label{bact}
\end{equation}
It is convenient to pick a gauge in which $N$ is constant. Since the path integral over $q$ is now Gaussian, it may be performed exactly, with the exponent being given by the classical action. The equation of motion, $\ddot{q}= \frac{2 \Lambda}{3} N^2,$ is solved by
\begin{equation}
\bar{q}(t) = \frac{\Lambda}{3}N^2 t^2 + \left( - \frac{\Lambda}{3}N^2 + q_1 \right) t \,,
\label{eq:qclass}
\end{equation}
with the boundary conditions $q_0=0$ and $q_1=a_1^2$, 
The corresponding classical action,
\begin{equation}
\bar{S}^{(0)}[q_1;0;N] = 2 \pi^2\left[ N^3 \frac{\Lambda^2}{36} + N\left(3 - \frac{\Lambda}{2}q_1\right) -\frac{3q_1^2}{4 N} \right],
\label{eq:S0cl}
\end{equation}
results in the propagator \footnote{In evaluating the path integral over $q$, we include all paths from $q[0]=0$ to $q[1]=q_1$, including those for which $q$ goes negative. Our methods rely on analyticity, hence we do not impose any barrier forcing $q$ to remain positive. Should the details of the theory near $q=0$  strongly affect the relevant semiclassical saddle point solutions, it seems to us this would necessarily imply sensitivity to the UV completion. In contrast, the saddle point solutions we study here all take the form of locally regular (albeit complex) solutions of the classical Einstein-$\Lambda$ equations, with modest curvature everywhere. In this case, geometrical higher derivative corrections to the low energy  Einstein-$\Lambda$ effective action are consistently small, at least on shell, and the results are therefore more likely to be reliable. }
\begin{equation}
G[q_1;0] = \sqrt{\frac{3 \pi i}{2 \hbar}} \int_{0^+}^\infty \frac{\mathrm{d}N}{\sqrt{N}} e^{i \bar{S}^{(0)}[q_1;0;N]/\hbar}\,,\label{eq:propMin}
\end{equation}
where the integration measure $1/\sqrt{N}$ arises from the Gaussian integral over $q.$ As mentioned above, our propagator satisfies 
\begin{equation}
\hat{H} G[q_1;0] = - i \hbar \delta(q_1)\,,\label{eq:GF}
\end{equation}
with $\hat{H}$ the Hamiltonian operator~\cite{Feldbrugge:2017kzv}. 

In order to discuss more general contours $\mathcal{C}$ for the integral over the lapse $N$,  such as that advocated in \cite{Dorronsoro:2017}, we will write the propagator as follows:\begin{equation}
G_\mathcal{C}[q_1;0] = \sqrt{\frac{3 \pi i}{2 \hbar}} \int_{\mathcal{C}} \frac{\mathrm{d}N}{\sqrt{N}} e^{ i \bar{S}^{(0)}[q_1;0;N]/\hbar}\,. \label{eq:prop}
\end{equation}

\subsection{Picard-Lefschetz theory}

The generalized propagator \eqref{eq:prop} is a highly oscillatory integral. We rely on Picard-Lefschetz theory to evaluate it, in a semiclassical approximation  -- for more details see \cite{Witten:2010cx,Feldbrugge:2017kzv}. One starts by analytically continuing the classical action $\bar{S}^{(0)}$ to the complex $N$-plane. The exponent is expressed in terms of its real and imaginary parts $h$ and $H$ (which are dimensionless) as
\begin{equation}
e^{i S/\hbar} = e^{h+iH}\,,
\end{equation}
where $h$ is known as the Morse function. The idea then is to deform the integration contour ${\cal C}$ into the complex plane, while keeping its end points fixed, in order to turn the oscillating integral into an absolutely convergent one which, moreover, can then be approximated as a saddle point integral. This is achieved by deforming the integration contour onto a set of steepest descent paths $\mathcal{J}_\sigma$ (also known as Lefschetz thimbles) associated to the saddle points of $h$ (each labeled by $\sigma$). In principle, one has to also prove that the ``contours at infinity" created by this deformation are negligible. This is usually not difficult (see ~\cite{Feldbrugge:2017kzv} for examples). Along a steepest descent contour the phase $H$ is constant so that the integral is no longer oscillatory. Paths of steepest descent follow the Morse function in a downwards flow until $h$ diverges to minus infinity. Thus an integral over a full thimble always runs between singularities of the Morse function, of this character. As long as $h$ diverges fast enough, which is typically a modest requirement, the integral along the corresponding Lefschetz thimbles is absolutely convergent since 
\begin{equation}
|G_\mathcal{C}[q_1;0] \leq \sum_\sigma \sqrt{\frac{3 \pi}{2 \hbar}} \int_{\mathcal{J}_\sigma} \left|\frac{\mathrm{d}N}{\sqrt{N}}\right| e^{h(N)}\,. 
\end{equation}
Not all saddle points and steepest descent paths contribute to the contour integral along any particular contour ${\cal C}$. A Lefschetz thimble $\mathcal{J}_\sigma$ is relevant if and only if the corresponding steepest {\it ascent} contour $\mathcal{K}_\sigma$ through the same saddle point $\sigma$ intersects ${\cal C}$. The reason for this is quite intuitive: the original integral is highly oscillatory and thus involves many cancellations. If it is to be replaced by a non-oscillatory integral the integrand must be smaller in magnitude than it is along the original contour. Hence, starting from the original contour we flow down to the Lefschetz thimble. In Fig.~\ref{fig:background} we illustrate the application of Picard-Lefschetz theory to the Lorentzian contour $\mathcal{C}_1=(0^+,\infty)$ and two alternate contours, $\mathcal{C}_2^-$ which runs from $N=-\infty$ to $N=+\infty$ just {\it below} the essential singularity at $N=0$ and $\mathcal{C}_2^+$ which runs from $N=-\infty$ to $N=+\infty$ just {\it above} the essential singularity at $N=0$. 

An important subtlety in Picard-Lefschetz theory is what to do when a thimble centred on one saddle point runs down to another saddle point. For example, thimble $\mathcal{J}_4$ -- the steepest descent contour from saddle point 4 -- coincides with the steepest ascent contour from saddle point 1, $\mathcal{K}_1$. The resolution of this dilemma is to add a small (complex) perturbation to the action which removes the degeneracy -- for example one can imagine giving Planck's constant a small complex phase. Such a perturbation breaks the degeneracy between $\mathcal{J}_4$ and $\mathcal{K}_1$ and causes $\mathcal{J}_4$ to just avoid saddle point 1. The perturbation can be taken to zero and in this limit does not affect the result. One sign of the phase causes $\mathcal{J}_4$ to narrowly miss saddle 1 and run off to infinity along the right ``side" of thimble $\mathcal{J}_1$. The other sign causes $\mathcal{J}_4$ to continue down the left ``side" of thimble $\mathcal{J}_1$. Either are perfectly valid definitions of the completion of thimble 4, and their application will yield exactly the same results. For simplicity, in what follows we shall adopt the second definition for $\mathcal{J}_4$, pictured in Figure \ref{fig:Map} below, and similarly define the completion of $\mathcal{J}_3$ to run to the origin along the right ``side" of $\mathcal{J}_2$.

\subsubsection{The causal propagator: integrating over positive lapse} \label{positivelapse}

The integration domain $\mathcal{C}_1$ only intersects the line of steepest ascent from one of the four saddle points: $\mathcal{K}_1$ corresponding to saddle point $1$ (see the left panel of Fig.~\ref{fig:background}). Observe that $\mathcal{C}_1$ (orange line) can be deformed into the Lefshetz thimble $\mathcal{J}_1$ (orange dashed line) without passing any singularity. Moreover, one can easily show that the additional arcs around the origin and at complex infinity required to complete the deformed contour, have a vanishing contribution \cite{Feldbrugge:2017kzv}. In the saddle point approximation, the propagator is then given by
\begin{equation}
G_{\mathcal{C}_1}[q_1;0] =c_1 e^{-\frac{12 \pi^2}{\hbar \Lambda} \, -i 4\pi^2 \sqrt{\frac{\Lambda}{3\hbar^2}} (q_1 - 3/\Lambda)^{3/2}}\,,
\end{equation}
where the constant $c_1$ includes the functional determinants and prefactors. In principle, for small $\hbar$ it can be expressed as a series in $\hbar$.  
The weighting $e^{-\frac{12 \pi^2}{\hbar\Lambda}}$ is the inverse of the famous Hartle-Hawking result $e^{\frac{12 \pi^2}{\hbar\Lambda}}$ and agrees with Vilenkin's tunneling proposal~\cite{Vilenkin:1982de}, as well as with arguments by Sorkin~\cite{Sorkin:2009ka}. Evaluation of the integration domain $(-\infty,0)$ leads to the Lefschetz thimble $\mathcal{J}_2$ giving an equivalent (but complex conjugate) result.

\subsubsection{Solutions of the homogeneous Wheeler-DeWitt equation}\label{sec:RealLine}

In an attempt to recover the Hartle-Hawking result, Diaz Dorronsoro \textit{et al.} \cite{Dorronsoro:2017} have instead proposed the integration domain $\mathcal{C}_2^-$. We have already explained why this contour cannot be claimed to be Lorentzian. Nevertheless, let us continue to analyze it.  From Figure \ref{fig:background} one sees that $\mathcal{C}_2^-$ is intersected by steepest ascent lines from all four saddles. From left to right, the contour intersects $\mathcal{K}_2, \mathcal{K}_3,\mathcal{K}_4$, and $\mathcal{K}_1$. We thus conclude that all four saddles contribute to the path integral. The corresponding deformed contour is indicated by the dashed orange line in the figure. Thus the path integral can be rewritten as a sum over all four thimbles,
\begin{equation}
G_{\mathcal{C}_{2}^-}[q_1;0] \approx |c_1| \, e^{\frac{12 \pi^2}{\hbar\Lambda}} \cos\left({4\pi^2\over \hbar} \sqrt{\frac{\Lambda}{3}} (q_1 - 3/\Lambda)^{3/2} +\varphi_1\right) + |c_2| \, e^{-\frac{12 \pi^2}{\hbar\Lambda}} \cos\left({4\pi^2\over \hbar} \sqrt{\frac{\Lambda}{3}} (q_1 - 3/\Lambda)^{3/2} +\varphi_2\right)\,,
\label{eq:dd}
\end{equation}
where $c_1=|c_1|e^{i\varphi_1}$,  $c_2=|c_2|e^{i\varphi_2}$ are coefficients to be expanded in powers of $\hbar$. $G_{\mathcal{C}_{2-}}[q_1;0]$ is real because the contributions from $\mathcal{J}_2, \mathcal{J}_3$ are complex conjugates of those from $\mathcal{J}_1 ,\mathcal{J}_4$. 

The lower saddle points, represented by the first term, dominate in the semiclassical expansion. Thus Diaz Dorronsoro \textit{et al.}'s contour recovers the Hartle-Hawking result at leading order in the exponential factor $e^{\frac{12 \pi^2}{\hbar\Lambda}}$. However, it also generates the second term in (\ref{eq:dd}) which represents a non-perturbative (and exponentially small) correction.  This is a minor correction for the background, but it will become problematic when we consider the perturbations. 

As we mentioned in the Introduction, there is another way of getting a real solution of the Wheeler-DeWitt equation,  from a truly Lorentzian contour, by integrating over purely real $-\infty <N<0^-$ and $0^+<N<+\infty$. This combination, representing the real part of our Lorentzian propagator, is equivalent to the continuous contour $\mathcal{C}_{2}^+$ which avoids the essential singularity at $N=0$ by passing above it (see the right panel in Fig.~\ref{fig:background}), because the small semicircle above the origin gives a vanishing contribution in the limit as $0^-$ and $0^+$ tend to 0. Note also that $\mathcal{C}_{2}^+$ only intersects the steepest ascent contours $\mathcal{K}_1$ and $\mathcal{K}_2$, from saddle points 1 and 2. It follows that the integral along $\mathcal{C}_{2}^+$ equals the sum of the Lefschetz thimbles $\mathcal{J}_2$ and $\mathcal{J}_1$, taken with appropriate signs. Hence the path integral along $\mathcal{C}_{2}^+$ is twice the real part of the path integral along $\mathcal{C}_1$. In the saddle point approximation, it is given by
\begin{equation}
G_{\mathcal{C}_{2}^+}[q_1;0]  = 2\text{Re}[G_{\mathcal{C}_1}[q_1;0]] \approx 2|c_2|e^{-\frac{12 \pi^2}{\hbar\Lambda}}\cos\left({4\pi^2\over \hbar} \sqrt{\frac{\Lambda}{3}} (q_1 - \frac{3}{\Lambda})^{3/2} +\varphi_2\right)\,.
\end{equation}
As explained above, $G_{\mathcal{C}_{2}^-}[q_1;0]$ and $G_{\mathcal{C}_{2}^+}[q_1;0]$ provide two independent, real solutions of the Wheeler-DeWitt equation. One might have hoped that one could subtract $G_{\mathcal{C}_{2}^+}[q_1;0]$  from $G_{\mathcal{C}_{2}^-}[q_1;0]$ in order to remove the two upper saddle points entirely. Unfortunately, this does not work, because the {\it entire} Lefschetz thimbles $\mathcal{J}_1$ and $\mathcal{J}_2$ contribute to $G_{\mathcal{C}_{2}^+}[q_1;0]$, whereas only the two outer ``sides" of these thimbles contribute to $G_{\mathcal{C}_{2}^-}[q_1;0]$. Furthermore, since the thimbles are not perfectly symmetrical, their ``outer" and ``inner" sides are not identical. This means there is no possible way to cancel the contributions of the upper two thimbles and hence to recover Hartle and Hawking's result.

\section{Perturbations}\label{sec:Pert}


We now turn our attention to the perturbations, treated in general relativistic linear perturbation theory. In \cite{Feldbrugge:2017fcc} we showed that the no boundary causal propagator generates an inverse Gaussian distribution for the perturbations, meaning large perturbations are favored. Here we shall review and extend this treatment to the wavefunction and lapse contour proposed by Diaz Dorronsoro \textit{et al.}, showing that it too includes unsuppressed perturbations.  Furthermore, we identify a new and larger source of unsuppressed perturbations coming from a branch cut which the Picard-Lefschetz thimble through the Hartle-Hawking saddles encounters. These results strengthen and generalize our result, allowing us to prove that no possible redefinition of the lapse contour can rescue the no boundary proposal. 

The second order action for a linearized (tensor) perturbation $\phi_l$ with principal quantum number $l$ is given in terms of the background squared scale factor $q(t)$ as
\begin{align}
S^{(2)}[q,\phi;N]&=
\frac{1}{2} \int_0^1 \left[ q^2 \frac{\dot{\phi}_l^2}{N}-Nl(l+2)\phi_l^2\right]\mathrm{d}t\ \cr
&=\frac{1}{2} \int_0^1 \left[ \frac{\dot{\chi}_l^2}{N}+N\left({\ddot q\over q}-{l(l+2)\over q^2}\right)\chi_l^2\right]\mathrm{d}t 
- {1\over 2 N}\left[{\dot{q}\over q}\chi_l^2\right]_0^1,
\label{eq:S2}
\end{align}
where we have re-expressed the dimensionless tensor metric perturbation $\phi_l$ in terms of the canonically normalized field $\chi_l =q\, \phi_l$.  Note that we have orthonormalized the modes on the unit sphere (thus no prefactor of $2\pi^2$ appears in the action). As explained in the introduction, to avoid needless complexity in the equations we only consider a single mode. It is straightforward to amend all the formulae we derive by replacing $l(l+1)(l+2)\phi_l^2$ with  $\sum_{lmn} l(l+1)(l+2)\phi_{lmn}^2$ where the $\phi_{1,lmn}$ are the expansion coefficients expressing in the final tensor perturbation in terms of orthonormal tensor spherical harmonics on the three sphere, with quantum numbers $l,m,n$~\cite{Gerlach:1978gy}. Since the treatment of each harmonic proceeds identically we will not write out this sum -- one may always think of setting all Fourier coefficients, bar one, to zero on the final three-geometry. For ease of notation, where there is no danger of confusion, we will also usually drop the subscript $l$. Note that the perturbation of the lapse $N$ is non-dynamical in the absence of matter and may be set to zero. 


If we neglect the backreaction of the linear perturbations on the background, such as is reasonable for small final amplitude $\phi_1$, then we can evaluate the path integral first for $q$ and then for $\phi$, using the classical solution for the background $q$ in the action (\ref{eq:S2}) for $\phi$. To integrate out the perturbations, we again just find the classical solution and use this to evaluate the classical action. The total semiclassical exponent is then given by $S^{(0)}[q_1;N]+ S^{(2)}[q_1,\phi_1;N]$. We perform the final ordinary integral over $N$ using saddle point methods. We shall not calculate any functional determinants in this paper, although this is perfectly possible. These should not alter any conclusions about the Picard-Lefschetz flow, nor the final semiclassical exponent, in any regime where the semiclassical expansion is valid. 

The no boundary path integral on a contour ${\cal C}$ is then given, in this leading semiclassical approximation, by 
\begin{equation}
G_{\cal C}[q_1,\phi_1;0] \propto \int_{\cal C}  {\mathrm{d}N \over \sqrt{D(N,q_1,\hbar)} } e^{i \bar{S}^{(0)}[q_1;N]/\hbar + i \bar{S}^{(2)}[q_1,\phi_1;N]/\hbar}\,,
\label{eq:PropClass}
\end{equation}
where $\bar{S}^{(0)}[q_1;N]$ is the classical action for the background solution $\bar{q}$ satisfying the final boundary condition $q(1)=q_1$ and the initial, no boundary condition $q(0)=0$ (see equation \eqref{eq:S0cl}). Likewise, $\bar{S}^{(2)}[q_1,\phi_1;N]$ is the classical action for the perturbation, in the background $\bar{q}$,  satisfying $\phi(1)=\phi_1$ as well as a second condition we shall define shortly. The quantity $D(N,q_1,\hbar)$ is the functional determinant which is in principle calculable in terms of the classical modes and as a series expansion in $\hbar$ (for a recent review see, {\it e.g.}, Ref.~\cite{Dunne:2007rt}). However, in this paper we shall focus on the semiclassical exponent, and shall not consider the functional determinant any further.

\subsection{Semiclassical path integral over the perturbations}\label{sec:STens}

In this section, we shall perform the path integral over the perturbations in the leading semiclassical approximation. That is, we shall fix the perturbation amplitude on the final three-geometry, $\phi_1$ and perform the path integral by the saddle point method, {\it i.e.}, by solving the equations of motion and computing the classical action. The boundary condition on the perturbations at $t=0$ is delicate because the background geometry is sufficiently singular for a range of real values of the lapse, that the perturbations obey a singular equation of motion. We shall find that, nevertheless, for generic complex $N$, the criterion of finite classical action selects a unique perturbation mode. 

At fixed $N$, the classical equation for $\chi$ following from (\ref{eq:S2}) is 
\begin{equation}
\ddot{\chi}= \left(\frac{\ddot{\bar{q}}}{\bar{q}} - \frac{N^2 l (l+2)}{\bar{q}^2} \right) \chi\,.
\label{eom4chi}
\end{equation} 
Near $t=0$, this becomes
\begin{equation}
\ddot{\chi} \approx - \frac{N^2 l (l+2)}{(q_1-\Lambda N^2/3)^2 }{\chi\over t^2} \equiv {\gamma^2-1\over 4}{\chi \over t^2} \,,
\end{equation} 
from which we see $\chi \sim t^{{1\over 2} (1\pm \gamma)}$, as $t\rightarrow 0$. Notice that the equation of motion for $\chi$ is singular and this results in some unusual properties of the perturbations, as we explain below. 

For small real $N$, we take $\gamma$ to be real and positive. Provided $N$ is real and smaller in magnitude than a particular value $N_-$, then both solutions for $\chi$ are monotonic in $t$ and both vanish at $t=0$. However, only one of them has finite action so it is natural to select that one as the saddle point solution. For real $N$ larger in magnitude than $N_-$ but smaller than another, larger value, $N_+$, $\gamma$ is imaginary and the solutions oscillate an infinite number of times as they approach $t=0$. In fact, both solutions have a finite regularized action, so the finite action criterion becomes ambiguous for $N$ in this range. Increasing the magnitude of $N$ beyond $N_+$, while keeping $N$ real, we see that $\gamma$ becomes real once again. However, as we explain shortly, in this latter regime, there are {\it no} finite action classical solutions. 

\begin{figure}[h]
\centering
\includegraphics[width=0.75\textwidth]{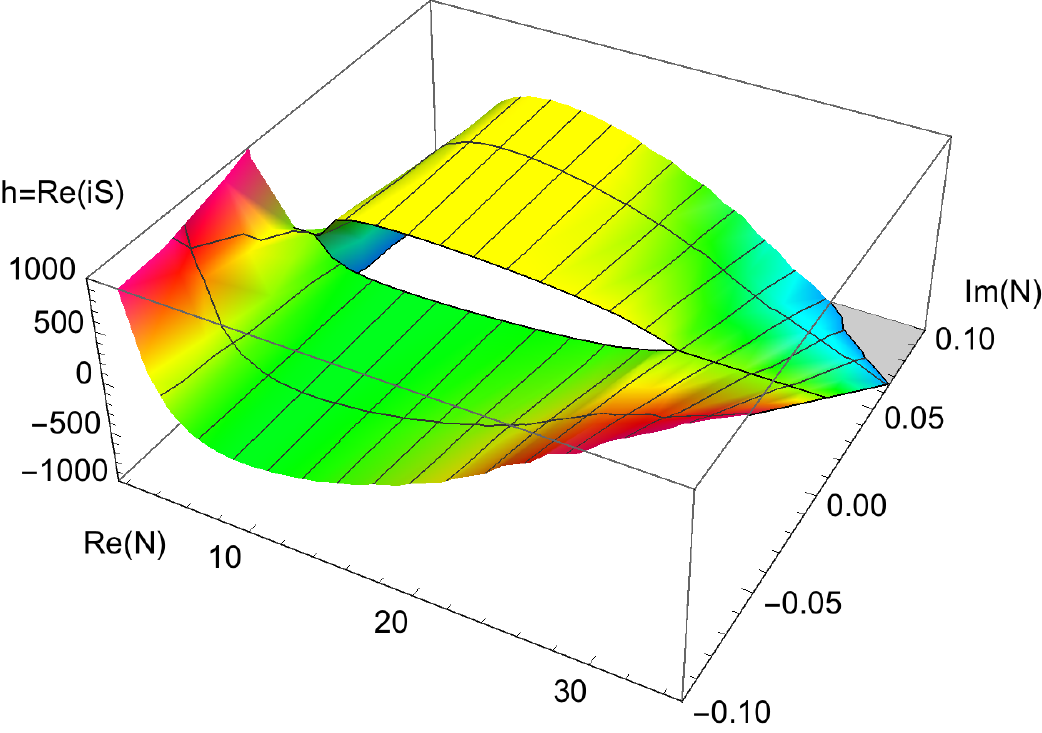}
\caption{The Morse function $h = Re(iS/\hbar)$ around a branch cut, in units where $\hbar=1$ and for the parameters $\Lambda = 3, q_1 = 101, l=10, \phi_1=1.$ At the cut, the Morse function reaches its maximum at $N_\star = 10$ coming from the upper half plane, and its minimum also at $N_\star,$ though approaching the cut from below.}
\label{fig:MorseCut}
\end{figure}

The two critical values are given by
\begin{align}
N_- &= \frac{3}{\Lambda} \sqrt{2 l(l+2) + q_1 \Lambda/3 - 
  2\sqrt{ l (l+ 2) (l(l+2) + q_1 \Lambda /3)}}\,, \\ 
N_+ &= \frac{3}{\Lambda} \sqrt{2 l(l+2) + q_1 \Lambda/3 +
 2\sqrt{   l (l+ 2) (l(l+2) + q_1 \Lambda/3)}}\,,
 \label{eq:npm}
\end{align}
with geometric mean $N_\star\equiv \sqrt{N_+ N_-}=\sqrt{3 q_1/\Lambda}$. It follows that we can take
\begin{align}
\gamma={\sqrt{(N_-^2-N^2)(N_+^2-N^2)}\over (N_\star^2-N^2)},
 \label{eq:gamnpm}
\end{align}
defined to be real and positive for small real $N$ and for other values of $N$ by analytic continuation. The branch cuts needed to define the square roots are conveniently placed along the real intervals $-N_+< N < -N_-$ and $N_-< N < N_+$. In Appendix B, we prove that $\Re[\gamma]$ is positive for all complex $N$ away from these cuts. On the upper side of the cuts, $\gamma$ is negative imaginary and on the lower sides it is positive imaginary. Away from the cuts, as is  evident from (\ref{eq:S2}), the action integral converges at $t=0$ only for the mode behaving as $t^{{1\over 2} (1+ \gamma)}$ as $t \rightarrow 0$. The complete solution of (\ref{eom4chi}) with this small $t$ behavior is
\begin{equation}
\chi(t) = \bar{q}(t)^{\frac{1}{2}} \left(\frac{t}{3q_1+(t-1)N^2 \Lambda}\right)^{\frac{\gamma}{2}} \left((3 q_1-\Lambda N^2)(1+\gamma)+2 \Lambda N^2 t \right)\,,
\label{exactsol}
\end{equation}
and the corresponding, correctly normalized classical solution is 
\begin{equation}
\phi(t)= \phi_1{ \chi(t) \over  \bar{q}(t)} {q_1\over \chi(1)}.
\label{phisol}
\end{equation}
This solution allows us to calculate the classical action from (\ref{eq:S2}). With an integration by parts and using the equations of motion, we find
\begin{equation}
S^{(2)}[q_1,\phi_1;N]=\left[\bar{q}^2 {\bar{\phi}\dot{\bar{\phi}} \over 2 N}\right]_0^1= {l(l+2)q_1\phi_1^2\over 4N(3 l(l+2)+q_1\Lambda)}\left(-3 q_1-N^2\Lambda +\gamma (N_\star^2-N^2)\right),
\label{classact}
\end{equation}
which is real where $\gamma$ is real, but gains a negative or positive imaginary part (meaning that the semiclassical exponent $i S/\hbar$ gains a positive or negative real part) as $N$ approaches the real axis from above or below the branch cuts. This behavior is illustrated in Figure \ref{fig:MorseCut}.

There is one additional important consideration: for real $N$, the background solution for the scale factor \eqref{eq:qclass} is real and quadratic in $t.$ For $N> N_{\star}$ (or $N < - N_\star$) the background solution starts at $q=0,$ then turns negative before crossing $q=0$ a second time, at $t_s = 1-\frac{3q_1}{\Lambda N^2}$, to eventually reach $q_1$ at $t=1$. Thus there is a second singularity in these real but off-shell-in-$N$ background geometries, as sketched in the right panel of Fig. \ref{fig:onoffshell}. 
It is obvious from (\ref{exactsol}) that if $\chi$ behaves as $ t^{{1\over 2} (1\pm \gamma)}$ near $t=0$, then it behaves as $( t_s-t)^{{1\over 2} (1\mp \gamma)}$ near $t=t_s$. Thus, for  real $\gamma$ and $N>N_\star$ then if the action integral  converges at $t=0$, it diverges at $t=t_s$, and vice versa. We conclude that for  $N>N_+$ or $N<-N_+$ {\it no} solution of the perturbation equations of motion has finite action.  Hence, in performing the integration over $N$ in the last step (\ref{eq:PropClass}) of our calculation, however we deform the contour ${\cal C}$, we cannot allow it to cross the real $N$-axis for real $N$ beyond the outer ends of the two branch cuts.

Finally, notice that at large $|N|$ in the complex $N$-plane, the background action $\bar{S}^{(0)} \sim N^3$ -- dominates over the perturbation action $\bar{S}^{(2)}\sim N$. The same holds in the small $|N|$ limit, where both the background and the perturbation diverge like $1/N$  (the background and the perturbation action have opposite sign).  As a consequence the asymptotic regions of convergence are preserved when we add linearized tensor perturbations.

\subsection{Integrating over the lapse $N$}

Having determined the classical action for the background (\ref{eq:S0cl}) and for the perturbations (\ref{classact}), we are now ready to evaluate the final integral over the lapse $N$, given in (\ref{eq:PropClass}). We have already explained the principles in the Introduction: here we shall give the details. 

\subsubsection{The saddle point contribution}

In the first approximation, we ignore backreaction from the perturbations on the background and simply evaluate the combined classical action ((\ref{eq:S0cl}) plus (\ref{classact})) at the relevant saddle points for the background. For simplicity, in this section we shall only discuss the saddles in the right half-plane: those in the left half-plane are simply related by symmetry. Assuming the radius of the final three-universe is greater than the de Sitter radius $\sqrt{3/\Lambda}$, the  two classical saddles for the background are given by 
\begin{equation}
N_s^\pm =  \frac{3}{\Lambda}\left[\sqrt{\frac{\Lambda}{3}q_1 - 1} \pm i \right].
\end{equation}
 
At this saddle points, the parameter $\gamma$ defined in Section \ref{sec:STens} is precisely equal to $l+1$, meaning that the tensor modes $\phi_l$ behave as $t^{l/2} $ near the singularity, which means they are regular there. In Appendix C we describe the relevant change of variables which exhibits this property. 

The values of the classical action at the upper and lower saddle points respectively are
\begin{equation}
\bar{S}^{(2)}(N_s^{\pm}) = \mp i {\phi_1^2 q_1\over 2} { l(l+2)\over l+1\pm i\sqrt{q_1\Lambda/3-1}}.
\end{equation}
There are two simplifying regimes. If the wavelength on the final three-geometry, $\sim \sqrt{q_1} l^{-1}$ is  well within the Hubble radius $\sqrt{3/ \Lambda}$, we obtain \begin{equation}
i {\bar{S}^{(2)}(N_s^{\pm}) \over \hbar}    \approx  \pm {\phi_1^2 q_1\over 2 \hbar} l, \quad  l\gg \sqrt{ \Lambda q_1\over 3},
\end{equation}
a result which is independent of $\Lambda$ and which agrees with the result of Appendix C. 

In the opposite limit, we obtain the result for the ``frozen" modes in the expanding de Sitter spacetime, which have passed out of the de Sitter Hubble radius and ceased to evolve. In this case, we obtain 
\begin{equation}
i {\bar{S}^{(2)}(N_s^{\pm})\over \hbar} \approx \pm \frac{3}{2\Lambda}l(l+1)(l+2)\phi_1^2 - i\sqrt{\frac{3q_1}{4\Lambda}}l(l+2) \phi_1^2 , \quad \sqrt{ \Lambda q_1\over 3} \gg l.
\label{frozenmodes}
\end{equation}
For the lower, Hartle-Hawking saddle point, the real part of the exponent exhibits the familiar scale-invariant inflationary power spectrum $\sim l^{-3}$ at large $l$, so that the real-space variance of the tensor modes is logarithmically divergent. Unfortunately, as we explained in the Introduction, the upper saddle is also relevant and it leads to an inverse Gaussian distribution meaning that the tensor modes are out of control. 

In view of this unsettling result, one should ask whether all the assumptions which went into calculating it are really valid. In particular, can we really trust it for large $\phi_1$, where the contribution of the upper saddle point outweighs the corresponding lower one? The calculation assumed linear perturbation theory, which requires that the perturbation amplitude is small throughout, {\it i.e.}, $|\phi(t)|\ll 1$ for all $t \in [0,1]$. However, there is a strong redshifting effect in a de Sitter background, and the amplitude of linearized tensor modes decreases inversely with the scale factor while it is inside the Hubble radius. Thus a mode which has just frozen at some large value of $q_1$ with amplitude $\phi_1$ has a much greater amplitude $\sim \phi_1 \sqrt{q_1}$ when it is followed back to the ``throat" of de Sitter spacetime. The condition that the mode has just frozen at $q_1$ reads $l\sim \sqrt{q_1\Lambda}$. The condition that the perturbation contribution to the final semiclassical exponent outweighs the background contribution is that $l^3 \phi_1^2$ exceeds unity (assuming $l$ is large). For this to be true, the initial amplitude $\phi_1\sqrt{q_1}$ must exceed $(l \Lambda)^{-{1\over 2}} $. This is possible, while maintaining the validity of linear theory at all times, if the frozen mode number $l$ exceeds $\Lambda^{-1}$. That requires that the de Sitter spacetime has undergone expansion by a factor $\Lambda^{-{3\over2}}$, {\it i.e.}, that $q_1$ exceeds this factor, which is a rather modest condition. Our conclusion is that it is perfectly possible to have the perturbations dominate in the semiclassical exponent, while remaining consistent with linear perturbation theory throughout the evolution of the perturbation modes. This is confirmed by the numerical calculations we shall report in Section V. In fact, those calculations show that nonlinear effects further enhance the discontinuity in the effective action across the real $N$-axis, created by integrating out the perturbations .

\subsubsection{The branch cut contribution}

Before turning to the detailed implications of the various contours of integration discussed in section \ref{sec:MiniSuperSpace}, we study the branch cut. We will specialize to Re$(N)>0,$ but analogous considerations apply for Re$(N)<0.$ As discussed above, the branch cut represents an impenetrable barrier to the integral over the lapse, since traversing it would mean running into regions where the perturbations are not well defined. Moreover, as shown in Fig. \ref{fig:Barrier}, some of the Lefschetz thimbles intersect the branch cuts. This means that when evaluating the path integral, in some cases we are  forced to distort the contour of integration around the branch cut. We thus need to know the contribution of the branch cut to the integral over the lapse. Note that because the perturbative action is infinite on the real $N$-axis outside of the cut (\textit{i.e.} for $N>N_+$), we are forced to deform the contour to pass on the inside of the cut.

Let us focus on a mode that has just frozen, \textit{i.e.}, a mode for which $l \sim \sqrt{q_1 \Lambda},$ and which has a large amplitude (but within the limits of perturbation theory) \textit{i.e.} $\phi_1 \sim l^{-\frac{1}{2}}.$ We work in the limit of large final scale factor. Similar calculations can be performed for other wavenumbers and amplitudes. 

We approximate the integral with the integration contour going around the branch cut in a clockwise direction. As we saw above, the Morse function is much higher above the cut than below, hence it is sufficient to consider the integral running just above the branch cut on the real $N$-axis, see also Fig. \ref{fig:MorseCut}. The maximum of the Morse function occurs at $N_\star^+,$ which is the location of the saddle point of the perturbative action \eqref{classact} evaluated on the upper side of the branch cut (note that the Morse function of the background action is zero on the real $N$-axis). 
The total exponent $iS/\hbar$ and its first two derivatives evaluated at $N_\star^+,$ keeping the leading real and imaginary terms in the limit of large $q_1,$ are given by
\begin{align}
\frac{i}{\hbar}S(N_\star^+) & = -i \sqrt{\frac{4\Lambda q_1^3}{3\hbar^2}} + \frac{3}{2\hbar\Lambda}(l(l+2))^{3/2}\phi_1^2\,,\\
\frac{i}{\hbar}S_{,N}(N_\star^+) & = 3\frac{i}{\hbar}\,, \\
\frac{i}{\hbar}S_{,NN}(N_\star^+) & = - \frac{i}{\hbar} \sqrt{\frac{\Lambda}{12q_1}}l(l+2)\phi_1^2 - \frac{\Lambda}{6\hbar}(l(l+2))^{1/2} \phi_1^2\,.
\end{align}  
At large scale factor, the phase in the last expression can be dropped. We can approximate the integral running just above the cut from left to right as an integral over real $x$ along the path $N_\star^+ + x,$
\begin{align}
e^{\frac{i}{\hbar}S(N_\star^+)}\int_{-\infty}^{0} \mathrm{d}x e^{- \frac{\Lambda}{12\hbar}(l(l+2))^{1/2} \phi_1^2 x^2}  = &   \frac{\sqrt{3\pi\hbar}}{\sqrt{\Lambda}(l(l+2))^{1/4}\phi_1} e^{\frac{i}{\hbar}S(N_\star^+)} \nonumber \\ \sim & \sqrt{\frac{\hbar}{\Lambda}} e^{-i \sqrt{\frac{4\Lambda q_1^3}{3\hbar^2}} + \frac{3}{2\hbar\Lambda}(l(l+2))^{3/2}\phi_1^2}\,. \label{Cutcontribution1}
\end{align}
From the point $N_\star$ the path of steepest descent runs straight up, and we may check that including this contribution does not significantly affect the integral arising from the cut itself. For this case we add an integration along a path $N_{\star}^+ + i y$ with positive real $y,$ obtaining essentially the same result as above,
\begin{align}
e^{\frac{i}{\hbar}S(N_\star^+)}\left[\int_{-\infty}^{0} \mathrm{d}x e^{- \frac{\Lambda}{12\hbar}(l(l+2))^{1/2} \phi_1^2 x^2} + i \int_0^{+\infty} \mathrm{d} y e^{-\frac{3}{\hbar}y}\right]= & \left(  \frac{\sqrt{3\pi\hbar}}{\sqrt{\Lambda}(l(l+2))^{1/4}\phi_1} + \frac{i\hbar}{3} \right) e^{\frac{i}{\hbar}S(N_\star^+)} \nonumber \\ \sim &  \sqrt{\frac{\hbar}{\Lambda}} e^{-i \sqrt{\frac{4\Lambda q_1^3}{3\hbar^2}} + \frac{3}{2\hbar\Lambda}(l(l+2))^{3/2}\phi_1^2}\,.\label{Cutcontribution2}
\end{align}
Hence the integral around the branch cut yields unsuppressed fluctuations, with weighting 
\begin{equation}
\sqrt{\frac{\hbar}{\Lambda}} e^{+ \frac{3}{2\hbar\Lambda}(l(l+2))^{3/2}\phi_1^2}\,.
\end{equation}

\begin{figure}[h] 
\begin{minipage}{0.48\textwidth}
\includegraphics[width=\textwidth]{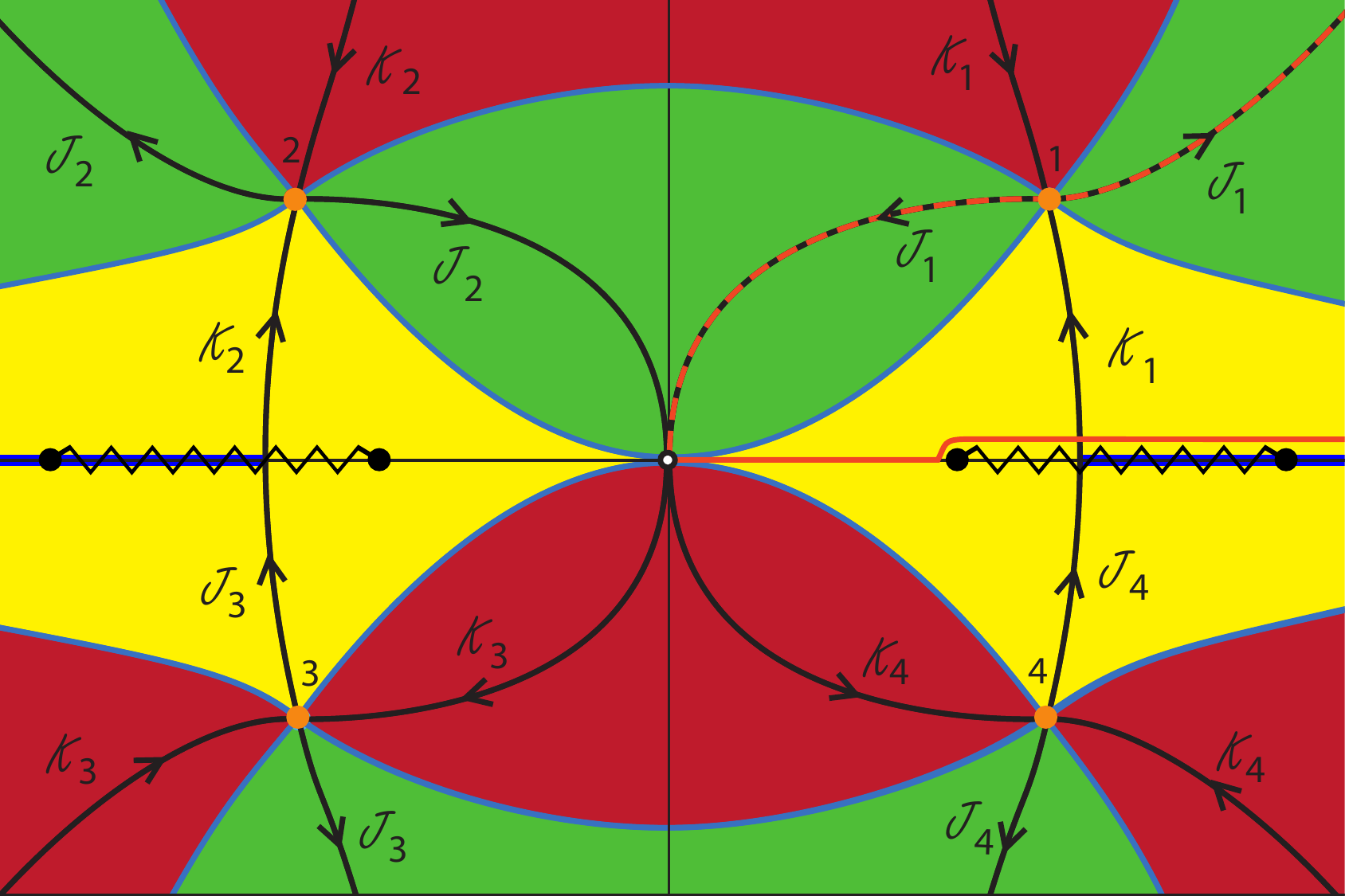}
\end{minipage}\quad
\begin{minipage}{0.48\textwidth}
\includegraphics[width=\textwidth]{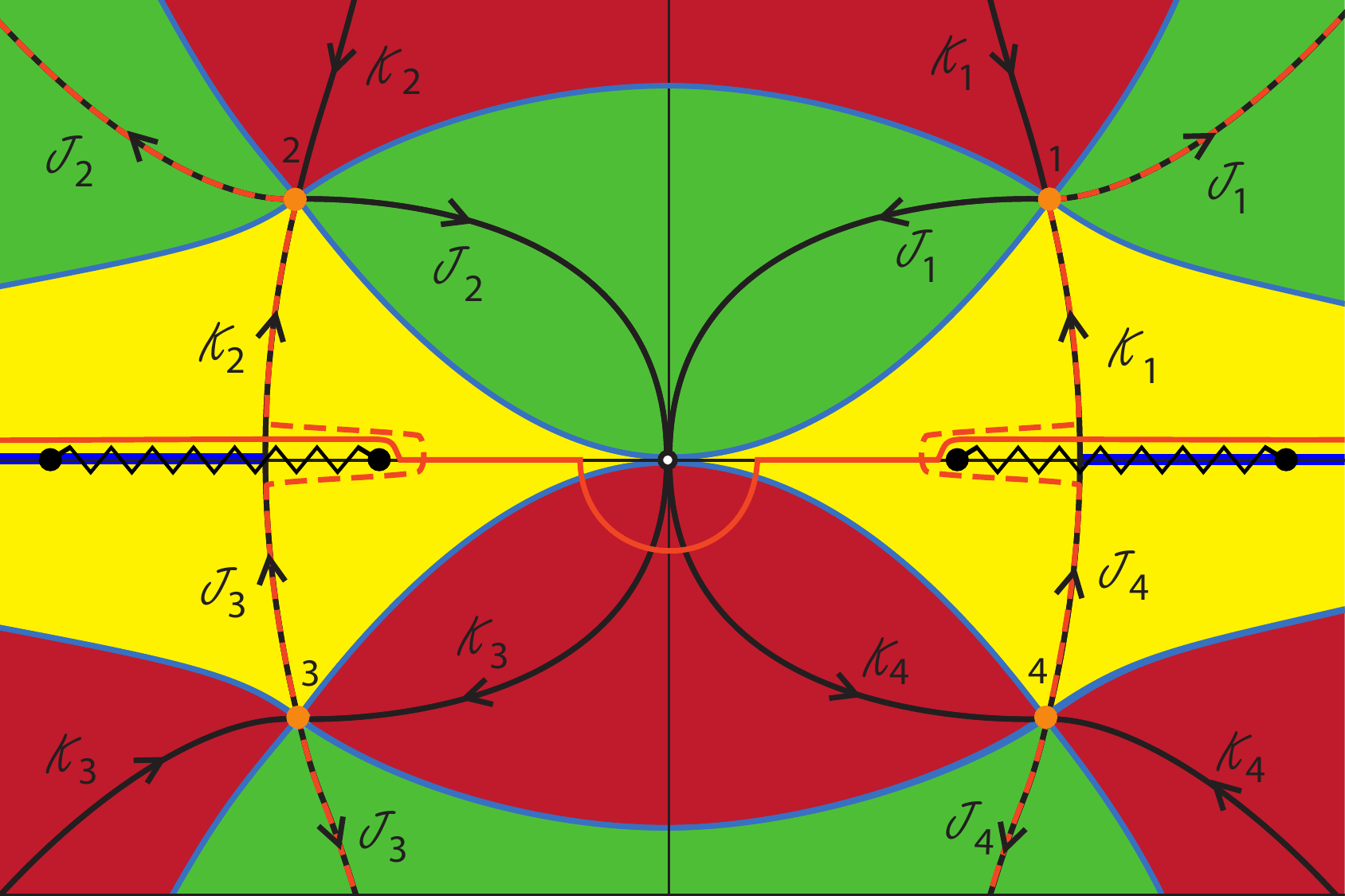}
\end{minipage}
\caption{Picard-Lefschetz theory for a $\Lambda$-dominated universe with gravitational waves. The solid orange and dashed orange lines are the original and deformed integration contours respectively, while the  zigzag lines denote the branch cuts. The lines denoted by $\mathcal{J}_i$ are lines of steepest descent and the lines denoted by $\mathcal{K}_i$ are lines of steepest ascent. Left panel: the integration path for the Lorenztian propagator, deformed to run above the cut.  Right panel: the integration domain prescribed by Diaz Dorronsoro {\it et al.} \cite{Dorronsoro:2017} with the original integration domain above the branch cuts. Note that the contour must be deformed to partially encircle the branch cuts in order to reach the Lefschetz thimbles.}
\protect
\label{fig:GW}
\end{figure}

\subsubsection{Integrating $N$ over positive values}

For the no boundary proposal, defined in terms of the propagator, we integrate over positive lapse $\mathcal{C}_1=(0^+,\infty).$ According to Picard-Lefschetz theory, we should distort the contour integral over $N$ to the relevant Picard-Lefschetz thimble $\mathcal{J}_1$ at the first stage of the calculation. We then need the solution for the classical background and perturbations, given in previous sections. Using the background action (\ref{eq:S0cl}) and the perturbation action (\ref{frozenmodes}), we obtain 
\begin{equation}
G_{\mathcal{C}_1}[q_1,\phi_1;0] \propto e^{i R(q_1,\phi_1)/\hbar}e^{-\frac{12\pi^2}{\hbar \Lambda}+ \frac{3}{2\hbar\Lambda}l(l+1)(l+2)\phi_1^2 }\,,
 \label{GC1above}
\end{equation}
with the phase given by the real part of the classical action 
\begin{align}
R(q_1,\phi_1)= -4 \pi^2 \sqrt{\frac{\Lambda}{3}} \left( q_1 - \frac{3}{\Lambda}\right)^{3/2} - \sqrt{\frac{3q_1}{4\Lambda}} l(l+2)\phi_1^2\,.
\label{eq:classPhase}
\end{align}
This is the result described in \cite{Feldbrugge:2017fcc}, where the background is suppressed as $e^{-12\pi^2/(\hbar\Lambda)},$ but the fluctuations are unsuppressed, so they are out of control. 


\subsubsection{Integrating $N$ from $-\infty$ to $+\infty$}

The integration domain $\mathcal{C}_2^+,$ deformed to pass above the essential singularity at the origin $N=0,$ gives twice the real part of the half line contour given (\ref{GC1above}) and (\ref{eq:classPhase}) above. Its implications are immediately obtained from Eq. \eqref{GC1above} above. We do not need to discuss this contour further.

Diaz Dorronsoro {\it et al.} propose to use the integration domain $\mathcal{C}_2^-,$ passing below the essential singularity at the origin $N=0$. As discussed above, asymptotically the contour must be deformed to run above the real $N$-axis to yield a convergent path integral, hence we will adopt this definition here. Picard-Lefschetz theory implies the relevance of all four saddle points, since the integration contour is intersected by all lines of steepest ascent (see the right panel in Fig. \ref{fig:GW}). For the background, the lower saddle points dominate over the upper saddle points leading to the Hartle-Hawking result. Obtaining this result for the background appears to have been the main goal in choosing this complex contour. When including the perturbation action $\bar{S}^{(2)}$, we need to deform the contour to avoid the branch cut on the side of the origin, as shown in the figure. Up to the first sub-leading order in the saddle point approximation, the path integral evaluates to
\begin{align}
G_{\mathcal{C}_2^-}[q_1,\phi_1;0]   \approx &2 \Re\left[e^{i {R(q_1,\phi_1)\over \hbar}}\left(C_4 e^{\frac{12\pi^2}{\hbar \Lambda}- \frac{3}{2\hbar \Lambda}l(l+1)(l+2)\phi_1^2} +(C_{b}+C_1e^{-\frac{12\pi^2}{\hbar \Lambda}})e^{+ \frac{3}{2\hbar \Lambda}l(l+1)(l+2)\phi_1^2}\right)\right]\,.
\label{GC2nonpert}
\end{align}
The functional determinants corresponding to the Hartle-Hawking and Picard-Lefschetz saddle points are $C_4$ and $C_1$. The term $C_{b}$ is the prefactor of the integral along the branch cut. The overall phase is again given by the real part of the classical action (see equation \eqref{eq:classPhase}).


The lower saddle points alone would have given the standard Bunch-Davies vacuum state. However, the branch cut and the upper saddle points lead to non-perturbative corrections, suppressed by one and two powers of the nonperturbative factor $e^{-12\pi^2/(\hbar\Lambda)}$ respectively. However, as both the perturbation amplitude $\phi_1$ and the wavenumber $l$ increase, as discussed below Eq.~(\ref{frozenmodes}), these corrections can dominate. The consequence is that the Bunch-Davies vacuum obtains corrections which are nonperturbative in the semiclassical ($\hbar$) expansion that are so large that the theory does not admit a sensible vacuum any more. Put differently, the no-boundary proposal does not imply the Bunch-Davies vacuum for perturbations, as was until recently believed. Rather, increased fluctuations receive an ever larger weighting, leading to a breakdown of the model. 

\bigskip
The instability can be related to the existence of the branch cuts on the real line in the perturbative action \eqref{eq:S2}. In the absence of such a singularity, the Morse function on the real line is strictly zero, and Picard-Lefschetz theory implies that any contour defined on the real line would have to flow down to lower values to be expressible as a manifestly convergent integral. In this case it would be impossible for the total weighting to become positive, and the fluctuations would not be able to surpass the background. In fact this makes perfect sense: quantum effects are suppressed compared to classical evolution, which would occur with probability $1$ (\textit{i.e.} weighting  $0$). The branch cut changes this. The Morse function no longer tends to zero as one approaches the real $N$-axis from above or below the cut -- rather it has a discontinuity leading to the instability discussed above. Another aspect of the problem is to notice that the resulting amplitude violates the correspondence principle, \textit{i.e.} classical physics is no longer recoverable in the limit $\hbar \rightarrow 0.$ For large fluctuations, the propagators \eqref{GC1above}, \eqref{GC2nonpert} do not satisfy this condition. Whichever point of view one prefers, the conclusion in all cases is that the no boundary proposal becomes untenable and that the idea of a smooth semi-classical beginning of the universe fails.

\section{Backreaction}\label{sec:Back}

The inverse Gaussian distribution of the tensor perturbations, described in the previous section, arises already within the limits of validity of general relativistic linear perturbation theory, signalling a clear problem with the no boundary proposal. However, one may wonder whether  backreaction of the gravitational waves might be significant in the regime where the upper saddle points start to dominate over the lower saddle points (assuming the contour $\mathcal{C}_{2}^-$ passing below the origin). To settle this question we studied the backreaction numerically, in two representative situations of interest: for the lowest modes, \textit{i.e.}, for the $l=2$ modes, we have evaluated the full Einstein equations numerically. For the higher $l$ modes we have solved the linear equation of motion for $\phi$ and included its backreaction at second order in the equation of motion for the scale factor $q$. As we will discuss below, these studies serve to reinforce the conclusions drawn in linear perturbation theory.

\subsection{The $l=2$ mode}
The $l=2$ modes are particularly interesting as a possible non-linear completion of the metric exists, in the form of the Bianchi IX line element
\begin{align}
ds_{IX}^2 = - N_p^2(t)dt_p^2 + \sum_m \left( \frac{l_m(t)}{2} \right)^2 \sigma_m^2\,,
\end{align}
where $N_p$ is the physical lapse function and $\sigma_1 = \sin\psi d\theta - \cos \psi \sin \theta d\varphi$, $\sigma_2 = \cos \psi d\theta + \sin \psi \sin \theta d \varphi$, and $\sigma_3 = - (d\psi + \cos\theta d\varphi)$ are differential forms on the three sphere such that $0 \leq \psi \leq 4 \pi$, $0 \leq \theta \leq \pi$, and $0 \leq \phi \leq 2 \pi.$ For ease of notation we will denote a derivative w.r.t. physical time $t_p$ by an overdot in this section (and only in this section). Employing the original definition of Misner \cite{Misner:1969hg}, we can re-write the three scale factors as 
\begin{align}
l_1(t_p) &= a(t_p) \exp \left[\frac{1}{2}\left(\beta_+(t_p) + \sqrt{3}\beta_-(t_p)\right)\right]\,, \\
l_2(t_p) &= a(t_p) \exp \left[\frac{1}{2}\left(\beta_+(t_p) - \sqrt{3}\beta_-(t_p)\right)\right]\,, \\
l_3(t_p) &= a(t_p) \exp \left[-\beta_+(t_p)\right]\,,
\end{align}
which makes it clear that $a$ is the average scale factor while the $\beta$s quantify anisotropic perturbations. In these coordinates the action becomes
\begin{align}
S = 2\pi^2 \int dt_p N_p a \left[  \frac{1}{N_p^2}\left( -3\dot{a}^2 +a^2 \left(\frac{1}{2}\dot{\phi}^2 + \frac{3}{4}\dot{\beta}^2_+ + \frac{3}{4}\dot{\beta}^2_- \right) \right) - \left( a^2  V(\phi) + U(\beta_+, \beta_-)\right)\right]\,,
\end{align}
where the full non-linear potential is given by
\begin{align} \label{anisotropypotential}
U(\beta_+, \beta_-)  & = - 2 \left( e^{ 2 \beta_+ } + e^{-\beta_+ - \sqrt{3}\beta_-} + e^{-\beta_+ + \sqrt{3}\beta_-} \right) + \left( e^{ -4 \beta_+ } + e^{2\beta_+ - 2\sqrt{3}\beta_-} + e^{2\beta_+ + 2\sqrt{3}\beta_-} \right)  \\ 
& = -3 + 6 \beta_+^2 + 6 \beta_-^2 + \dots \label{anisotropypotentialexpanded}
\end{align}
Varying with respect to the lapse $N_p$ (and working in the gauge $\dot{N}_p=0$) we obtain the Friedman constraint equation
\begin{align} \label{Friedman}
3\dot{a}^2 =a^2 \left( \frac{3}{4}\dot{\beta}^2_+ + \frac{3}{4}\dot{\beta}^2_- \right)  +  N_p^2  U(\beta_+, \beta_-)\,,
\end{align}
while the equations of motion for $a,\beta_+,\beta_-$ are given by
\begin{align}
\frac{\ddot{a}}{a} + \frac{1}{2}\frac{\dot{a}^2}{a^2} + \frac{3}{8}\left( \dot{\beta}^2_+ + \dot{\beta}^2_- \right) - \frac{N_p^2}{6a^2} U(\beta_+, \beta_-)  & = 0\,, \\
  \ddot{\beta}_\pm + 3\frac{\dot{a}}{a}\dot{\beta}_\pm  + \frac{2}{3}\frac{N_p^2}{a^2} U_{,\beta_\pm} & = 0\,. \label{betap} 
\end{align}
Expanding the last equation to linear order we obtain
\begin{equation}
\ddot{\beta}_\pm + 3\frac{\dot{a}}{a}\dot{\beta}_\pm  + 8 \frac{N_p^2}{a^2} \beta_\pm  = 0\,.
\end{equation}
A comparison with Eq. \eqref{eom4chi} confirms that the $\beta$s are non-linear versions of the $l=2$ modes -- more specifically, they are non-linear versions of two $l=2$ modes which are such that they preserve the Bianchi IX symmetry \cite{Grishchuk:1975ec}. 
To match with our earlier normalization conventions, one has to re-scale
\begin{align}
\beta_\pm = \frac{1}{\sqrt{3}\pi} \phi_{\pm}\,,
\end{align}
where $\phi_{\pm}$ denote two separate $l=2$ modes. The structure of the potential $U$ shows that when going beyond linear order, the equations of motion lead to direct couplings between these two $l=2$ modes. 

\begin{figure}[h]
\begin{minipage}{0.5\textwidth}
\includegraphics[width= \textwidth]{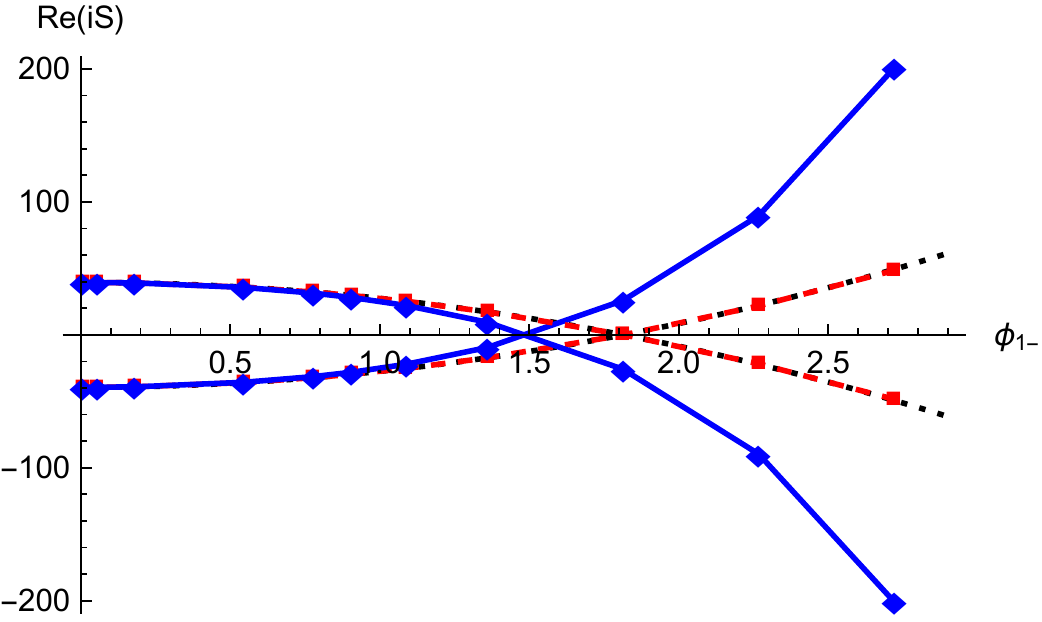}
\end{minipage}%
\begin{minipage}{0.5\textwidth}
\includegraphics[width=0.9\textwidth]{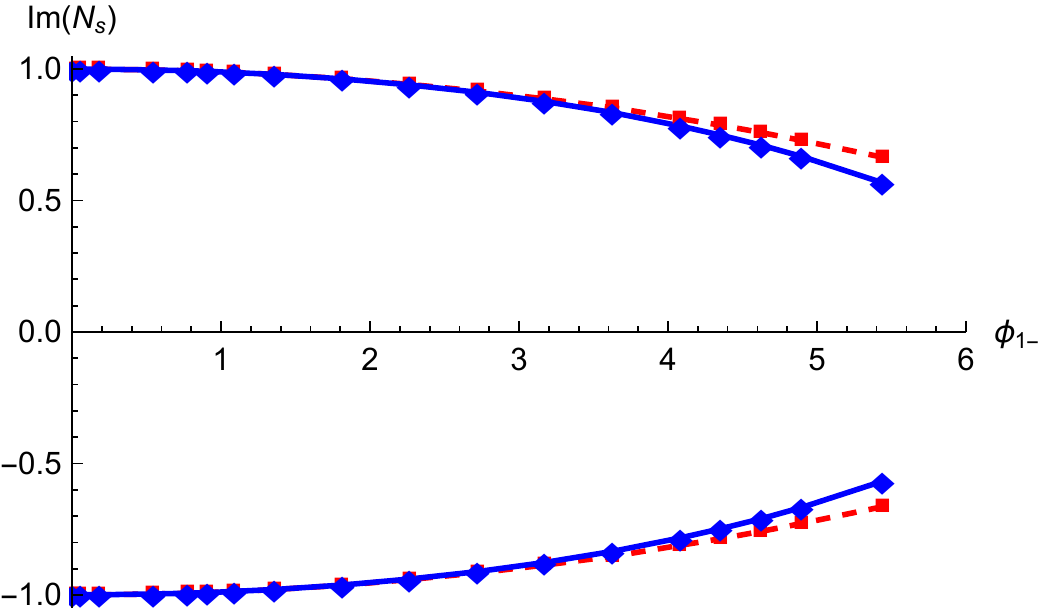}
\end{minipage}%
\caption{These graphs show the weighting at the saddle points (left panel) and the imaginary part of the saddle point locations (right panel) as a function of the $l=2$ anisotropy mode amplitude $\phi_{1-},$ for $\Lambda=3.$ In the plot of the action (left) the line starting at $+12\pi^2/\Lambda=+4\pi^2$ for $\phi_{1-}=0$ corresponds to the saddle points in the lower half $N$ plane, while the line starting at $-4\pi^2$ corresponds to the saddle points in the upper half plane. In black (mostly hidden behind the red line) are the linear results without backreaction, in red the results including backreaction but still in linear perturbation theory, and in blue the results stemming from solving the fully backreacted Einstein equations. For values of $\phi_{1-}$ below $1$ the linear and non-linear results agree to high precision, while one can see that at larger values of the anisotropy the non-linear corrections enhance the instability of the fluctuations, and move the saddle points further towards the real $N$-axis. Note that the weighting of the upper saddle points surpasses that of the lower ones when backreaction is still entirely negligible. Moreover, the non-linear effects of the full Einstein equations imply that the (unstable) upper saddle points come to dominate already for smaller amplitudes of the fluctuations.}
\label{fig:nonlinear}
\end{figure}

In the present section we work in a gauge where $N_p=1$ and where one then has to determine the value of the time coordinate of the final hypersurface on which the boundary conditions $q_0=0,q_1=a_1^2, \phi_\pm=\phi_{1\pm}$ are satisfied. This is done using the shooting method discussed in \cite{Bramberger:2017rbv}. In this method, the (generally complex valued) second time derivatives of $\phi_\pm$ at the no boundary point $a=0$ are adjusted using an optimization algorithm such that at a final time $t_f$ the desired real values $q_1, \phi_{1\pm}$ are simultaneously reached. The total time interval $\int N_p \mathrm{d}t_p = t_f$ can then also be related to the lapse function $N$ using the change of coordinates $N_p \mathrm{d}t_p = N q^{-1/2} \mathrm{d}t,$
\begin{equation}
N = \int_0^1 N \mathrm{d} t = \int_0^{t_f} a(t_p)\mathrm{d}t_p\,.
\end{equation}

Our results are shown in Fig. \ref{fig:nonlinear}. For ease of comparison with linear perturbation theory, we only show results for the case where a single $l=2$ mode (here chosen to be $\phi_{1-}$) takes on a non-trivial value on the final hypersurface. The left panel shows how the weighting of the saddle point solution (for saddles in the upper half plane) increases as the perturbation amplitude is increased. The opposite behavior is seen for the saddle points in the lower half plane. As is evident from the figure, backreaction at second order in perturbation theory is utterly negligible. Even more importantly, the effects of the instability are even stronger when non-linear terms are included, and the dominance of the upper saddle point over the lower ones occurs already for smaller values of $\phi_{1-}$ than in the linear theory. Also, as shown in the right panel, the saddle point moves faster towards the real $N$-axis in the non-linear theory. These results consolidate our analytic results, and indicate that the inclusion of non-linear backreaction only reinforces the instability that we have identified.

\subsection{Backreaction in $\phi$ of higher $l$ modes}

To quadratic order in the gravitational wave modes, the equations of motion corresponding to the total action $S=S^{(0)}+S^{(2)}$ are
\begin{align}
0=&\,\ddot{q}- \frac{2 N^2}{3} \Lambda + \frac{\dot{\phi}^2}{3 \pi^2} q\,, \label{eq:shoot1}\\
0=&\,\ddot{\phi} +2 \frac{\dot{q}}{q} \dot{\phi} + \frac{N^2}{q^2} l(l+2)\phi\,. \label{eq:shoot2}
\end{align}
The term $\frac{\dot{\phi}^2}{3 \pi^2} q$ encodes the backreaction of the perturbations $\phi$ on the scale factor $q$, ignored in the analytic calculations of the previous sections.

\begin{figure}[h]
\begin{minipage}{0.49\textwidth}
\includegraphics[width=\textwidth]{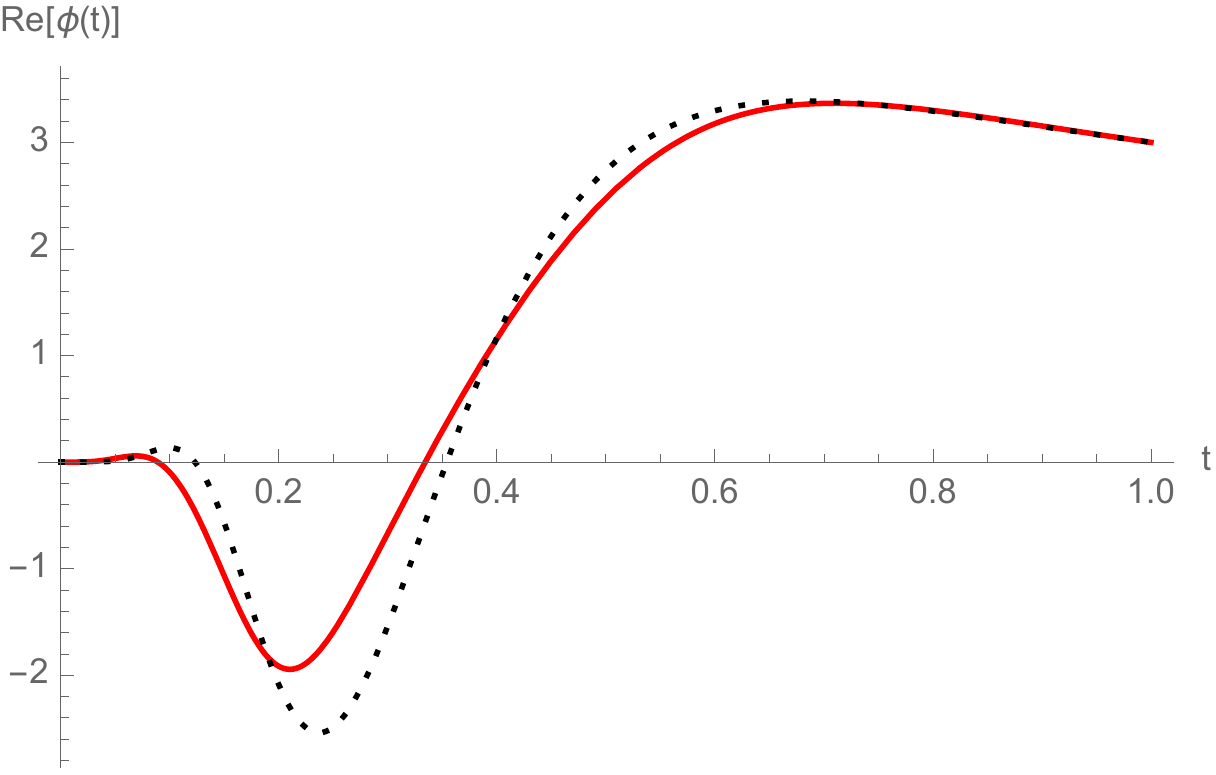}
\end{minipage}\ 
\begin{minipage}{0.49\textwidth}
\includegraphics[width=\textwidth]{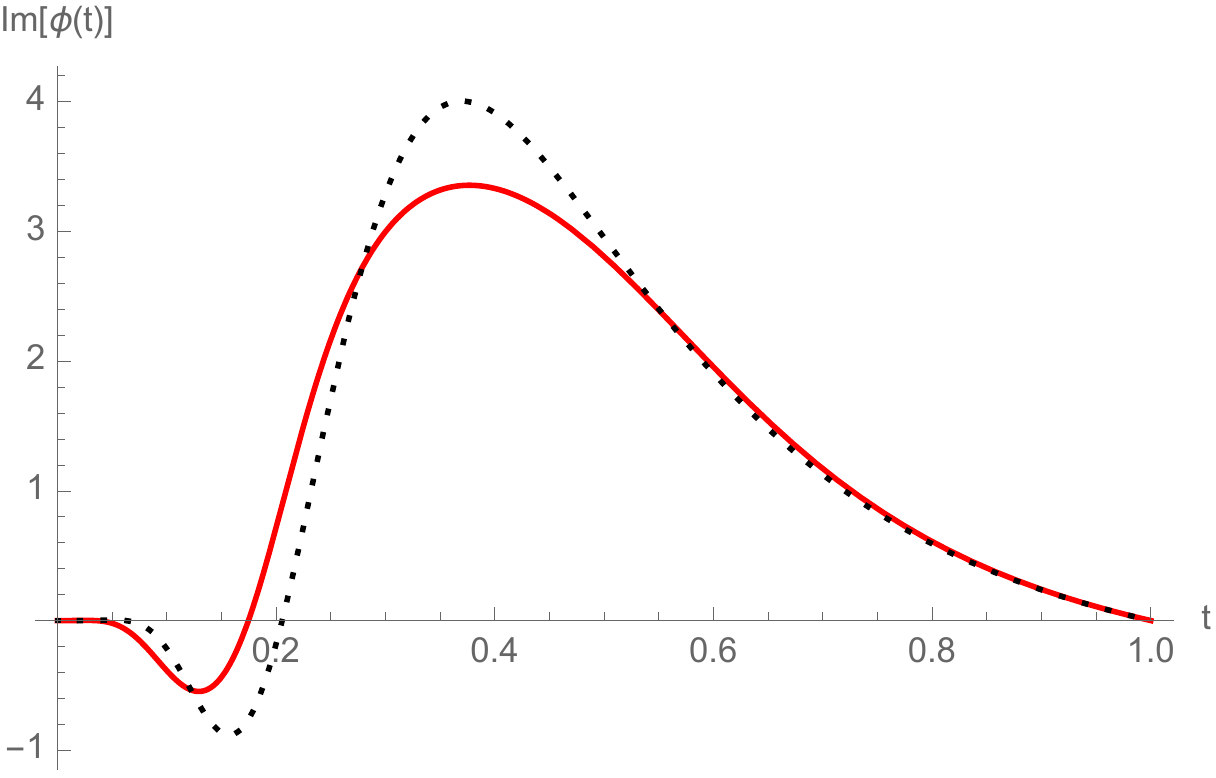}
\end{minipage}
\caption{The real and imaginary parts of the mode function $\phi(t)$ in the saddle point $N_{1}$ with the boundary conditions $q_0=0,\phi_0=0$, $q_1=101$, and $\phi_1=3$ for mode $l=10$. The black dashed lines correspond to the analytic result without backreaction. The red lines correspond to the numerical analysis with backreaction. We observe that the backreaction for $\phi_1=3$, which violates the perturbation theory condition $|\phi(t)| > 1$, leads to a change in the mode functions of approximately $10\%$, with no qualitative change. For the boundary condition $0 \leq \phi_1 \leq 0.8$, which do satisfy the condition $|\phi(t)| < 1$ for all $t\in [0,1]$, the correction due to backreaction is completely negligible.}
\protect
\label{fig:modes}
\end{figure}

We use a numerical shooting method to solve the equations of motion  \eqref{eq:shoot1}  and \eqref{eq:shoot2}  with ``no boundary" boundary conditions $q(0)=\phi(0)=0,\,q(1)=q_1$, and $\phi(1)=\phi_1$ for a given spherical wavenumber $l$ and lapse $N.$ We start with the analytic solution \eqref{eq:qclass} of the scale factor ignoring backreaction. We then solve \eqref{eq:shoot2} numerically for an arbitrary $\phi(1),$ and normalize the solution to enforce $\phi(1)=\phi_1$. The resulting approximate solution for $\phi$ is used in the equation of motion \eqref{eq:shoot1} for $q(t)$,  likewise solved as a one dimensional shooting problem. The procedure is  then iterated until both solutions converge. Fig.~\ref{fig:modes} compares the results with and without backreaction included, for a saddle point solution with $\phi_1=3$, for which the perturbation contribution to the Morse function at the saddle outweighs that for the unperturbed background. 

 \begin{figure}[h]
\centering
\includegraphics[width=0.75 \textwidth]{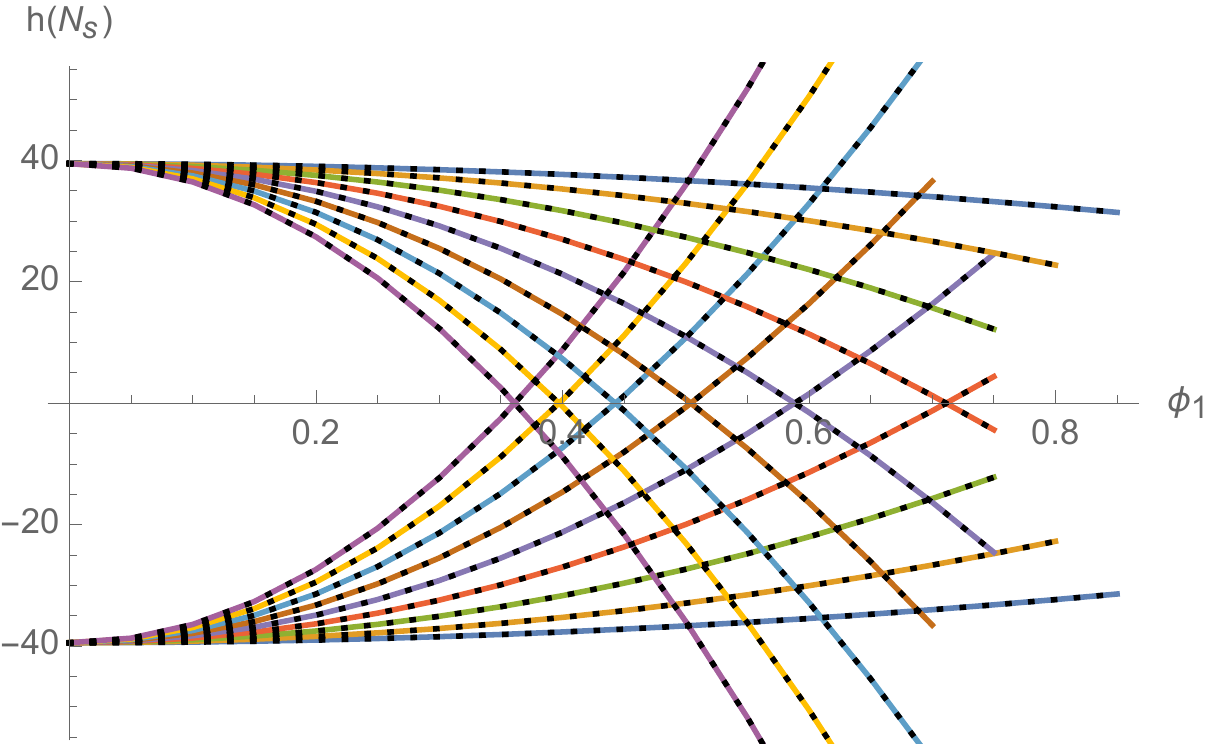}
\caption{The Morse function evaluated at the saddle points $h(N_s)$ as a function of the boundary condition $\phi_1$ for which $|\phi(t)|\leq 1$ for all $t$. The lines starting from $h = -4\pi^2$ and $h=4\pi^2$ correspond to the upper and the lower saddle points. The blue to purple lines correspond to $l=2,3,\dots,10$. The colored lines are the numerical simulations with backreaction. The dotted lines are the analytic results without backreaction. We observe that the backreaction of the perturbation $\phi$ on the scale factor $q$ is very small for the boundary condition $0 \leq \phi_1 \leq 0.8$. Consequently, the analytic calculations which neglect backreaction are accurate when the upper saddle point starts to dominate over the lower saddle point for $l\geq 5$.}
\label{fig:BackReaction}
\end{figure}

For more modest values of the final perturbation, $0 \leq \phi_1 \leq 1$, we find that the backreaction is small and does not significantly affect the location of the saddle points of the Morse function in the complex $N$-plane. The saddle points move the most for the high $l$ modes. For $l=10,$ $q_1=101$ and $\phi_1=1,$ for example, the saddle points are located at $N_s= \pm 10.0232 \pm 0.97904 i$, compared to the background saddle points $N_s=\pm 10 \pm i$. In agreement with our analytic arguments, the value of the Morse function at the saddle points does, however, change significantly as $\phi_1$ is increased (see Fig. \ref{fig:BackReaction}). The lines starting from $h=-4 \pi^2$ and $h=+4 \pi^2$ correspond respectively to the upper and the lower saddle points. For the modes $l=2,3$ and $l=4$ the lower saddle point always dominates over the upper saddle point in the regime of validity of linear perturbation theory. This coincides with the analysis of the $l=2$ mode using full Einstein equations, discussed above, which showed that 
 the upper saddle only dominates at values of $\phi_1$ greater than $\sim 1.5$). For $l \geq 5$, the upper saddle points dominate over the lower saddle points within the regime of validity of linear theory. 

\begin{figure}[h]
\centering
\includegraphics[width=0.75 \textwidth]{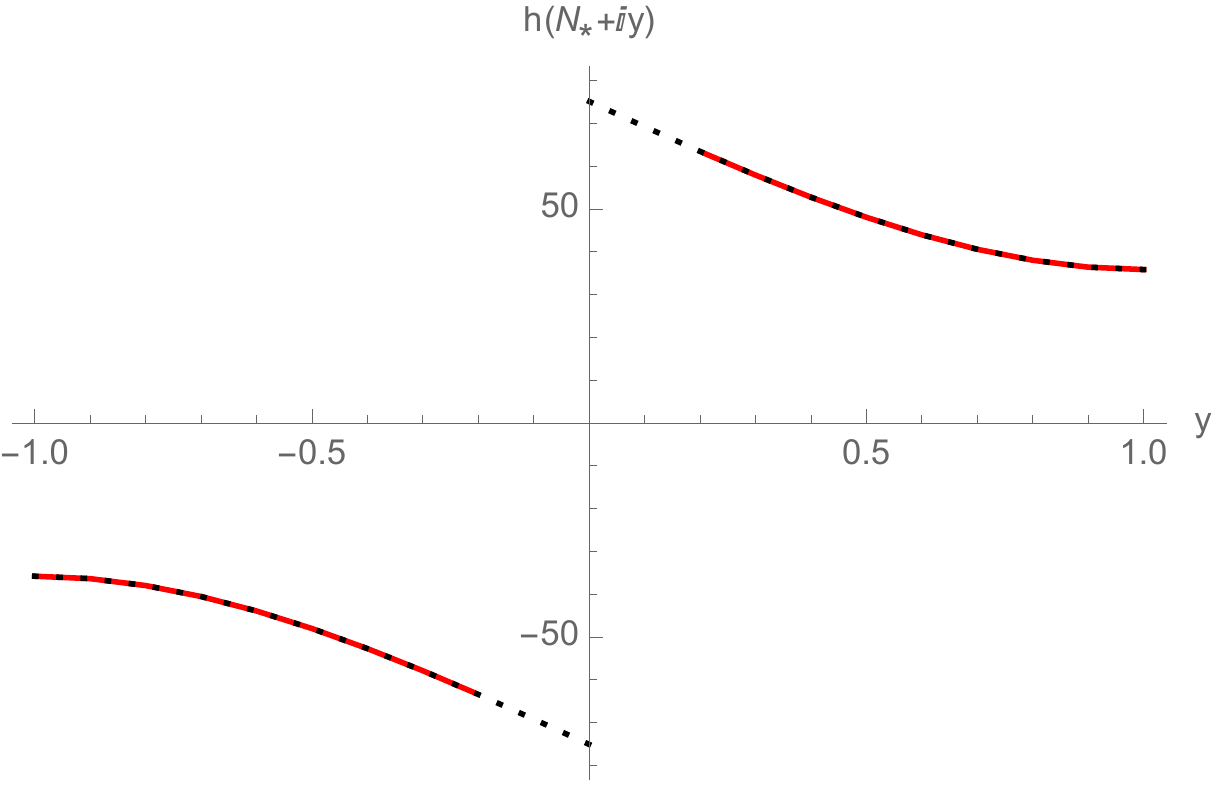}
\caption{The Morse function $h$ evaluated at $N=N_\star+i y$ with the boundary conditions $\Lambda = 3, q_1=101, \phi_1=0.5$ for the mode $l=10$. The dashed line is the analytic result without backreaction. The red line is the numerical calculation including backreaction. We plot the backreaction calculation for the points for which $|\phi(t)| < 1$ ({\it i.e.} for which linear theory is reliable) and the backreaction on the $q$ variable is small, {\it i.e.}, $\frac{\dot{\phi}^2q}{3 \pi^2} < \left| \frac{2N^2 \Lambda}{3} \right|$ for all $t$. Note that the backreaction is extremely small near the saddle point for these boundary conditions. The jump at $y=0$ illustrates the branch cut. The backreaction becomes significant when $N$ approaches the real axis. However, the backreaction does not appear to remove the branch cut.}
\label{fig:BranchCut}
\end{figure}

As explained in section \ref{sec:Pert}, it is possible for the upper saddles to dominate (and even to acquire a positive real exponent) due to the existence of  branch cuts on the real line in the effective action for $N$, given in \eqref{eq:S2} and \eqref{classact}. Fig.~\ref{fig:BranchCut} shows that including the effects of backreaction does not remove this branch cut. As we approach the point $N_{\star}$ from above and below the real $N$-axis, one can see the jump in the Morse function at $N_\star$ is maintained.

\begin{figure}[h]
\begin{minipage}{0.4\textwidth}
\includegraphics[width=\textwidth]{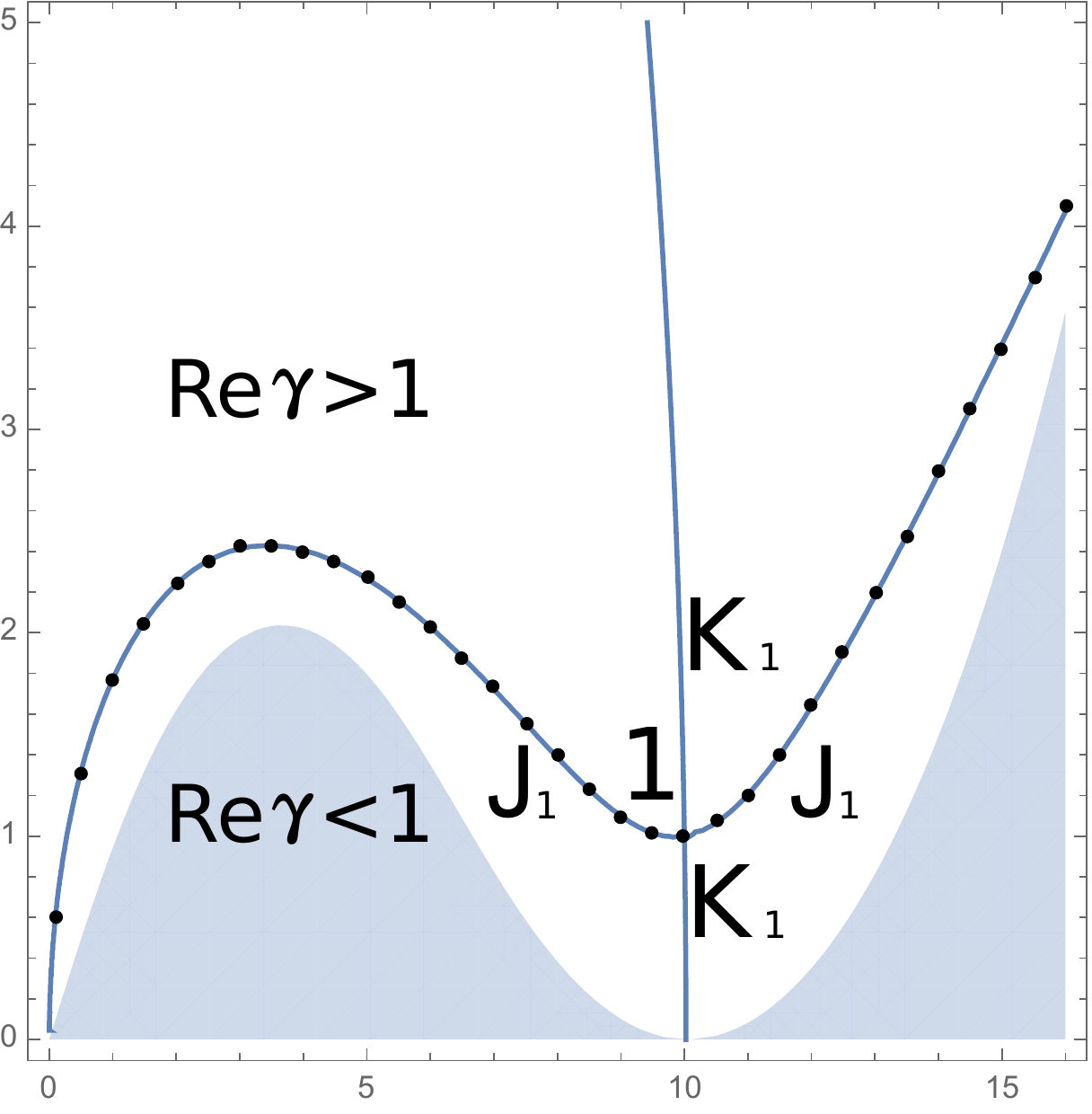}
\end{minipage}\\
\begin{minipage}{0.49\textwidth}
\includegraphics[width=\textwidth]{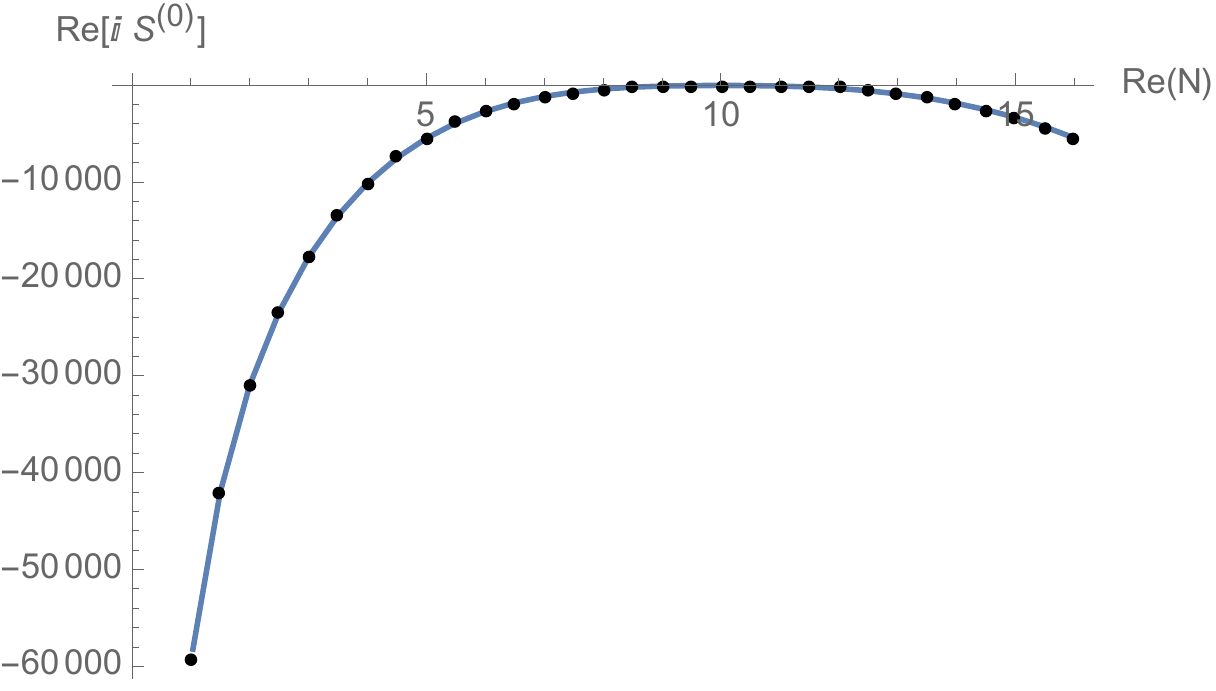}
\end{minipage}
\begin{minipage}{0.49\textwidth}
\includegraphics[width=\textwidth]{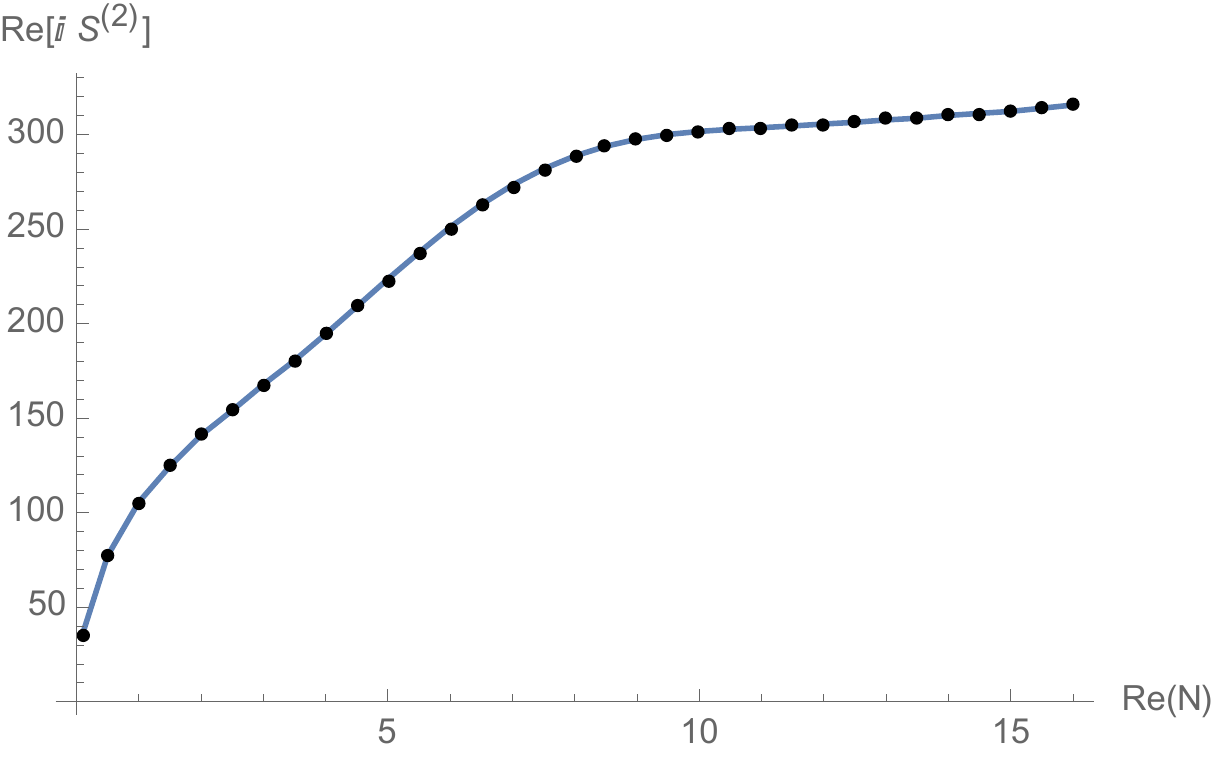}
\end{minipage}
\caption{The Morse function along the thimble for the boundary conditions $\Lambda = 3, q_1=101, \phi_1=1$ for the mode $l=10$. For these boundary conditions the action of the background and perturbations are comparable in the saddle points. Upper panel: first quadrant of the complex $N$-plane with the lines of steepest ascent and descent of the upper saddle point. The shaded/white regions, defined by $\text{Re}[\gamma]$ less/greater than $1$ respectively, denote the regions in which the finite action condition selects a mode which diverges/vanishes on the initial boundary. The points on the thimble $\mathcal{J}_1$ indicate the values of the lapse $N$ at which we have evaluated the importance of backreaction in the lower panels. Lower panels: the Morse functions for the background and perturbation actions $\bar{S}^{(0)}$ and $\bar{S}^{(2)}$. The blue lines denote the analytic Morse function while the black dots denote the numerical results including backreaction. Note that nonlinear backreaction is completely negligible along the thimble.}
\protect
\label{fig:AlongThimble}
\end{figure}

In general, we find that the effects of nonlinear backreaction are significant only near the real $N$-axis. This is perhaps not surprising, since we know the theory is quite singular there: for example it is well known that the Bianchi IX model studied in the previous section exhibits chaotic behavior for real metrics as the singularity $q=0$ is approached. However, we find that nonlinear backreaction is insignificant along the Lefschetz thimble $\mathcal{J}_1$ associated with the upper saddle point, {\it i.e.} the thimble relevant to the strictly Lorentzian path integral. The upper panel of Fig.~\ref{fig:AlongThimble} illustrates the first quadrant of the complex $N$ plane. In the white region, defined by $\text{Re}[\gamma]>1$, the finite action condition forces the mode function to vanish on the initial boundary, i.e. $\phi_0=0$. In this region the shooting method described above may be used. In contrast, in the shaded region, defined by $\text{Re}[\gamma]\leq 1$, the finite action condition selects a mode function $\phi$ which diverges on the initial boundary and a different method should be used.

Note that the part of the Lefschetz thimble shown in the Figure lies entirely in the white region. In order to study the significance of backreaction, we selected $32$ regularly spaced points along the thimble. In the lower two panels we plot the Morse functions given by the analytic calculations and the numerical calculations including backreaction, respectively, for each of the background and fluctuation actions. We observe that backreaction is negligible along the Lorentzian Lefschetz thimble $\mathcal{J}_1$. Thus, deforming the contour from real fields to the Lefschetz thimble appears to render the path integral significantly more calculationally tractable. 

\subsection{Backreaction in $\chi$}

\begin{figure}[h]
\centering
\includegraphics[width=0.75 \textwidth]{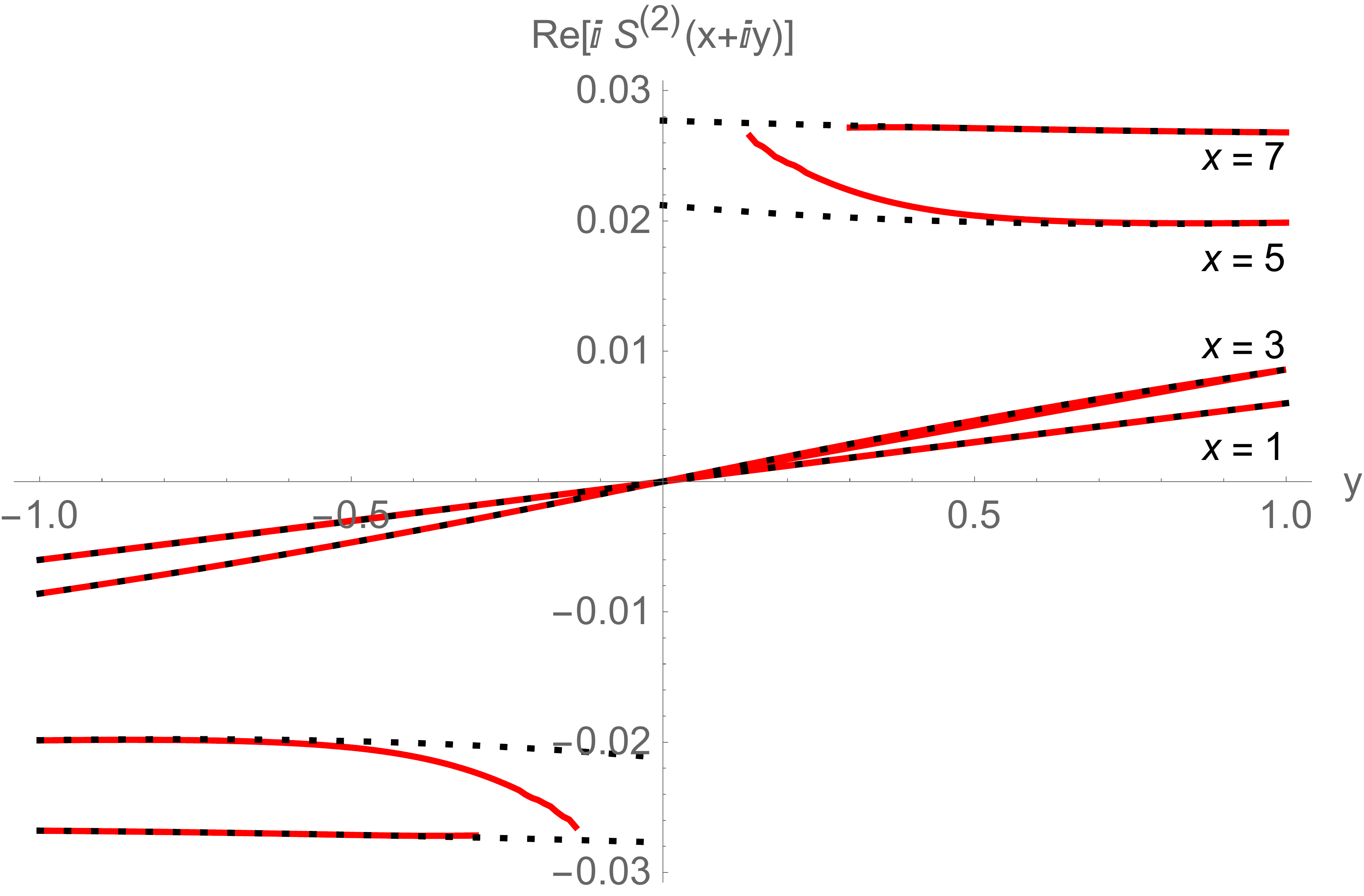}
\caption{The Morse function for the perturbations is plotted along a line $N=x+i\,y$ in the complex $N$-plane, crossing the real $N$-axis $y=0$ in the vertical direction. The parameters and boundary values are $\Lambda=3,q_0=0,q_1=101,\, l=10$, $\chi_0=0,$ and $\chi_1=1$. The black dashed line is the analytic result of perturbation theory without backreaction. The red line is the numerical result including backreaction up to second order in the perturbations.}

\label{fig:CrossingRealAxis}
\end{figure}

In the previous sections we studied the significance of backreaction in terms of the mode function $\phi$. As can be seen in the upper panel of Fig.~\ref{fig:AlongThimble}, the shooting method for $\phi$ only works for part of the complex $N$ plane. Near the real line, in particular, the shooting method for $\phi$ breaks down. In order to study the effects of backreaction near the real $N$-axis to the left of the branch cut we change coordinates to $\chi = q \phi$ as was discussed in section \ref{sec:Pert}. To quadratic order in the gravitational wave modes, the equations of motion corresponding to the action $S = S^{(0)} + S^{(2)}$ are 
\begin{align}
0=&\, \ddot{q} - \frac{2N^2}{3}\Lambda  + \frac{q}{2\pi^2}\left(\frac{\dot{\chi}}{q} - \frac{\dot{q}}{q^2}\chi\right)^2\\
0=&\ddot{\chi} - \left(\frac{\ddot{q}}{q} - \frac{N^2 l(l+2)}{q^2}\right)\chi
\end{align}
By solving these equations with the shooting method described in the previous section we can study the effects of backreaction in the shaded regions of the upper panel of Fig.~\ref{fig:AlongThimble} since the finite action condition implies $\chi_0=0$ (see Fig.~\ref{fig:CrossingRealAxis}). Observe that for $\text{Re}[N]=1,3$ the Morse function does not appear to be discontinous across the real $N$-axis ({\it i.e.} it suffers no branch cut) and backreaction remains negligible. When crossing the real line further to the right, for $\text{Re}[N]=5,7,$ we do find a branch cut and likewise notice significant backreaction. These results indicate that the presence or absence of a branch cut in the effective action for $N$, on the real $N$-axis, appears at least qualitatively consistent with indications from perturbation theory. This finding is significant for the discussion of the next section.

\section{No Contour Works}

In Section \ref{sec:principles} we reviewed the physical principles for integrating over positive lapse ($\mathcal{C}_1$) when defining the Lorentzian propagator. In contrast, Diaz Dorronsoro {\it et al.}~\cite{Dorronsoro:2017} proposed a different, intrinsically complex contour $\mathcal{C}_2^-$ running below the origin over $-\infty <N<\infty$, with the motivation of obtaining a real wavefunction satisfying the Wheeler-DeWitt equation. (For an earlier discussion of various lapse contours for the background, see, {\it e.g.}, Ref.~\cite{Halliwell:1988ik}). We have already shown that with either choice, one cannot avoid contributions from saddles 1 and 2 (shown in Fig.~\ref{fig:Map}), both of which yield an unsuppressed, inverse Gaussian distribution for the perturbations. The goal of this section is to extend this analysis and show that {\it no} integration contour avoids this problem. 

Before passing to the proof of this general result, let us also mention the new complication we have detailed in this paper. When the condition of finite action is imposed upon the perturbation modes, a large part of the real $N$-axis must be excluded from the definition of any possible contour, due to the presence of branch cuts and the non-existence of finite action perturbation modes at large $|N|.$

\subsection{Time reparametrization invariance}

Let us consider a generalized path integral, where the lapse $N$ is integrated over some complex contour ${\cal C}$ in the complex $N$-plane. Time reparametrization invariance severely limits the points at which ${\cal C}$ may start and end. Defining $\langle 1| 0 \rangle_{\cal C}  \equiv \int_{\cal C} dN \langle 1| e^{-i H N/\hbar}| 0\rangle$, as a natural generalization of (\ref{prop}), it follows that $H \Psi= \int_{\cal C} i \hbar (d/dN)  \langle 1| e^{-i H N/\hbar}| 0\rangle $, {\it i.e.}  one obtains surface terms at the start and end of ${\cal C}$. If one insists that the Hamiltonian annihilates the amplitude, it follows that all surface terms must vanish: that is, ${\cal C}$ must start and end at a point where the Morse function $h(N)$ tends to minus infinity (we assume the $h(N)$ dependence dominates over the prefactor). There is a subtlety at the point $N=0$ because, as discussed above Eq.~(\ref{prop2}), the integrand $ \langle 1| e^{-i H N/\hbar}| 0\rangle$ is singular in the small $N$ limit, with the prefactor and the exponent diverging in just such a way that the combined surface term generates a delta function. This is a universal property of quantum propagators, ensuring that they are Green's functions. As we reviewed in Section \ref{sec:principles}, for geometrodynamics this introduces a primitive notion of causality.  Nevertheless, it remains true that the only acceptable contours for ${\cal C}$ are those which start and end at points where the Morse function tends to minus infinity. Furthermore, if the integral over $N$ is to converge, ${\cal C}$ must approach these singular points along paths which can be deformed into paths of steepest descent. Near $N=0$, the background action (\ref{eq:S0cl}) diverges as $-A/N$ with $A$ positive. It follows that ${\cal C}$ must approach the origin along a path which can be deformed to run into the origin along the positive imaginary axis, since that is the unique curve of steepest descent which ends at $N=0$. In contrast, when we consider the singular point at $N=\infty$, the classical action is proportional to  $A N^3$, with $A$ positive. So the point at infinity may be approached in three inequivalent ways: the exponent $i N^3$ yields steepest descent paths along $N=|N| e^{i \theta}$ with $\theta=\pi/6$, $5 \pi/6$ or $3\pi/2$. (As we shall discuss later, in general there is a branch cut emanating from the origin. In this case we must distinguish the angle $3\pi/2$ from $-\pi/2$, relative to the positive $N$-axis.)

\subsection{Picard-Lefschetz theory}

\begin{figure}
\centering
\includegraphics[width=0.75\textwidth]{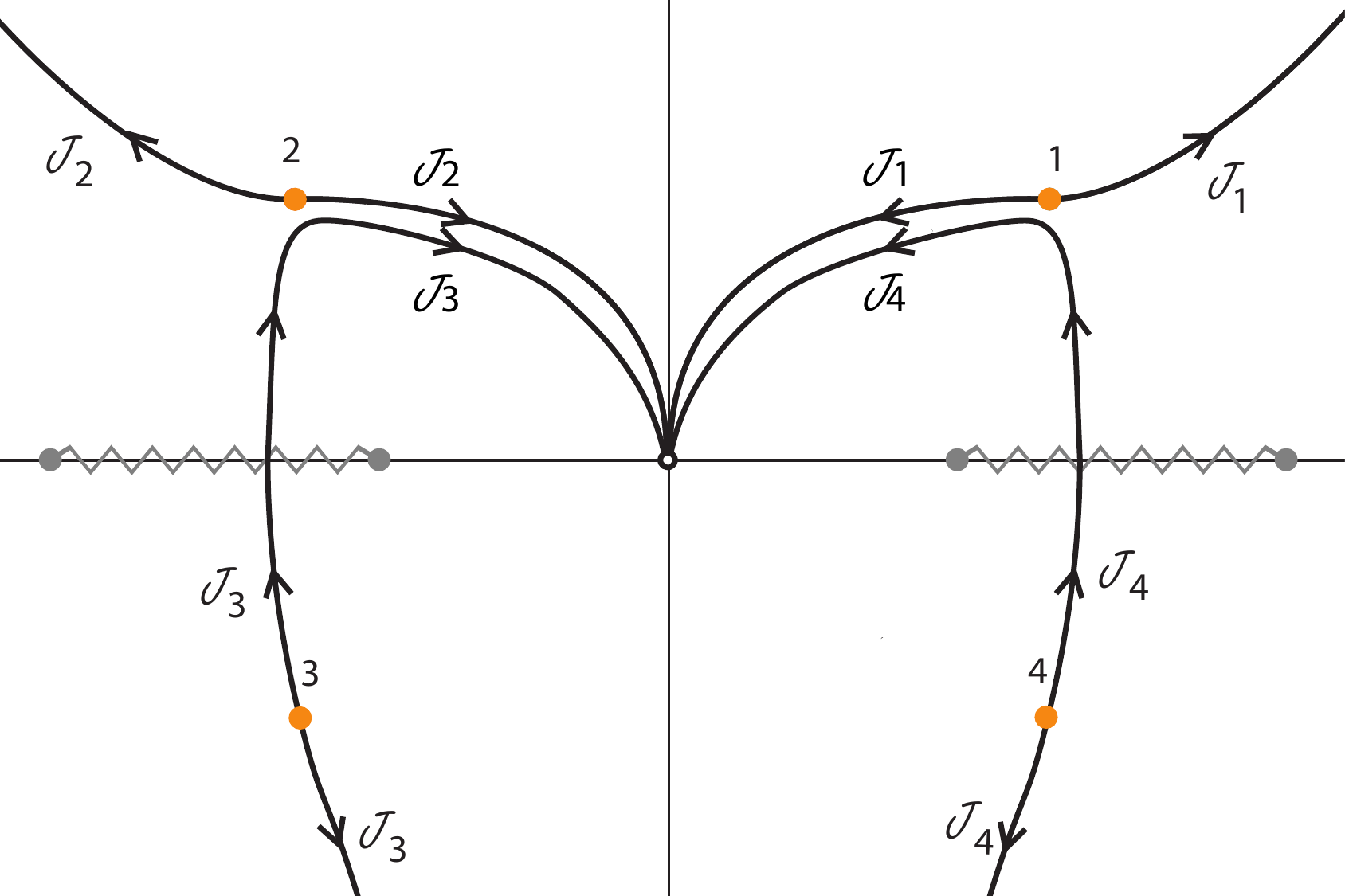}
\caption{The Picard-Lefschetz thimbles (solid dark lines, labelled $J_1-J_4$) for our problem, in the complex $N$-plane. Any convergent integral starting and ending at a singularity of the Morse function can be represented as a sum of these thimbles. Also shown are the branch cuts (jagged grey lines) which the no boundary perturbations introduce on the real $N$-axis. }
\label{fig:Map}
\end{figure}

Let us begin by ignoring the branch cuts on the real $N$-axis. Then it is a basic result of Picard-Lefschetz theory that any contour ${\cal C}$ running between singular points of the Morse function and yielding a finite integral may be deformed, using Cauchy's theorem, into a sum of Picard-Lefschetz ``thimbles".  In order to remove any degeneracies between thimbles (see Section II of Ref.~\cite{Feldbrugge:2017kzv}), we introduce a small deformation so that thimble 3 is completed by passing just below the steepest descent curve connecting saddle point 2 to the origin, and similarly for thimble 4 and thimble 1. These thimbles are illustrated in Fig.~\ref{fig:Map}. Any contour ${\cal C}$ connecting singularities of the Morse function, as explained above, may now be deformed into a sum of thimbles, taken with appropriate signs.  Notice that, with this definition of the thimbles, in the limit that the deformation is taken to zero, thimble 4 and thimble 1 share that part of thimble 1 which connects saddle point 1 to the origin. And the thimble 3 shares a similar part of thimble 2. In this way {\it every thimble includes a contribution from saddle point 1 or 2}, both of which yield unsuppressed perturbations. This is the first part of the proof. 

For example, our Lorentzian propagator is given by thimble 1. Its real part is given by the sum of thimble 1 and thimble 2, both taken `left to right,' and its imaginary part is proportional to the difference. The contour of Diaz Dorronsoro {\it et al.}  is deformable to thimbles 1,2,3 and 4 taken with signs such that the steepest descent contours connecting saddle points 1 and 2 to the origin are cancelled. But the steepest descent contours connecting saddle points 1 and 2 to infinity are included. Other contours include thimbles 3 or 4, either taken alone or combined into a contour which `runs around' the origin, starting from negative imaginary values. Likewise we can combine thimble 3 and thimble 2, taken with a sign such that the steepest descent contour connecting saddle point 2 to the origin is cancelled. And so on.

Now let us include the two branch cuts on the real $N$-axis. Clearly, their introduction affects thimbles 3 and 4. Any contour which gives contributions from saddles 3 or 4 -- the Hartle-Hawking saddles -- as, for example, that of Diaz Dorronsoro {\it et al.}, necessarily involves contributions from either or both branch cuts, which also give unsuppressed perturbations. 

Finally, we note in passing that in general the singularity at $N=0$ may also be the terminus of a branch cut extending to infinity. In pure de Sitter, the path integral over the background yields a square root of $N$ and the perturbations, as noted above,  yield a functional determinant. So one can consider a nontrivial contour for the integral over $N$ coming in from infinite negative imaginary values and completely encircling this branch cut. It will be deformable to a combination of thimbles 1,2, 3 and 4.  Again, it necessarily acquires contributions from both saddles 1 and 2, and the branch cuts. This completes our proof that no choice of contour for $N$ avoids the problem of unsuppressed perturbations. 

\section{Discussion}

The introduction of a new, more precise formulation of semi-classical quantum gravity should, we believe, have wide implications. 

Among the questions which now appear accessible is whether or not there is a quantum version of de Sitter spacetime. In Refs.~\cite{Polyakov:2007mm,Polyakov:2012uc}, Polyakov has expressed doubts, claiming in particular that there is strong interplay between IR and UV effects leading to various divergences. While there are considerable differences between our approaches,
his words certainly resonate with our findings. Namely, our predicted distribution of the tensor modes diverges, a problem which worsens in the UV, as a result of a nonlinear (and completely nonlocal) interplay between these modes and the quantized background, {\it i.e.}, the IR.  Because the background has the ``wrong sign" kinetic term, convergence of the Feynman path integral over backgrounds in effect chooses the ``wrong sign" Wick rotation for the perturbations, giving them an inverse Gaussian distribution and implying they are completely out of control. The general theorem we proved in Ref.~\cite{Feldbrugge:2017fcc} identified the fundamental topological nature of this problem. 

Our work indicates that a ``no boundary" formulation of quantum de Sitter spacetime does not exist. However, the question remains whether there is any other viable formulation. In particular, one might try to define it by considering the Lorentzian propagator between two real classical three-geometries, both large and roughly spherical, so that there would be two possible intervening classical four-geometries, one a ``bounce" including the throat of de Sitter and the other not including the throat. In the former case, one may have to worry about strong nonlinear backreaction, in that the quantum mechanical perturbations present in the contracting phase of global de Sitter spacetime blueshift as the size of the three-sphere shrinks, potentially causing a ``big crunch" at the throat and terminating the semiclassical description. The methods reported in this paper, especially concerning semiclassical backreaction, seem ideal for approaching these questions. In particular, we are very interested in understanding implications for the future quantum evolution of our universe, which now appears to be entering a potentially eternal de Sitter phase. Is it possible to describe such a phase semiclassically? Is such a phase ultimately stable against quantum mechanical perturbations? Finally, given that our universe emerged from the big bang filled with radiation and matter, how does the presence of these other forms of energy affect the quantum mechanical behaviour of the background and the perturbations? The work of Refs.~\cite{Gielen:2015uaa,Gielen:2016fdb} on a radiation-dominated ``perfect bounce" indicated no problems with unsuppressed perturbations, such as those identified here. We are therefore keen to explore further the possible advantages of such a scenario. 

Second, our work may well have implications for cosmological inflation. In the usual descriptions, one treats the background classically, and quantizes only the perturbations. However, as is well known, it is not possible to consistently couple classical degrees of freedom with quantum degrees of freedom, so this usual description can at best be only an approximation. Second, the usual treatment of inflation {\it assumes} the quantum mechanical perturbations start out in their local adiabatic vacuum -- the so-called Bunch-Davies state. Here too, this is at best an approximation since it applies only to short wavelength modes. There is the additional problem that such modes are generically sub-Planckian deep in the inflationary era, so without a clear treatment of quantum gravitational phenomena, they cannot be precisely specified. 

The promise of the no boundary proposal was to present a ``completion" of the inflationary scenario, along with the hope that all these questions could be resolved within a low-energy, effective description of Einstein gravity coupled to quantum fields. We believe that our work has now excluded that option. Therefore, it raises fundamental questions for inflation: How is the theory to be completed? How does inflation avoid non-perturbative corrections to the Bunch-Davies vacuum, of the type we have shown to exist for the contour proposed by Diaz Dorronsoro {\it et al.} in their attempt to rescue Hartle and Hawking's  proposal?

Although the main outcome of this work has so far been negative, we find it very exciting that we can at last formulate semi-classical quantum cosmology  in a precise enough way to identify clear problems. In our view, the bigger the problem, the more instructive the clue it provides. It is particularly surprising to find that quantum effects on the universe on large scales can be so significant and can so drastically influence the description of phenomena on the smallest scales, as we found. One cannot help but hope that herein lies a clue to understanding the dark energy and resolving the biggest ``fine tuning" puzzle in physics. Quantum field theory tells us that the dark energy is dominated by the UV vacuum energy, yet it is also what ultimately sets the IR cutoff by limiting our causal horizon and thus the largest scale anyone shall ever see. Again, our work provides a clue as to how the UV and the IR are connected in quantum gravity, in a way we have yet to fully unravel. 

Finally, we mentioned in the opening of this paper that the no boundary concept has been very fruitful in mathematical physics, in contexts such as holography, as well as conformal and topological field theory. We believe our more precise formulation should also be useful in these contexts.


\acknowledgements
We thank Claudio Bunster, Alice Di Tucci, Juan Diaz Dorronsoro,  Laurent Freidel, Gary Gibbons, Jaume Gomis, Jonathan Halliwell, Jim Hartle, Stephen Hawking, Thomas Hertog, Ted Jacobson, Oliver Janssen, Rafael Sorkin, Sumati Surya and Bill Unruh, as well as the participants of the workshop ``Bounce Scenarios in Cosmology" for stimulating discussions. JLL would like to thank Perimeter Institute for warm hospitality while this work was completed. Research at Perimeter Institute is supported by the Government of Canada through Industry Canada and by the Province of Ontario through the Ministry of Research and Innovation.

\appendix
\section*{Appendix A: Illustration of higher dimensional Picard-Lefschetz theory}

In this appendix, we use a simple two-dimensional, conditionally convergent, oscillatory integral to illustrate some features of Picard-Lefschetz theory. The integral may be computed iteratively in two ways: by integrating first over one variable and then the other, or vice versa. In the former case, as we shall see, we obtain a one dimensional conditionally convergent integral with a nonsingular integrand. In the latter case, the first integration yields an integrand which is singular on the original contour for the second variable. In order to give the integral meaning we must exclude the singular point or, equivalently, deform the original contour to avoid the singularity. An alternative to making sense of the original integral is to use higher dimensional Picard-Lefschetz theory and Cauchy's theorem to distort the two original integration contours so that the two-dimensional integral becomes absolutely convergent {\it before} either integral is performed. As discussed in Section II, this ensures that the result of the integral is independent of the order in which the two integrals are then taken. More than this, as we shall see, distorting the contours to a two-dimensional steepest descent surface ensures that no one dimensional integral generates a singularity which might then complicate subsequent integrals.  

We consider the integral
\begin{align} \label{a1}
I=\int_{-\infty}^{\infty} dN \int_{-\infty}^{\infty} dz \, e^{i (N-1)z^2 +i N^2},
\end{align}
where both $N$ and $z$ are taken over all real values. 
To begin with, we compute $I$ by integrating first over $N$. This is a simple Gaussian integral which yields
\begin{align} \label{a2}
I=e^{i{\pi/4}} \sqrt{\pi} \int_{-\infty}^{\infty} dz \, e^{-i (z^2+{1\over 4} z^4) } ={1\over 2} e^{i({1\over 2}-{\pi\over 8})} \pi^{3\over 2} H_{-{1\over 4}}^{(2)}\left({1\over 2}\right),
\end{align}
the last being a standard result. 
Alternatively, we may compute $I$ by integrating first over $z$. But there is an immediate problem: the $z$ integral, taken along the real $z$-axis, fails to converge at $N=1$. 
So we can exclude the point at $N=1$, which is of zero measure, to obtain
 \begin{align} \label{a3}
I=e^{i{\pi/4}} \sqrt{\pi} \int_{1}^{\infty} {dN\over \sqrt{N-1}} e^{i N^2}+ e^{-i{\pi/4}} \sqrt{\pi} \int_{-\infty}^{1} {dN\over \sqrt{1-N}} e^{i N^2}.
\end{align}
Equivalently, we may define the integral over a continuous contour for $N$ following the real axis from $-\infty$ to $\infty$ but avoiding $N=1$ on an infinitesimal semicircle passing above it (only if we pass above is the $z$ integral in (\ref{a1}) convergent). In the limit that the semicircle radius is taken to zero, we also obtain (\ref{a3}).
One can readily check that (\ref{a3}) agrees with (\ref{a2}). 

While both derivations are correct, the second one is complicated by the occurrence of the singularity, requiring a contour for $N$ which is either discontinuous or moves off the real axis. A similar phenomenon occurs in our more complicated path integral examples where the integration over the perturbations generates singularities in the resulting effective action for the lapse $N$ which likewise occur on the original, defining integration contour. In that case, the singularities are more severe, occurring in the {\it exponent} of the integrand, as explained in the Introduction. In both cases, however, if we wish to calculate the integral along the original $N$ contour having performed partial integrations, care is needed to determine exactly how the contour should avoid the singularities. 

Let us now see how Picard-Lefschetz theory enables one to deform the integration contour to steepest descent contours {\it before} either integral is performed, in order to avoid such problems. As we shall see in this example, this avoids singularities of the type just discussed being generated, leaving one with a completely unproblematical absolutely convergent integral which can be calculated iteratively, in any convenient order. Examining (\ref{a1}), we see that we can make the integral over $N$ convergent by rotating the $N$ contour, setting $N=e^{i {\pi\over 4}} \sqrt{2} n = (1+i) n$, with $n$ real (the factor of $\sqrt{2}$ is merely to simplify the algebra). Note that this contour for $N$ never passes through the point $N=1$. Turning now to the $z$ integral, we set $z=x+i y$ and find the steepest descent contour $y(x)$ passing through the saddle point at $z=0$. As explained, {\it e.g.}, in Section II of Ref.~\cite{Feldbrugge:2017kzv}, the imaginary part of the exponent is constant along this contour. This immediately yields the equation $(n-1)(x^2-y^2)-2 n x y =0$. One of the two solutions gives the steepest descent contour, in which the real part of the exponent (the Morse function $h$) is monotonically decreasing away from the saddle (the other solution, which we ignore, gives the steepest ascent contour). Setting $n=\nu+{1\over 2}$, the steepest descent contour is
\begin{align} \label{a4}
y={1+2\nu -\sqrt{2+8 \nu^2}\over 1-2\nu} x
\end{align}
and the Morse function on this contour is
 \begin{align} \label{a5}
h&=-2(1+4 \nu^2){\sqrt{2+8\nu^2} -1 -2 \nu\over (1-2\nu)^2}x^2 - 2 \left(\nu+{1\over 2}\right)^2.
\end{align}
Notice that the slope of the contour in the complex $z$-plane changes from a negative to a positive value as $\nu$ (or $n$) runs from $-\infty$ to $\infty$, so that the coefficient of $x^2$ is always negative, whatever the (real) value of $\nu$.

Having found the steepest descent contour in the complex $z$-plane, we may now integrate over $z$, along the steepest descent contour (\ref{a4}), {\it without} generating any singularity in $\nu$:
\begin{align} \label{a6}
I&=\sqrt{2}e^{i {\pi\over 4}}\int_{-\infty}^{\infty}  dx\left(1+i {dy\over dx}\right) \int_{-\infty}^{\infty}  d\nu e^{-2[(1+4 \nu^2){\sqrt{2+8\nu^2} -1 -2 \nu\over (1-2\nu)^2}x^2 +(\nu+{1\over 2})^2]}\cr
&=\sqrt{\pi}e^{i {\pi\over 4}} \int_{-\infty}^{\infty}  d\nu \left(1+i{1+2\nu -\sqrt{2+8 \nu^2}\over 1-2\nu} \right)
{|1-2 \nu|\over \sqrt{1+4 \nu^2} \sqrt{\sqrt{2+8\nu^2} -1 -2 \nu} } e^{- 2 \left(\nu+{1\over 2}\right)^2}.
\end{align}
While this expression is hardly elegant, it is completely nonsingular along the integration contour for $\nu$. It is easy to integrate numerically, for example, and yields a value identical to that of (\ref{a2}). 

The key point illustrated by this example is that when we use higher-dimensional Picard-Lefschetz theory to convert the original integral, which is only conditionally convergent, into one which is absolutely convergent, partial integrals do not typically generate any singularities. However, if we instead leave $N$  fixed while performing the $z$ integral, we obtain singularities on the real $N$-axis. Care is then needed to determine how the $N$ integration contour should be taken around them. 

\appendix
\section*{Appendix B: Proof that $\Re[\gamma]>0$ almost everywhere in the cut $N$-plane}

A crucial role in our analysis is played by the constant $\gamma$ describing the behavior of the perturbation modes near $t=0$. Near $t=0$, we have $\phi\sim t^{{\gamma-1\over 2}}$ and the Lagrangian density ${\cal L}(t) \sim t^{\gamma -1}$. So, if  the real part of $\gamma$ is positive, the singularity at $t=0$ is integrable. We shall now prove that $\Re[\gamma]>0$ almost everywhere in the complex $N$-plane, the exception being the closed real intervals $N_-\leq N\leq N_+$ and $-N_+ \leq N \leq -N_-$, where $N_-$ and $N_+$ are positive constants, defined in (\ref{eq:npm}).  

In Section \ref{sec:Pert}, we show that
\begin{align} \label{b1}
\gamma=\sqrt{N_-^2-N^2\over N_\star^2-N^2} \sqrt{N_+^2-N^2\over N_\star^2-N^2}\equiv \sqrt{Z_--Z\over Z_\star-Z} \sqrt{Z_+-Z \over Z_\star-Z},
\end{align}
where $N_-<N_\star<N_+$, and we define $Z=N^2$. Consider the first factor. It is the square root of a M\"obius map, namely a one-to-one mapping of the complex $Z$-plane onto the complex $Y$-plane, where $Y=(Z_--Z)/(Z_\star-Z)$. This map takes the real interval $Z_-<Z<Z_\star$ onto the negative real $Y$-axis, the upper half $Z$-plane to the lower half $Y$-plane and vice versa. Setting $S=\sqrt{Y}$ takes the entire $Z$-plane to the right half of the $S$-plane, {\it i.e.}, $\Re[S]\geq 0$, with the upper half $Z$-plane being mapped to the lower right quadrant of the $S$-plane, and the lower half $Z$-plane to the upper right quadrant of the $Z$-plane. Except for the closed interval $Z_-\leq Z\leq Z_\star$, which is mapped to the imaginary $S$-axis, every point in the complex $Z$-plane is mapped to a point with $\Re[S]>0$, with the sign of $\Im[S]$ being opposite to the sign of $\Im[Z]$. Now consider the second factor, written as the complex number $T$.  Similar arguments show that, except for the real interval $Z_\star \leq Z \leq Z_+$, every point in the complex $Z$-plane is mapped to $\Re[T]>0$, with the sign of $\Im[T]$ being the same as the sign of $\Im[Z]$. Therefore $\Re[ST]=\Re[S]\Re[T]-\Im[S]\Im[T]>0$ except on $Z_-\leq Z\leq Z_+$, where $ST$ is pure imaginary. This proves our claim. 

\appendix
\section*{Appendix C: No boundary amplitude for perturbed $S^3$ with $\Lambda=0$}

In subsection D of the Introduction, we discussed a very simple example of a ``no boundary" gravitational path integral. Namely, we set $\Lambda=0$ in the action (\ref{bact}) governing the background and impose a final three-geometry consisting of a sphere of radius squared $q_1$. The background solution is $q = q_1t$, and the classical background action is $2 \pi^2(-3 q_1^2/(4 N) +3 N)$. This has two saddle points, at $N=N_s^{\pm}=\pm i q_1/2$. This means that the saddle points are actually on the imaginary axis. For either saddle, the line element is $q_1(dt^2/(4t)+t d\Omega_3^2)=q_1(dr^2+r^2d\Omega_3^2)$, where $r\equiv \sqrt{t}$, {\it i.e.}, just that for a flat Euclidean ball of radius $R\equiv \sqrt{q_1}$. 

Thus the saddle in the upper (lower) half plane has semiclassical exponent $-6 \pi^2 R^2/\hbar$ (or $+6 \pi^2 R^2/\hbar$) respectively, where $R$ is the radius of the ball in reduced Planck units. By studying the flow, one can easily see that the upper saddle is on a Picard-Lefschetz thimble connecting the origin $N=0$ (approached from above)  to $N=+i \infty$. Therefore this saddle point is relevant to the {\it Euclidean} path integral for quantum gravity, in the absence of a cosmological constant. The reason the Euclidean path integral is meaningful is that if $\Lambda=0$, the positive imaginary $N$-axis is in fact a direction of steepest descent. Note, however, that since the action is odd in $N$ one can only take $N$ to run over the positive imaginary axis, not the entire imaginary axis.  The same saddle is {\it also} relevant to the {\it Lorentzian} path integral, with a defining contour $0^+<N<\infty$, which also makes sense. The steepest ascent flow from the saddle is easily shown to be a circle: setting $N=x+i y$, the flow is $y^2={1\over 4} q_1^2-x^2$. So the steepest ascent contour meets the real $N$-axis at $N=q_1/2$, where the background classical action is real.

Returning to our example of n subsection D of the Introduction, we analyzed the perturbation equations of motion. From the discussion below equation (\ref{schop}), it follows that the scaling exponent governing the perturbation $\gamma=l+1$, so that the tensor modes are proportional to $t^{l/2}=r^{l}$. This is exactly the right scaling with $r$ to make them analytic at $r=0$. (If a tensor quantity has $l$ indices and is expressible in terms of a tensor product of $l$ Cartesian coordinates, it must necessarily scale as $r^l$.)  The perturbation exponent is easily calculated from equation (\ref{schop}). 

Putting everything together, we find the causal propagator to create a perturbed three-sphere of radius $R$ in reduced Planck units ``from nothing," when $\Lambda=0$, is
\begin{equation}
G^{\Lambda=0}[q_1,\phi_1;0]\sim e^{(-6 \pi^2 + l \phi_1^2/2 )R^2/\hbar},
\end{equation} 
up to prefactors. As discussed in the Introduction, performing the semiclassical path integral over the perturbations generates a branch cut on the real $N$-axis, over $q_1/(2\sqrt{l(l+2)}<N<\infty$ (and similarly for negative $N$). On the upper side of this branch cut, $\gamma=-i \sqrt{4 l(l+2) N^2/q_1^2-1}$ and the real part of the semiclassical exponent (the Morse function) is positive: one finds $\Re[iS^{(2)}(N)/\hbar]=  {1\over 2} (q_1 \phi_1^2/\hbar) \sqrt{l(l+2)-4}$ at $N=q_1/2$, where the steepest ascent contour from the background saddle meets the $N$-axis. This is greater than the Morse function from perturbations at the saddle, for all $\phi_1$ and $l$. As in the de Sitter example, we see that when we integrate over the perturbations the net semiclassical exponent can become positive, at least when backreaction is ignored. Based on the results of Section V, we do not expect the inclusion of nonlinear backreaction to change this conclusion.

The fact that this calculation, like that for de Sitter, yields an unacceptable inverse Gaussian distribution for the perturbations implies, we believe, that one should not consider ``no boundary" amplitudes of this type, even for $\Lambda=0$. 

\appendix
\section*{Appendix D: Behavior of solutions and classical action as $N\rightarrow N_\star$}


At $N_\star=\sqrt{3 q_1/\Lambda}$ the classical background solution (\ref{eq:qclass}) takes the simple form $\bar{q}(t)=q_1 t^2$. The equation of motion of the canonically normalised perturbation $\chi\equiv \bar{q}(t)\phi$, equation (\ref{eom4chi}), has two solutions
\begin{equation}
\chi_{\star \pm}=  t\left(t \mp  iN_\star\sqrt{l(l+2)}/q_1 \right) e^{\pm iN_\star\sqrt{l(l+2)}/(q_1t)}.
\end{equation} 
The choice of mode is fixed by the finite action condition and depends on how we approach the point $N_\star$.  When approached from above, the finite action solution is $\phi(t)=\phi_1 \chi_{\star +}(t)/ (t^2 \chi_{\star+}(1))$. The corresponding classical action is 
\begin{equation}
\bar{S}^{(2)}(N_\star) = \frac{\bar{q}^2 \bar{\phi} \dot{\bar{\phi}}}{2N} \bigg|_{t=0}^1 = - \frac{i N_\star}{i+\sqrt{l(l+2)} \, N_\star/q_1} \frac{l(l+2)}{2}  \phi_1^2\,.
\label{actnst}
\end{equation}
When approached from below, the perturbation and action are given by the complex conjugate expressions. In both cases, the action agrees with $\bar{S}^{(2)}[q_1;\phi_1;N]$, given in equation \eqref{classact}, evaluated just above or below $N=N_\star$ in the complex $N$-plane, taking into account the analytic properties of $\gamma(N)$ explained in Appendix B. 
\bibliographystyle{utphys}
\providecommand{\href}[2]{#2}\begingroup\raggedright\endgroup

\end{document}